\documentclass[a4paper,11pt]{article}
\pdfoutput=1
\usepackage{jheppub}
\usepackage{amsmath}
\usepackage{amssymb}
\usepackage{dcolumn}	
\usepackage{bm}			
\usepackage{bbm} 		
\usepackage{multirow}
\usepackage{blkarray}
\usepackage{slashed}
\usepackage{bbding}
\usepackage{pifont}
\usepackage[usenames,dvipsnames]{xcolor}
\definecolor{hgreen}{rgb}{0,.3,0}
\definecolor{hred}{rgb}{.3,0,0}
\definecolor{hblue}{rgb}{0,0,.3}
\definecolor{LightGray}{gray}{0.95}
\makeatletter
\def\endfmffile{%
	\fmfcmd{\p@rcent\space the end.^^J%
		end.^^J%
		endinput;}%
	\if@fmfio
	\immediate\closeout\@outfmf
	\fi
	\ifnum\pdfshellescape>\z@
	\immediate\write18{mpost \thefmffile}%
	\fi}
\makeatother

\usepackage{pbox}
\usepackage{caption}
\usepackage{subcaption}
\usepackage{xcolor}
\usepackage[utf8]{inputenc}

\newcommand{\pid}{\hat{\pi}}
\newcommand{\fpid}{f_{\pid}}

\newcommand{\gd}{\hat{g}}

\newcommand{\bomega}{\bm{\omega}}
\newcommand{\bmass}{\bm{M}}
\newcommand{\byuk}{\bm{Y}}
\newcommand{\byukt}{\widetilde{\bm{Y}}}
\newcommand{\bA}{\bm{A}}
\newcommand{\bAt}{\widetilde{\bm{A}}}
\newcommand{\bmpsi}{\bm{m}_\psi}
\begin{document}

\preprint{CERN-TH-2021-150}

\title{A Theory of Dark Pions}

\author[a]{Hsin-Chia Cheng,}
\affiliation[a]{Center for Quantum Mathematics and Physics (QMAP), Department of Physics,\\ University of California, Davis, USA}
\author[b,c]{Lingfeng Li,}
\affiliation[b]{\mbox{Jockey Club Institute for Advanced Study, Hong Kong University of Science and Technology, Hong Kong}}
\affiliation[c]{\mbox{Department of Physics and Brown Theoretical Physics Center, Brown University, Providence, USA}}
\author[d,e]{Ennio Salvioni,}
\affiliation[d]{Theoretical Physics Department, CERN, Geneva, Switzerland}
\affiliation[e]{Dipartimento di Fisica e Astronomia, Universit\`a di Padova, Italy}
\date{\today}
\abstract{
We present a complete model of a dark QCD sector with light dark pions, broadly motivated by hidden naturalness arguments. The dark quarks couple to the Standard Model via irrelevant $Z$- and Higgs-portal operators, which encode the low-energy effects of TeV-scale fermions interacting through Yukawa couplings with the Higgs field. The dark pions, depending on their $CP$ properties, behave as either composite axion-like particles (ALPs) mixing with the $Z$ or scalars mixing with the Higgs. The dark pion lifetimes fall naturally in the most interesting region for present and proposed searches for long-lived particles, at the LHC and beyond. This is demonstrated by studying in detail three benchmark scenarios for the symmetries and structure of the theory. Within a coherent framework, we analyze and compare the GeV-scale signatures of flavor-changing meson decays to dark pions, the weak-scale decays of $Z$ and Higgs bosons to hidden hadrons, and the TeV-scale signals of the ultraviolet theory. New constraints are derived from $B$ decays at CMS and from $Z$-initiated dark showers at LHCb, focusing on the displaced dimuon signature. We also emphasize the strong potential sensitivity of ATLAS and CMS to dark shower signals with large multiplicities and long lifetimes of the dark pions. As a key part of our phenomenological study, we perform a new data-driven calculation of the decays of a light ALP to exclusive hadronic Standard Model final states. The results are provided in a general form, applicable to any model with arbitrary flavor-diagonal couplings of the ALP to fermions.
}

\maketitle


\section{Introduction and the model\label{sec:intro}}

A light, confining hidden sector coupled feebly to the Standard Model (SM) is in general an interesting possibility for new physics, often referred to as a hidden valley (HV)~\cite{Strassler:2006im}. More sharply, it can be part of the answers to outstanding questions of the SM. The (little) hierarchy problem may be solved by models of neutral naturalness~\cite{Chacko:2005pe,Burdman:2006tz,Cai:2008au}, where the partners of the top quark are not charged under SM color but a dark color symmetry, and dark confinement around the GeV scale is a generic prediction~\cite{Craig:2015pha}. If the hierarchy problem is solved by cosmological relaxation, a confining hidden sector may be the origin of the backreaction potential that stops the relaxion~\cite{Graham:2015cka}. Dark strong dynamics can also provide attractive scenarios for dark matter, with several plausible candidates found among the dark hadrons. 

If at least some of the dark hadrons decay to SM particles, the feeble coupling connecting the hidden and visible sectors generally implies macroscopic lifetimes. Thus, hallmark signatures of HV models are given by long-lived particles (LLPs), which have been a topic of rapidly increasing interest at the Large Hadron Collider (LHC)~\cite{Alimena:2019zri} and beyond~\cite{Beacham:2019nyx}. At the LHC, searches for LLPs hold a strong discovery potential, provided that dedicated and innovative strategies can be implemented at the level of event selection and analysis. This is especially true for ``dark jet'' or ``dark shower'' topologies, where the decay of a heavy particle (such as a $Z$ or Higgs boson) to the hidden sector produces jets made of light dark hadrons. The associated phenomenology deserves further attention.\footnote{A very recent appraisal can be found in Ref.~\cite{Knapen:2021eip}.} Areas where important progress is needed include maximizing the dark shower coverage of existing detectors (primarily ATLAS, CMS and LHCb), understanding the interplay with low-energy production processes such as flavor-changing neutral current (FCNC) meson decays, as well as comparing to the sensitivity of proposed LLP-specific experiments. In this paper we study a model of dark QCD with light pseudo Nambu-Goldstone bosons (pNGBs), namely dark pions, coupled to the SM through irrelevant $Z$ and Higgs portals. This {\it theory of dark pions} provides a new coherent framework to address the above questions.

The low-energy spectrum of the hidden (or dark) sector depends on the number of light quark flavors, $N$, charged under the dark QCD (assumed to have $SU(N_d)$ as the color group) and having masses below the strong scale $\Lambda$. If $N = 0$, dark glueballs are at the bottom of the spectrum~\cite{Juknevich:2009ji}. An example is the Fraternal Twin Higgs model~\cite{Craig:2015pha}, where the lightest dark glueball is expected to mix with the Higgs boson, giving rise to phenomenology that has been extensively studied~\cite{Juknevich:2009gg,Craig:2015pha}. For $N = 1$, the low-energy spectrum contains several mesons with masses around $\Lambda$, the lightest being a ($\eta'$-like) pseudoscalar, a vector, and a scalar~\cite{Farchioni:2007dw}. For instance, this scenario has been thoroughly analyzed in a realization of the tripled top framework for supersymmetric neutral naturalness~\cite{Cheng:2018gvu} that features electroweak-charged top partners, where the dark mesons mix dominantly with the $Z$ boson~\cite{Cheng:2019yai}. When $N \geq 2$ (but still below the conformal window, $N \lesssim 4N_d$), one expects chiral symmetry breaking and $N^2 - 1$ associated pNGBs, which in a slight abuse of notation we call {\it dark pions}, $\hat{\pi}$, for any $N$. As familiar from the SM, the dark pions can be much lighter than the rest of the hadrons, whose masses are at or above the dark QCD scale: $m_{\hat{\pi}} \ll \Lambda$. Here we focus on the multi-flavor case. The lifetimes of the dark pions depend on the amount and pattern of explicit isospin breaking, yielding a larger parameter space to explore compared to the one-flavor theory. With respect to the latter, notable differences are that dark meson production and decay are less tightly connected, and dark vector mesons dominantly decay to dark pions if the phase space is open, whereas for $N = 1$ they decay to SM particles.

The light dark quarks must be singlets under the SM if their masses are below $O(100)$ GeV, to satisfy collider bounds. The interactions connecting them to the visible sector dictate the phenomenology. Here we focus on the interesting possibility of irrelevant portals obtained by integrating out heavy states~\cite{Han:2007ae,Schwaller:2015gea,Alekhin:2015byh,Freytsis:2016dgf,Renner:2018fhh,Cheng:2019yai,Contino:2020tix}, which have been less studied compared to renormalizable ones, even though they have solid theoretical motivations. A concrete example is given by the scenario of Ref.~\cite{Cheng:2019yai}, where the mediation is provided by heavy fermions charged under both the SM electroweak (EW) and hidden color gauge symmetries, allowing for renormalizable Yukawa interactions between dark-colored quarks and the SM Higgs doublet. The supersymmetric partners of the heavy fermions play the role of scalar top partners, hence the mediation scale is naturally around TeV.

{\bf The model.} Drawing from the above discussion, the theory of dark pions considered in this work contains $N > 1$ flavors of Dirac fermions $\psi_i$, transforming in the fundamental representation of the dark color $SU(N_d)$, but singlets under all SM gauge symmetries. In addition, $N$ EW-doublet Dirac fermions $Q_i = (Q_u \;Q_d)^T_i$ with hypercharge $1/2$ are included, which also transform in the fundamental representation of $SU(N_d)$. This field content allows for Yukawa couplings involving the SM Higgs doublet,
\begin{equation} \label{eq:LUV}
- \mathcal{L}_{\rm UV} = \overline{Q}_L  \byuk \psi_R  {H} + \overline{Q}_R \byukt\psi_L {H} + \overline{Q}_L  \bmass Q_R +  \overline{\psi}_L \bomega \psi_R + \text{h.c.}\,,
\end{equation}
where $\byuk$, $\byukt$, $\bmass$, and $\bomega$ are $N\times N$ matrices in flavor space. The mass matrices $\bmass$ and $\bomega$ can be diagonalized with real and positive diagonal elements by separate unitary transformations on the $Q_{L,R}$ and $\psi_{L,R}$ fields, respectively, so we assume this form without loss of generality. The coupling matrices $\byuk,\, \byukt$ can be complex in general, with $N^2 + (N - 1)^2$ independent phases, decreasing to $(N - 1)^2$ if one of $\byuk, \byukt,$ or $\bomega$ vanishes. In addition there is always the strong $CP$ phase of dark QCD, which will be consistently set to zero in this work. The masses $M$ of the heavy dark quarks are taken to be larger than $\Lambda$ (and around TeV). The masses of the light dark quarks, which receive independent contributions of order $\omega$ and $Y \widetilde{Y} v^2/M$, where $v$ is the Higgs vacuum expectation value (VEV), are assumed to be much smaller than $\Lambda$. Hence, the dark QCD has $N$ light flavors. If $N \lesssim 4N_d$~\cite{Appelquist:1996dq}, at low energies the light quarks are confined and form a condensate. The $SU(N)_L \times SU(N)_R$ chiral symmetry is spontaneously broken to the diagonal $SU(N)_V$, resulting in $N^2-1$ pNGB dark pions.

The $\byuk = y_t \bm{1}$, $\bm{\widetilde{Y}} = 0$ limit of Eq.~\eqref{eq:LUV}, together with appropriate TeV-scale supersymmetry breaking, embodies a solution to the little hierarchy problem \`a la tripled top~\cite{Cheng:2018gvu,Cheng:2019yai}.\footnote{In the tripled top model there are two $SU(3)$ dark color gauge groups, each with one flavor of SM-singlet fermions which can be very light~\cite{Cheng:2019yai}. However, variations where the dark QCD has more than one light flavors are straightforward to construct, for example, by identifying the two dark color groups.} This sets a well-motivated target for the chiral structure and mediation strength, with $Y \sim y_t \sim 1$ and $M \sim \mathrm{TeV}$. On the other hand, the same Lagrangian ~\eqref{eq:LUV} was employed in the relaxion solution to the hierarchy problem~\cite{Graham:2015cka},\footnote{See also later studies of the $\omega, M < \Lambda$~\cite{Antipin:2015jia} and $\omega, Y \widetilde{Y}v^2/M < \Lambda < M$~\cite{Beauchesne:2017ukw} scenarios.} to generate a backreaction potential without running into difficulties with the SM strong $CP$ problem. In both cases $N = 1$ was originally chosen for the sake of minimality, but not necessity. The fact that this setup emerges naturally in very different approaches to the Higgs naturalness puzzle makes it a compelling choice for a benchmark theory of dark pions.

The properties of the individual dark pions depend on the symmetries and structure of the dark sector. If the $CP$ symmetry is preserved, the dark pions are classified into odd and even states: for example, with $N = 2$ the $\pid_1$ and $\pid_3$ are $CP$-odd $(J^{PC} = 0^{-+})$ while the $\pid_2$ is $CP$-even $(0^{--})$, where the index corresponds to $SU(2)$ generators. Therefore, this theory provides a coherent framework to study both $CP$-odd and -even light scalars feebly coupled to the SM. The $CP$-odd dark pions decay to SM particles through the $Z$ portal, i.e., by mixing with the longitudinal component of the $Z$ boson through dimension-$6$ operators. They behave as ALPs with an effective decay constant parametrically given by 
\begin{equation}\label{eq:ALP_sketch}
f_a \sim \min\left\{ \frac{M^2}{Y^2 \fpid}\,,\, \frac{M^2}{ \widetilde{Y}^2 \fpid} \right\}\, ,
\end{equation} 
where $\fpid$ is the dark pion decay constant, defined in analogy with the SM pion decay constant $f_{\pi} \approx 93\;\mathrm{MeV}$. The precise form of Eq.~\eqref{eq:ALP_sketch} is derived later, but one can already see that for $Y \sim 1$, $M\sim$~TeV and $\fpid \sim$~GeV, the $CP$-odd dark pions have $f_a \sim$~PeV. This highlights how the ALP decay constant does not necessarily correspond to a physical scale (no threshold exists near the PeV in our model), but is a combination of parameters of the underlying theory if the ALP is composite. The $CP$-even states decay to SM particles through the Higgs portal, i.e., by mixing with the Higgs boson through dimension-$5$ operators. As we show in detail later, the mixing angle is parametrically
\begin{equation} \label{eq:scalar_sketch}
\sin \theta \sim \frac{2\pi \fpid^2 Y \widetilde{Y} v}{Mm_h^2 }\, ,
\end{equation}
where $m_h$ is the Higgs boson mass. Since small dark pion masses are well motivated in our setup, and the dark pions couple to all SM fermions including quarks, the phenomenological analysis presented here requires a detailed description of ALP decays for an ALP mass $m_a \lesssim 3\;\mathrm{GeV}$, where exclusive hadronic SM final states must be considered. We obtain this by means of a novel calculation that extends the data-driven methods proposed in Ref.~\cite{Aloni:2018vki}. We emphasize that the results, reported in Appendix~\ref{app:ALP_decays_general}, apply to any ALP with {\it arbitrary} flavor-diagonal couplings to SM fermions. For the decays of light scalars, we make direct use of previous calculations~\cite{Winkler:2018qyg}.

As can be gleaned from Eqs.~\eqref{eq:ALP_sketch} and~\eqref{eq:scalar_sketch}, dark pion theories with $Y \sim \widetilde{Y}$ and with $ Y \gg \widetilde{Y}$ are very different, because in the former the dimension-$5$ Higgs portal dominates, whereas in the latter the dimension-$6$ $Z$ portal is most important. In addition, in general the dark sector contains $CP$ violation, which leads to mixing of different states and induces couplings of all dark pions to both $Z$ and Higgs portals, with relative strengths determined by the model parameters. On the other hand, some dark pions may be stable, if they are charged under an exact subgroup of the $SU(N)_V$ flavor symmetry. Here we consider benchmark scenarios that demonstrate quantitatively all of these features. Related previous work includes studies of light dark pions coupled to the SM through different heavy mediators~\cite{Freytsis:2016dgf,Renner:2018fhh}, and a general analysis of elusive dark sectors with non-renormalizable portals~\cite{Contino:2020tix}.\footnote{In a broader perspective, see also studies of heavier dark pions with masses above the EW scale~\cite{Kribs:2018oad,Kribs:2018ilo}. In addition, dark pions have been extensively examined as dark matter candidates, stabilized by symmetries within the hidden sector, e.g. in Refs.~\cite{Essig:2009nc,Bhattacharya:2013kma,Cline:2013zca,Hochberg:2014kqa,Harigaya:2016rwr,Kopp:2016yji,Berlin:2018pwi,Beauchesne:2018myj,Bernreuther:2019pfb,Contino:2020god}. In this work we offer only some brief comments about the possibility of dark pion dark matter.}

The presentation is organized as follows. In Section~\ref{sec:theory} we write the low-energy effective field theory (EFT) for dark quarks interacting with the SM, obtained by integrating out the heavy EW-charged fermions $Q$. We then discuss constraints from the invisible $Z$ and $h$ widths, and indirect bounds including EW precision observables. In Section~\ref{sec:theory_hadrons} the EFT for dark quarks is matched to an EFT for dark pions coupled to the SM, using both current algebra arguments and a chiral Lagrangian for the dark sector (the latter is reported in Appendix~\ref{app:dark_ChPT}). The dark pion EFT is used to determine the complete set of decay widths and branching ratios to SM particles. For $CP$-odd dark pions we make use of the new, general calculation in Appendix~\ref{app:ALP_decays_general}, where data-driven methods are exploited to evaluate ALP decays for $m_a \lesssim 3\;\mathrm{GeV}$. In Section~\ref{sec:benchmarks} several benchmark scenarios are presented, based on different symmetries that the theory may possess.  Lifetimes and decay patterns are calculated for representative parameters in each benchmark scenario, forming the basis for the phenomenological applications discussed in the next two sections. In Section~\ref{sec:FCNC} we study dark pion production from FCNC meson decays. The decay rates for $B \to K^{(\ast)}\hat{\pi}$ and $K\to \pi \hat{\pi}$ in our model are carefully derived. We obtain relevant constraints from displaced decays of $CP$-odd dark pions to dimuons at the CHARM, LHCb, and for the first time, CMS experiments. In Section~\ref{sec:dark_showers} we consider LHC dark shower signals initiated by $Z$ decays, which have been overlooked so far in experimental searches. We focus on $\hat{\pi} \to \mu^+ \mu^-$ decays, recasting a published LHCb search to obtain new constraints on the parameter space. We also comment briefly on the promising potential sensitivity of ATLAS and CMS. Finally, in Section~\ref{sec:UV} we return to the ultraviolet (UV) model and examine the LHC reach on direct production of the EW-doublet fermion mediators. The current bound and future reach are obtained from searches for supersymmetric EWinos, which share the same experimental signatures (provided the dark jets are mostly invisible). Our conclusions and some directions for further exploration are provided in Section~\ref{sec:conclusions}.
\enlargethispage{20pt}

\section{Effective theory for dark quarks\label{sec:theory}}
Starting from the Lagrangian (\ref{eq:LUV}) and assuming $\bmass \gg  \byuk v,  \byukt v$, where $\langle H\rangle= (\,0\,,\; v/\sqrt{2}\,)^T$ with $v \approx 246\; \text{GeV}$, we can integrate out the $Q$ fields at tree level and obtain the EFT
\begin{align}
\mathcal{L}_\text{EFT}&=\frac{1}{2}\,\overline{\psi}_R\byuk^\dag \bmass^{-2 }\byuk \left[|H|^2i\slashed{D} + i \gamma^\mu H^\dagger  D_\mu H \right]\psi_R+\text{h.c.}\nonumber\\
&+\frac{1}{2}\,\overline{\psi}_L\byukt^\dag \bmass^{-2}\byukt\left[|H|^2i\slashed{D} + i \gamma^\mu H^\dagger D_\mu H \right]\psi_L+\text{h.c.} \label{eq:eft} \\
&- \overline{\psi}_L \bomega \psi_R  +\overline{\psi}_L\byukt^\dag \bmass^{-1} \byuk \psi_R|H|^2 + \text{h.c.}\;, \nonumber
\end{align}
where we have retained operators up to dimension $6$. In general, this effective Lagrangian contains the same number of complex phases that appear in the UV, except if either $\byuk$ or $\byukt$ vanish, in which case the counting is reduced to $(N - 1) (N - 2 )/2$ [the ``apparently missing'' phases then appear in additional operators that were not included in Eq.~(\ref{eq:eft})].

The first terms in square brackets in the first two lines of Eq.~\eqref{eq:eft} renormalize the dark quark kinetic terms after inserting the Higgs VEV. These small corrections are neglected in the following, unless otherwise noted. The second terms in square brackets in the first two lines generate interactions of the $\psi$ with the $Z$ boson. The third line gives rise to the mass matrix,
\begin{equation} \label{eq:seesawmass}
{\bmpsi}= \bomega - \frac{v^2}{2} \byukt^\dag \bmass^{-1} \byuk  \,,
\end{equation}
where the last term is induced by the seesaw mechanism. For general $\byuk$ and $\byukt$ the mass eigenstates $\psi'$ are obtained via unitary transformations $\psi_{L,R} = U_{L,R} \psi^\prime_{L,R}\;$, and their diagonal mass matrix is
\begin{equation}
{\bm{m}_{\psi'}} = U_L^\dag {\bmpsi} U_R\, .
\end{equation}
Barring cancellations, the $\psi^{\prime}_i$ are light if both terms in Eq.~\eqref{eq:seesawmass} are small compared to $\Lambda$. This occurs most naturally if there is an (approximate) chiral symmetry acting on $\psi_L$ (or $\psi_R$) to suppress both $\bomega$ and $\byukt$ (or $\byuk$). For example, that is the case in the tripled top model, where $\byukt=0$~\cite{Cheng:2019yai,Cheng:2018gvu}. The third line of Eq.~\eqref{eq:eft} also generates the leading couplings of the dark quarks to the Higgs.
 
 \subsection{Constraints from $Z$ and Higgs invisible decays}
The first, important constraints on the parameter space are obtained by assuming that the dark hadrons mostly go undetected at colliders, so that the bounds on the $Z$ invisible width from LEP and on the Higgs invisible width from LHC apply. The EFT in Eq.~\eqref{eq:eft} induces $Z$ decays to dark quarks via dimension-6 operators,
\begin{equation} \label{eq:Z_darkfermions}
\Gamma ( Z\to \psi' \overline{\psi}^{\prime} ) \simeq \frac{N_d m_Z^3}{96 \sqrt{2} \pi G_F}   \left\{ \text{Tr} \big[  (\byuk^\dag \bmass^{-2} {\byuk})^2 \big] +  (\byuk \to \byukt) \right\},
\end{equation}
where the small dark quark masses were neglected. For $\bmass = M \bf{1}$, this gives a branching ratio
\begin{equation}\label{eq:Z_BR}
\text{BR} (Z \to \psi' \overline{\psi}^{\prime} ) \approx  1.8 \times 10^{-4} \, \bigg(\frac{ N_d \text{Tr} (  {\byuk}\byuk^\dag {\byuk}  \byuk^\dag )  +  (\byuk \to \byukt) }{3} \bigg) \left( \frac{1\, \text{TeV}}{M}\right)^4 .
\end{equation}
The LEP measurement of the $Z$ invisible width requires $\Delta \Gamma_Z^{\rm inv} < 2$~MeV at $95\%$ CL~\cite{ALEPH:2005ab}, and from Eq.~\eqref{eq:Z_darkfermions} we obtain
\begin{equation} \label{eq:Z_invisible_bound}
M \gtrsim 0.7\;\mathrm{TeV} \bigg(\frac{ N_d \text{Tr} (  {\byuk}\byuk^\dag {\byuk}  \byuk^\dag )  +  (\byuk \to \byukt) }{3} \bigg)^{1/4} .
\end{equation}
If $\byuk \sim \byukt$ parametrically, the leading interaction of the dark sector with the Higgs boson is the dimension-$5$ operator in the third line of Eq.~\eqref{eq:eft}, yielding
\begin{equation} \label{eq:h_psipsi_general}
\Gamma(h\to  \psi' \overline{\psi}^{\prime})\simeq \frac{N_d m_h}{8\sqrt{2}\pi G_F}\text{Tr} \left[ \byuk^\dag \bmass^{-1}\byukt\,  (\byuk^\dag \bmass^{-1} \byukt)^\dagger \right].
\end{equation}
For $\bmass = M \bf{1}$ the associated branching ratio is
\begin{equation}
\text{BR} (h \to \psi' \overline{\psi}^{\prime}) \approx 2.2 \times 10^{-2}\, \bigg(\frac{ N_d \text{Tr}( \byuk \byuk^\dag \byukt \byukt^\dag )}{3 \times 10^{-4}}\bigg) \left( \frac{1\, \text{TeV}}{M}\right)^2, 
\label{eq:hffwidth}
\end{equation}  
where we have taken $Y \sim \widetilde{Y} \sim 0.1$ as reference value for the Yukawas. Satisfying the current invisible Higgs width constraint, $\text{BR} (h \to \text{inv}) < 0.13$ at $95\%$ CL~\cite{ATLAS:2020cjb}, requires
\begin{equation} \label{eq:h_inv}
M \gtrsim 0.4\;\mathrm{TeV} \bigg(\frac{ N_d \text{Tr}( \byuk \byuk^\dag \byukt \byukt^\dag )}{3 \times 10^{-4}}\bigg)^{1/2}.
\end{equation}
Note that for $Y \sim \widetilde{Y} \sim 1$ the bound is $M\gtrsim 40\;\mathrm{TeV}$. The above $Z, h \to \mathrm{invisible}$ bounds are applied widely in the rest of the paper, as we focus mainly on GeV-scale dark pions, for which assuming invisible dark jets is a reasonable first approximation. Nonetheless, it should be kept in mind that these bounds may be weakened or lifted in regions of parameter space where most dark pions are short lived.

A quick glance at Eqs.~\eqref{eq:Z_invisible_bound} and~\eqref{eq:h_inv} indicates that the product $Y \widetilde{Y}$ is much more severely constrained than $Y^2$ or $\widetilde{Y}^2$. Given a new physics scale $M$, scenarios where $Y \sim \widetilde{Y}$ parametrically are subject to a coupling constraint about one order of magnitude stronger than scenarios with $\widetilde{Y}$ or $Y \sim 0$. This will have an important impact on the phenomenology, as the dark pion lifetimes scale with the fourth power of the Yukawa couplings. In addition, if $\widetilde{Y} \sim 0$~[$Y \sim 0$] the dominant decay of the Higgs to the dark sector is either to $\hat{g}\hat{g}$ via the one-loop operator 
\begin{equation}
\mathcal{L} = c_Q\text{Tr}\left(\byuk^\dag \bmass^{-2}\byuk+\byukt^\dag \bmass^{-2 }\byukt \right)\frac{\alpha_d}{24\pi}|H|^2\hat{G}^A_{\mu\nu}\hat{G}^{A\mu\nu}~,
\end{equation}
where $c_Q = 1$ arises from integrating out the $Q$ at one loop,\footnote{More generally, $c_Q$ may receive contributions from additional states, e.g. scalars in the tripled top model~\cite{Cheng:2019yai,Cheng:2018gvu}.} or to $\psi' \overline{\psi}^{\prime}$ via the first~[second] line of Eq.~\eqref{eq:eft}. These processes are too suppressed to lead to current constraints from $h\to \mathrm{invisible}$, but are discussed here for completeness. The one-loop Higgs decay to dark gluons gives
\begin{equation}
\Gamma(h \to \gd \gd) \simeq \frac{ c_Q^2 (N_d^2 - 1) \alpha_d^2  m_h^3}{2304 \sqrt{2} \pi^3 G_F} \left[ \text{Tr} (\byuk^\dag \bmass^{-2}\byuk+\byukt^\dag \bmass^{-2 }\byukt )\right]^2 ,
\end{equation}
resulting in a branching ratio for $\bmass = M \bm{1}$,
\begin{equation}
\text{BR}( h \to \gd \gd) \approx 1.3 \times 10^{-4}\, c_Q^2  \bigg( \frac{\alpha_d (m_h/2)}{0.2}\bigg)^2 \bigg( \frac{ (N_d^2 -1)  \big[ \text{Tr} ( \byuk \byuk^\dag +  \byukt \byukt^\dag )\big]^2 }{8} \bigg) \left( \frac{1\, \text{TeV}}{M}\right)^4.
\end{equation}
If $\byukt=0$ we obtain, using the dark quark equation of motion (EOM),
\begin{equation} \label{eq:h_yt0}
\Gamma(h\to \psi' \overline{\psi}^{\prime})\simeq \frac{N_d m_h }{32 \sqrt{2}\pi G_F}\text{Tr} \left[ \bomega \byuk^\dag \bmass^{-2} {\byuk} (\bomega \byuk^\dag \bmass^{-2} {\byuk})^\dagger \right] 
\end{equation}
and for $\bomega \approx m_{\psi'} \bf{1}$ and $\bmass = M \bm{1}$, the branching ratio is
\begin{equation}
\text{BR} (h \to \psi' \overline{\psi}^{\prime}) \approx 1.4 \times 10^{-5}\; \bigg(\frac{{ N_d \text{Tr}( \byuk \byuk^\dag \byuk \byuk^\dag  )}}{3} \bigg) \bigg( \frac{m_{\psi'}}{0.5\,\text{GeV}}\bigg)^2 \left( \frac{1\, \text{TeV}}{M}\right)^4 .
\end{equation}  
In the opposite case $\byuk = 0$, one replaces $\byuk \to \byukt$ and $\bomega \to \bomega^\dagger$ in Eq.~\eqref{eq:h_yt0}.
 
 \subsection{Indirect constraints} 
 
At one loop, integrating out the heavy fermions $Q$ in Eq.~\eqref{eq:LUV} also generates higher-dimensional operators built only of SM fields, which can be subject to relevant constraints. The most important one is $(H^\dagger\hspace{-1.5mm} \stackrel{\leftrightarrow}{D_\mu} \hspace{-1mm} H)^2$, encoding a contribution to the EW $T$ parameter~\cite{Peskin:1990zt,Peskin:1991sw}. In fact $T$ is most easily calculated in the UV theory, by applying, e.g., the results of Ref.~\cite{Anastasiou:2009rv}. The derivation of a general analytical expression is rather cumbersome, but the calculation simplifies if the dark Yukawas are diagonal, $\bm{Y} = \mathrm{diag}_{i}\,y_i$ and $\bm{\widetilde{Y}} = \mathrm{diag}_i\,\tilde{y}_i\,$: 
\begin{equation}
\widehat{T} \simeq \frac{N_d}{16 \pi^2}  \sum_{i\, =\, 1}^N \frac{v^2}{3 M_i^2} \Big( y_i^4 + \tilde{y}_i^4 + \frac{1}{2} y_i^2 \tilde{y}_i^2 \Big),
\end{equation}
at leading order in the large$\,$-$M_i$ expansion and taking real couplings for simplicity. The general case including flavor mixing can be treated numerically in a straightforward manner. It is useful to compare the $T$ parameter and $Z\to \mathrm{invisible}$ constraints, in the simple scenario $\bm{M} = M \bm{1}$, $\bm{Y} = Y \bm{1}$ and $\bm{\widetilde{Y}} = 0$,
\begin{align}
M \gtrsim&\;   0.9\,\mathrm{TeV}\; Y^2 \Big(\frac{N_d N}{6}\Big)^{1/2}   , \qquad (T\;\mathrm{parameter}) \\
M  \gtrsim&\;  0.8\,\mathrm{TeV}\;Y \Big(\frac{N_d N}{6}\Big)^{1/4}   , \qquad\; (Z\to \mathrm{invisible})
\end{align}
where for the former we have used the rough estimate $\widehat{T} \lesssim 10^{-3}$ and the latter follows from Eq.~\eqref{eq:Z_invisible_bound}. Since the two are comparable for $Y \sim O(1)$, and additional beyond-SM contributions can a priori alter the interpretation of the $T$ constraint, in most of our discussion we stick to the more robust invisible $Z$ width bound. When $Y \sim \widetilde{Y}$, both are subleading to the invisible $h$ branching ratio constraint.

The operators $|H|^2 B_{\mu\nu} B^{\mu\nu}$ and $|H|^2 W_{\mu\nu}^i W^{\mu\nu \,i}$ are also generated at one loop. However, since the $Q_d$ are electrically neutral and the $Q_u$ are charged but do not couple to the Higgs, we expect the operators to come in the linear combination $|H|^2 (g^2 W_{\mu\nu}^i W^{\mu\nu \,i} - g^{\prime\, 2}  B_{\mu\nu} B^{\mu\nu})$ which gives a vanishing contribution to the $h\gamma\gamma$ coupling.

$CP$ violation in the dark sector could feed into the visible sector, inducing electric dipole moments (EDMs) for SM particles. The strongest limit comes from the electron EDM~\cite{Andreev:2018tn}. Corrections to the electron EDM arise  through the loop-suppressed operator $\mathcal{O}_{B \widetilde{B} } = |H|^2 B_{\mu\nu} \widetilde{B}^{\mu\nu}$, which in turn contributes at one loop to the EDM (similar considerations apply to $|H|^2 W^i_{\mu\nu} \widetilde{W}^{\mu\nu\, i}$). Inspection of the relevant diagrams shows that $\mathcal{O}_{B\widetilde{B}}$ does not arise at one loop. Furthermore, if $\widetilde{Y} = 0$ or $Y = 0$ the two-loop contributions turn out to be strongly suppressed by an extra $\sim \omega^2/M^2$ factor.  If both $Y$ and $\widetilde{Y}$ are non-vanishing we estimate $c_{B\widetilde{B}} \sim N_d Y^2 \widetilde{Y}^2 g^{\prime\,2} / [ (4\pi)^4 M^2]$, leading to a constraint $M \gtrsim 1.5\;\mathrm{TeV}\; Y \widetilde{Y} $ for $N_d = 3$~\cite{Panico:2018hal}. This is much weaker than the Higgs invisible branching ratio bound, $M \gtrsim 40\;\mathrm{TeV}\; Y \widetilde{Y} $ from Eq.~\eqref{eq:h_inv}. In summary, we find that EDMs do not provide additional constraints in this model.

 \section{Effective theory for dark hadrons\label{sec:theory_hadrons}}
At energies below $\Lambda$, the $SU(N_d)$ gauge group confines and the dark quarks and gluons form hadrons. For $N \geq 2$, the lightest hadrons are pNGBs of the $SU(N)_L \times SU(N)_R \to SU(N)_V$  symmetry breaking and belong to the adjoint representation of $SU(N)_V$. As the simplest example and representative case for phenomenological studies, in this work we focus on $N=2$. The three dark pions $\hat{\pi}_{a}$ are defined in the basis where the light quark mass matrix is diagonal,
\begin{equation}
\hat{\pi}_a \sim i (\overline{\psi}_L^{\,\prime} \sigma_a \psi_R^\prime -  \overline{\psi}_R^{\,\prime} \sigma_a \psi_L^\prime) = \overline{\psi}^{\,\prime} \hspace{-0.5mm} i \sigma_a \gamma_5 \psi^\prime \,,
\label{eq:piondefinition}
\end{equation}
where $\sigma_a$  are the Pauli matrices. Importantly, the $\hat{\pi}_2$ has $J^{PC} = 0^{--}$ whereas $\hat{\pi}_{1,3}$ have $0^{-+}$, as can be derived from Eq.~\eqref{eq:piondefinition} using $\overline{\psi}_2^{\,\prime} P_{L,R} \psi_1^\prime \stackrel{C}{\rightarrow} \overline{\psi}_1^{\,\prime} P_{L,R} \psi_2^\prime\,$. 
Note that in the absence of a $U(1)$ flavor symmetry, i.e., if $\byuk$, $\byukt$ are not diagonal, $\hat{\pi}_1$ and $\hat{\pi}_2$ are distinct states. Their degeneracy will be lifted by  $\byuk$, $\byukt$ interactions, as demonstrated later by explicit examples. We do not discuss in detail the dark flavor-singlet $\hat{\eta}^{\,\prime}$, which at small $N_d$ receives a large mass from the dark $U(1)_A$ anomaly.

\begin{figure}[t]
\begin{center}
\includegraphics[width=0.49\textwidth]{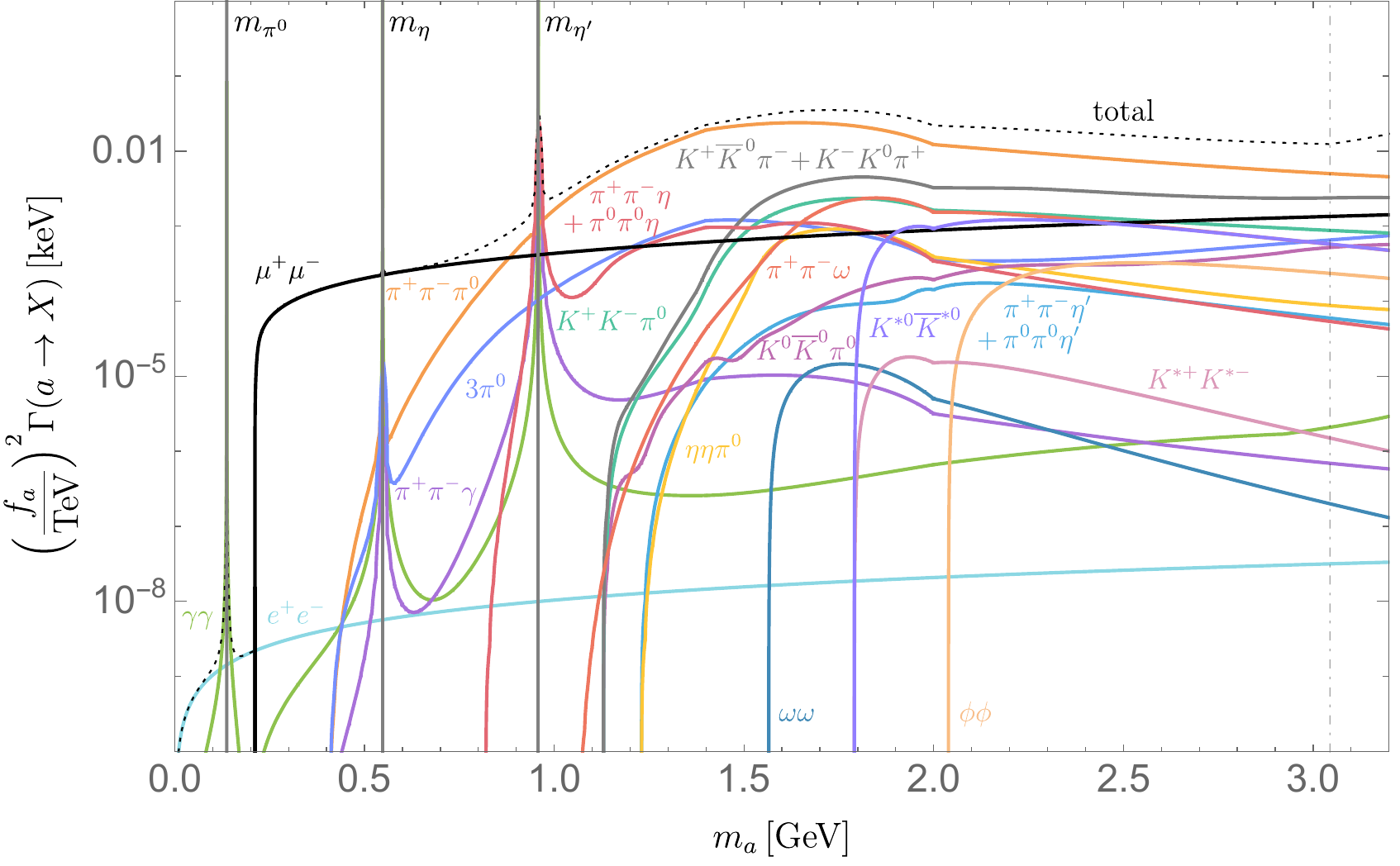}
\includegraphics[width=0.50\textwidth]{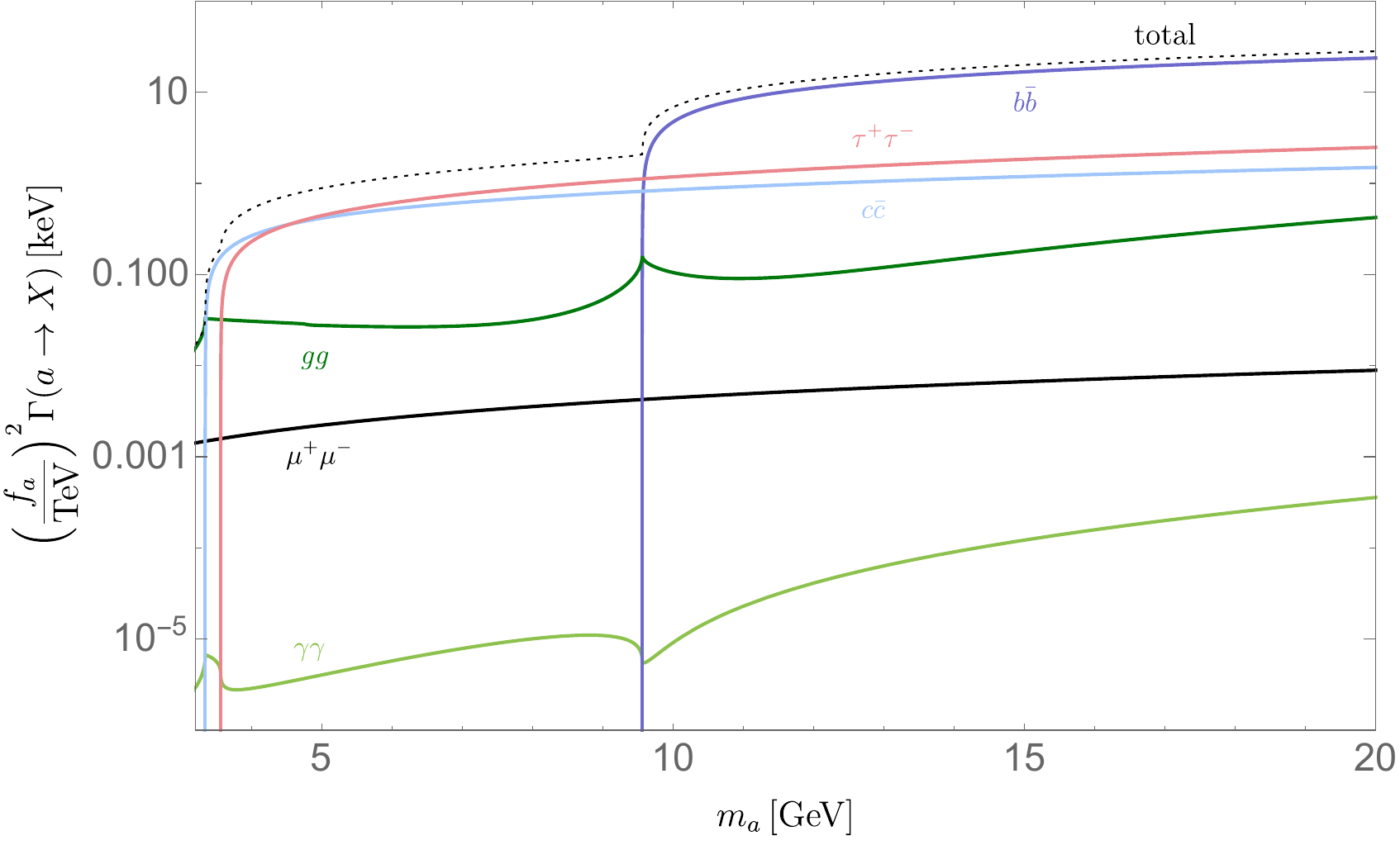}\vspace{3.mm}
\includegraphics[width=0.49\textwidth]{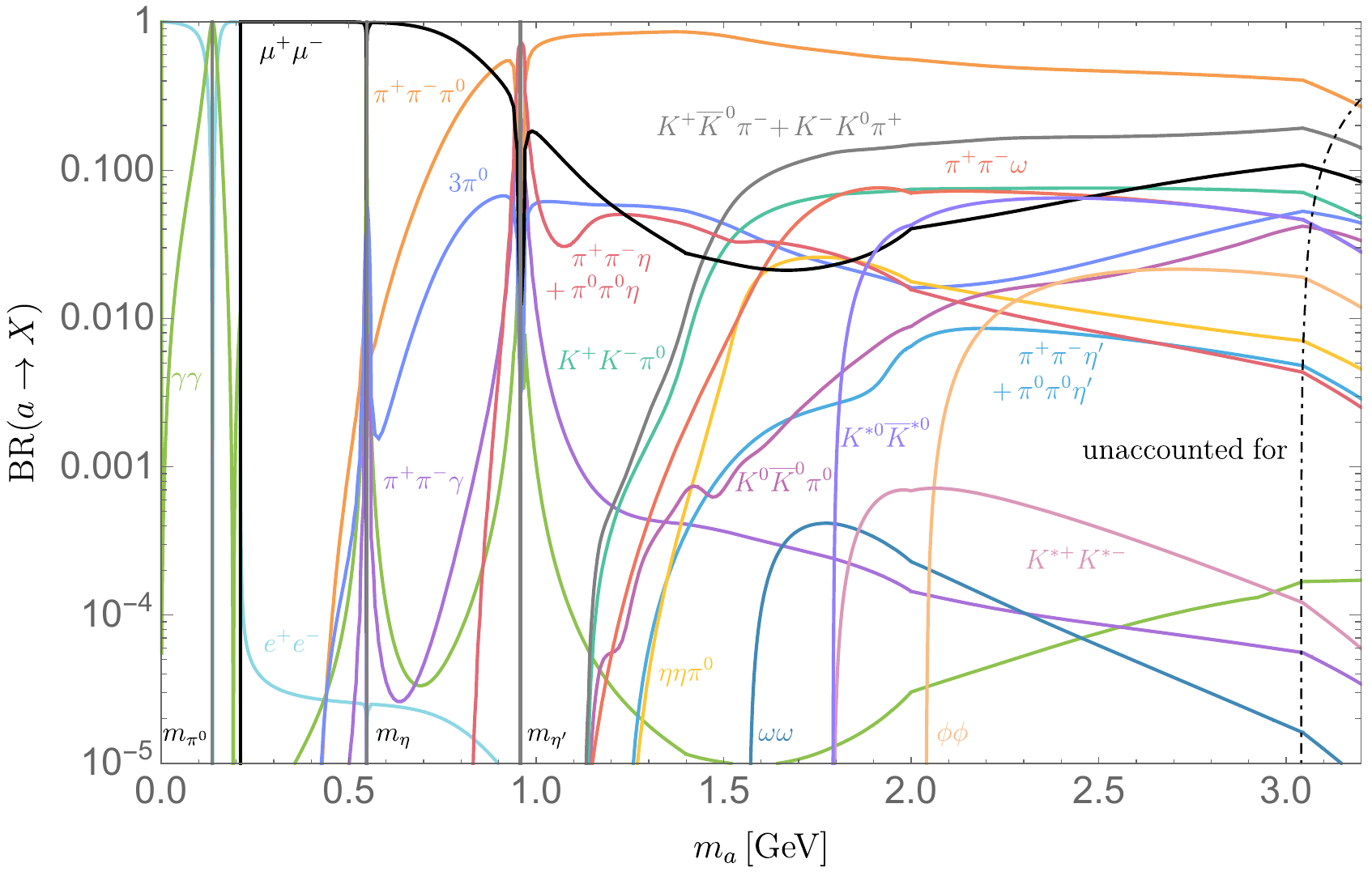}\hspace{0.75mm}
\includegraphics[width=0.495\textwidth]{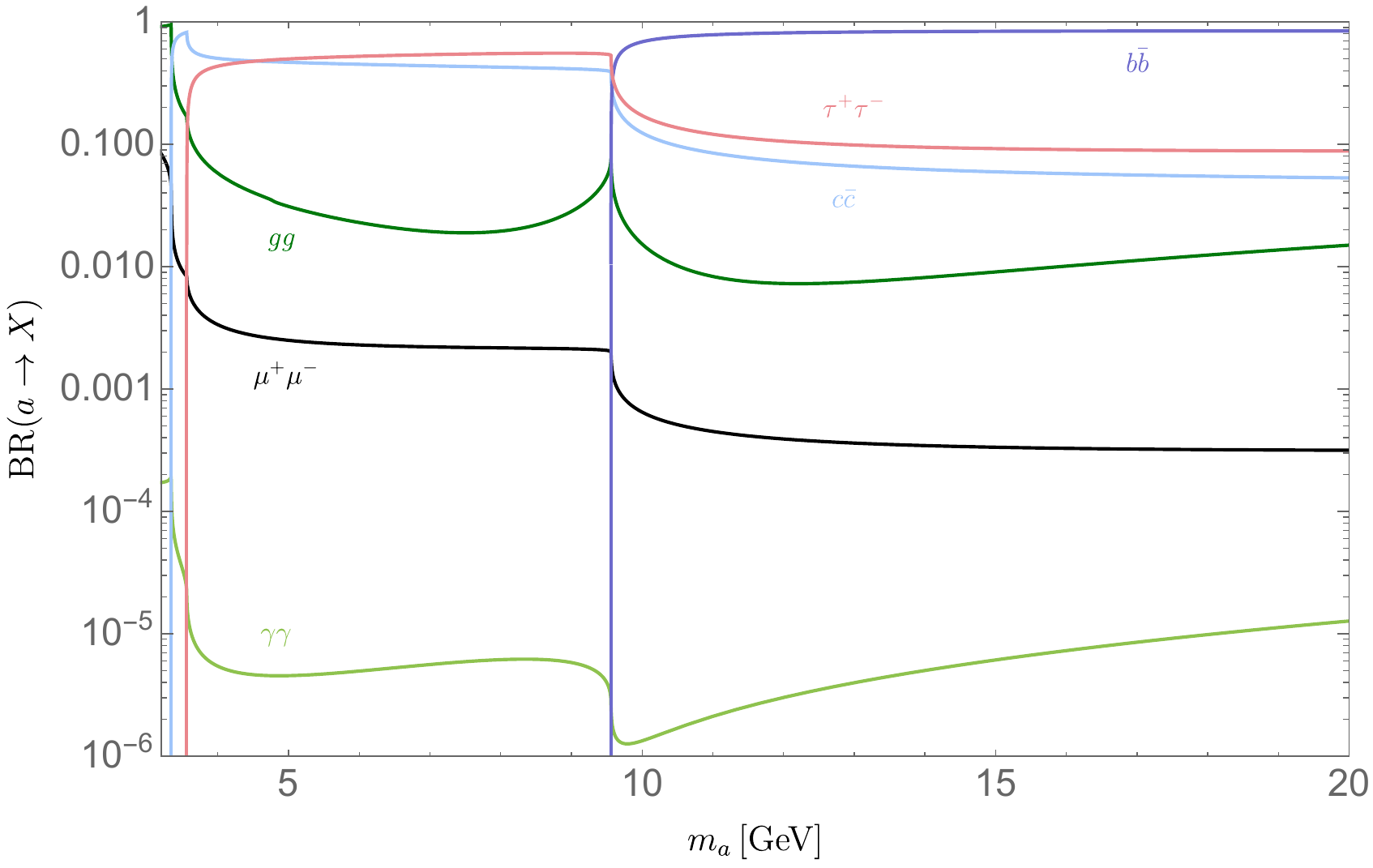}
\caption{\label{fig:ALP_widths} Decay widths {\it (top)} and branching ratios {\it (bottom)} of a light ALP coupled to SM fermions with $c_f = T_{Lf}^3$. In the top left panel the vertical dot-dashed line indicates the $m_a = m_a^{\rm trans}$ where our evaluation of the total hadronic width transitions from $\sum_i\Gamma (a \to {\rm excl}_{\,i})$ to $\Gamma(a\to gg)$. Correspondingly, in the bottom left panel the dot-dashed curve displays the branching ratio that is not captured by the considered exclusive modes. Note that at $m_a^{\rm trans}$ the NLO QCD correction to $\Gamma(a\to gg)$ in Eq.~\eqref{eq:pihat_gg} is $235\, \alpha_s(m_a^{\rm trans})/(12\pi) \approx 1.6$ times the leading order, suggesting a sizable residual uncertainty for this width.} 
\end{center}
\end{figure}

The couplings of the dark pions to the $Z$ boson can be derived from the interactions in the dark quark EFT of Eq.~\eqref{eq:eft},
\begin{equation}
- \frac{g_Z v^2}{4} \big( \overline{\psi}_R \byuk^\dag \bmass^{-2}\byuk \gamma^\mu \psi_R + \overline{\psi}_L \byukt^\dag \bmass^{-2}\byukt \gamma^\mu \psi_L \big) Z_\mu\,,
\label{eq:Z_interaction}
\end{equation}
where $g_Z=\sqrt{g^2+g'^2}$. We rewrite this as
\begin{equation}\label{eq:zint}
- \frac{g_Z}{2} \big( \overline{\psi}_R^{\,\prime} \, \bm{A} \,  \gamma^\mu \psi_R^\prime  +  \overline{\psi}_L^{\,\prime} \, \widetilde{\bm{A}} \, \gamma^\mu \psi_L^{\prime} \big) Z_\mu = - \frac{g_Z}{4} \sum_{q\,=\,0}^{3}  \left\{ \text{Tr} [\sigma_q(\bm{A} + \widetilde{\bm{A}}) ] j_{q}^\mu + \text{Tr} [ \sigma_q(\bm{A} - \widetilde{\bm{A}}) ] j_{5q}^\mu \right\} Z_\mu\, ,
\end{equation}
where the dimensionless matrices $\bm{A}$ and $\widetilde{\bm{A}}$ are defined as
\begin{equation}
\bA \equiv \frac{v^2}{2} U_R^\dag \byuk^\dag \bmass^{-2}\byuk  U_R\,, \qquad \bAt \equiv \frac{v^2}{2}  U_L^\dag \byukt^\dag \bmass^{-2}\byukt  U_L \,, 
\end{equation} 
and $\sigma_0 \equiv \mathbf{1}_2$. In addition,
\begin{align}
 j_q^\mu = j_{L q}^\mu + j_{R q}^\mu\, , \quad j_{5q}^\mu= j_{R q}^\mu - j_{L q}^\mu \,, \qquad  j_{L,R \,q}^{\mu}&=\overline{\psi}^{\,\prime}_{L,R}\gamma^\mu\frac{\sigma_q}{2}\psi^\prime_{L,R}\,.
\end{align}
The pions are excited by the axial vector current. We define their decay constant from
\begin{equation} \label{eq:f_def}
\langle 0|j_{5a}^{\mu} (0) |\pid_b (p)\rangle = -\, i \delta_{ab} \fpid\,p^\mu\,, 
\end{equation}
with normalization corresponding to $f_\pi \approx 93$ MeV in the SM. Thus the last term on the right-hand side of Eq.~\eqref{eq:zint} yields a tree-level $\hat{\pi}_a\,$-$\,Z$ mixing, and the partial width for the decay to a pair of SM fermions $f$ is 
\begin{equation}
\Gamma(\pid_{a} \stackrel{Z}{\to} f \bar{f}\,)=\frac{N_c^f }{4\pi} \left| \mathrm{Tr} [\sigma_a( \bm{A} - \widetilde{\bm{A}})] \right|^2{G_F^2 a_f^2 m_f^2 \fpid^2 m_{\pid_{a}}} \Big(1-\tfrac{4m_f^2}{m_{\pid_{a}}^2} \Big)^{1/2}\,,
\label{eq:darkpiondecay0}
\end{equation}
where $a_f = T_{L f}^{3}$ and $N_c^f = 3\,(1)$ for quarks~(leptons). It is important to note that, in the absence of $CP$-violating phases, $\hat{\pi}_{1,3}$ decay through the single $Z$ exchange but $\hat{\pi}_2$ does not, because
\begin{equation}
\mathrm{Tr} [\sigma_2( \bm{A} - \widetilde{\bm{A}})] = i [ (A - \widetilde{A})_{12} - (A - \widetilde{A})_{12}^\ast ] = 0\,,
\end{equation}
where we have used the hermiticity of $\bm{A},  \widetilde{\bm{A}}$. For light dark pions it is in fact convenient to integrate out the $Z$ boson, obtaining
\begin{equation}
\frac{g_Z^2 f_{\hat{\pi}}}{8m_Z^2}\mathrm{Tr}[\sigma_a(\bm{A} - \bm{\widetilde{A}})] \partial_\mu \hat{\pi}_a \bar{f} \gamma^\mu (v_f - a_f \gamma_5) f\,. \qquad (v_f = T_{Lf}^3 - 2 s_w^2 Q_f\,, \; a_f = T_{Lf}^3)
\end{equation}
Due to the conservation of the vector current, the interaction relevant to describe dark pion decays is
\begin{equation} \label{eq:Leff_quarks}
 - \frac{\partial_\mu \hat{\pi}_b}{f_a^{(b)}}  c_f  \bar{f}\gamma^\mu \gamma_5 f\,,\qquad \frac{1}{f_a^{(b)}} \equiv  \frac{\fpid}{2 v^2} \mathrm{Tr}\,[\sigma_b(\bA - \bAt)], \qquad c_f = T_{Lf}^3\,,
\end{equation}
where $f_a^{(b)}$ is the {\it effective decay constant} of $\hat{\pi}_b\,$. Equation~\eqref{eq:Leff_quarks} enables us to apply the new calculations presented in Appendix~\ref{app:ALP_decays_general}, where for arbitrary (flavor-diagonal) ALP-SM fermion couplings we perform the matching to the SM chiral Lagrangian, augmented with exchange of scalar, vector, and tensor resonances above $1$~GeV, and by extending data-driven methods pioneered in Ref.~\cite{Aloni:2018vki} we evaluate the ALP decay widths to an extensive set of exclusive hadronic SM final states. The results are reported in Fig.~\ref{fig:ALP_widths}, which is one of the main novelties of this work. The lifetime is also shown in Fig.~\ref{fig:f2_comparison}, see Appendix~\ref{app:ALP_decays_general}.

To gain some initial insight on the scales we take, e.g., $\widetilde{Y}  = 0$, giving the parametric scaling
\begin{equation} \label{eq:naive_fa}
f_a^{(1)} \sim f_a^{(3)}  \sim \frac{M^2}{Y^2 f_{\hat{\pi}}} \sim 10^3\;\mathrm{TeV} \; \left( \frac{M/Y}{\mathrm{TeV}} \right)^2 \left( \frac{\mathrm{GeV}}{f_{\hat{\pi}}} \right)
\end{equation} 
where $CP$ conservation was assumed for simplicity. As the constraint from $Z \to \mathrm{invisible}$ gives roughly $M/Y \gtrsim \mathrm{TeV}$, see Eq.~\eqref{eq:Z_invisible_bound}, for $f_{\hat{\pi}} \sim \mathrm{GeV}$ the $CP$-odd dark pions can be regarded as light ALPs with effective decay constants $\gtrsim \mathrm{PeV}$. 

In fact, from the quark-level EFT in Eq.~\eqref{eq:eft} we can directly derive that the dark pions couple to the Higgs current, namely $\mathcal{L} \supset i H^\dagger \hspace{-1.5mm}\stackrel{\leftrightarrow}{D_\mu} \hspace{-1.5mm} H \partial^\mu \hat{\pi}_b/f_a^{(b)}$.\footnote{In turn, this can be rewritten in terms of dark pion-SM fermion couplings using the leading order EOM for the hypercharge gauge field, $i H^\dagger \hspace{-1.2mm}\stackrel{\leftrightarrow}{D_\mu} \hspace{-1.2mm} H = - 2 \sum_{\Psi \in \mathrm{chiral}} Y_\Psi \overline{\Psi} \gamma_\mu \Psi - (2/g') \partial^\nu B_{\nu\mu}$, noticing that the piece involving $B$ vanishes upon integration by parts.} Because the dark pions are appropriate degrees of freedom only at energies below $\Lambda$ and the latter is smaller than the EW scale in most of our parameter space, the use of the broken EW phase is warranted and such an effective description is not fully justified. Nonetheless it affords us a first brief discussion of FCNC meson decays~\cite{Hall:1981bc,Frere:1981cc,Freytsis:2009ct}, by applying the leading-log results of Ref.~\cite{Gavela:2019wzg}. The flavor-changing couplings of the dark pions to quarks arise at one loop,
\begin{equation} \label{eq:FCNC_LL}
- g_{ij}^{a (b)} \partial_\mu \hat{\pi}_b \bar{d}_{L j} \gamma^\mu d_{L i} + \mathrm{h.c.}, \qquad g_{ij}^{a(b)} = - \frac{ g^2 }{ 4 f_a^{(b)}} \sum_{q\, \in\, u,c,t} \frac{V^\ast_{qj} V_{qi} }{16\pi^2} \frac{m_q^2}{m_W^2} \log \frac{M^2}{m_q^2}\,. 
\end{equation}
Note that the appropriate mass scale to cut off the logarithm is $M \sim \mathrm{TeV}$ -- the largest physical threshold here -- and not $f_a^{(b)}$, which is a combination of parameters with dimension of a VEV and does not correspond to the mass of any particle. In addition, owing to the modest separation between $M$ and $m_t$, finite pieces are expected to be important. Both expectations are confirmed by the explicit calculation in Section~\ref{sec:FCNC}. There, we show that current meson FCNC constraints are at the level $f_a \sim 10^3\; \mathrm{TeV}$, as obtained from $B\to X_s \hat{\pi}$ decays (where $X_s$ denotes a strange hadron state) with long-lived $\hat{\pi} \to \mu^+ \mu^-$ at CHARM, LHCb and CMS for $m_{\hat{\pi}} \gtrsim 2m_\mu\,$, and from searches for $K^+ \to \pi^+ \hat{\pi}$ with invisible $\hat{\pi}$ at E949 and NA62 for smaller dark pion masses.

The dark pions can also decay through tree-level Higgs exchange. To derive the decay width, the starting point are the following interactions in Eq.~\eqref{eq:eft},
\begin{equation} \label{eq:pi2_decay}
\overline{\psi}_L^{\,\prime} \bm{B} \psi_R^\prime h + \mathrm{h.c.} = \frac{1}{2} \overline{\psi}^{\,\prime} \big[ \bm{B} + \bm{B}^\dagger + (\bm{B} - \bm{B}^\dagger) \gamma_5 \big] \psi^\prime h\,, \qquad \bm{B} \equiv v\, U_L^\dagger \byukt^\dagger \bm{M}^{-1} \byuk U_R \,,
\end{equation}
where we have already rotated to the quark mass eigenstate basis and the coupling matrix $\bm{B}$ is dimensionless. The piece of Eq.~\eqref{eq:pi2_decay} containing $\gamma_5$ is relevant for dark pion decay, and we rewrite it as
\begin{equation}
- \frac{1}{2}  \sum_{q \,=\, 0}^{3} \mathrm{Tr} \big[ i \sigma_q  (\bm{B} - \bm{B}^\dagger) \big] \overline{\psi}^{\,\prime} \frac{ i \sigma_q }{2} \gamma_5 \psi^\prime h \,.
\end{equation}
Finally, recalling Eq.~\eqref{eq:piondefinition} we have
\begin{equation} \label{eq:pion_2}
\langle 0| \overline{\psi}^{\,\prime} \frac{i\sigma_a}{2} \gamma_5 \psi^\prime (0) |\pid_b (p)\rangle = -\,  \delta_{ab} \fpid \frac{ m^2_{\hat{\pi}_a}}{\mathrm{Tr}(\bm{m}_{\psi^\prime})} \,, 
\end{equation}
which allows us to calculate the decay width mediated by a single Higgs exchange,
\begin{equation} \label{eq:pi2_width}
\Gamma(\pid_{a} \stackrel{h}{\to} f \bar{f}\,) = \frac{N_c^f}{4\pi}  \left| \mathrm{Tr} [ i\sigma_a (\bm{B} - \bm{B}^\dagger)] \right|^2 \frac{G_F}{4\sqrt{2}} \, m_f^2\, \frac{ f_{\hat{\pi}}^2 \hat{B}_0^2 m_{\hat{\pi}_a} }{m_h^4 } \frac{  \Big(1 - \tfrac{4m_f^2}{m^2_{\hat{\pi}_a} } \Big)^{3/2} }{ \Big( 1 - \tfrac{m^2_{\hat{\pi}_a}}{m_h^2} \Big)^2 }\;,
\end{equation}
where we have employed the relation $m^2_{\hat{\pi}_a} = \hat{B}_0 \mathrm{Tr}(\bm{m}_{\psi^\prime})$, valid at leading order in the dark sector chiral perturbation theory (ChPT), see Appendix~\ref{app:dark_ChPT}. It is immediate to see that if $CP$ is conserved, the trace in Eq.~\eqref{eq:pi2_width} can be non-vanishing only for $a = 2$, since $(i\sigma_2)^\ast = i \sigma_2$ whereas $(i\sigma_{1,3})^\ast = - i \sigma_{1,3}\,$. Note that the interference between the $Z$- and $h$-mediated amplitudes vanishes in the $\hat{\pi}_a \to f\bar{f}$ process.

Comparing Eqs.~\eqref{eq:darkpiondecay0} and~\eqref{eq:pi2_width}, for $Y \sim \widetilde{Y}$ we find $\Gamma(\pid \stackrel{h}{\to} f \bar{f}\,) / \Gamma(\pid \stackrel{Z}{\to} f \bar{f}\,) \sim M^2 \hat{B}_0^2 / m_h^4$ which is $O(1)$ for typical choices $ M \sim 1\;\mathrm{TeV}$ and $\hat{B}_0 \sim 10\;\mathrm{GeV}$ (dimensionally, we expect $\hat{B}_0 \sim 4\pi f_{\hat{\pi}}$). However, for hierarchical Yukawas the ratio is suppressed by $\widetilde{Y}^2/Y^2$ or viceversa, and the pions decaying via the $Z$ mediation can have much shorter lifetimes than those decaying via the Higgs exchange.

\begin{figure}[t]
\begin{center}
\includegraphics[width=0.49\textwidth]{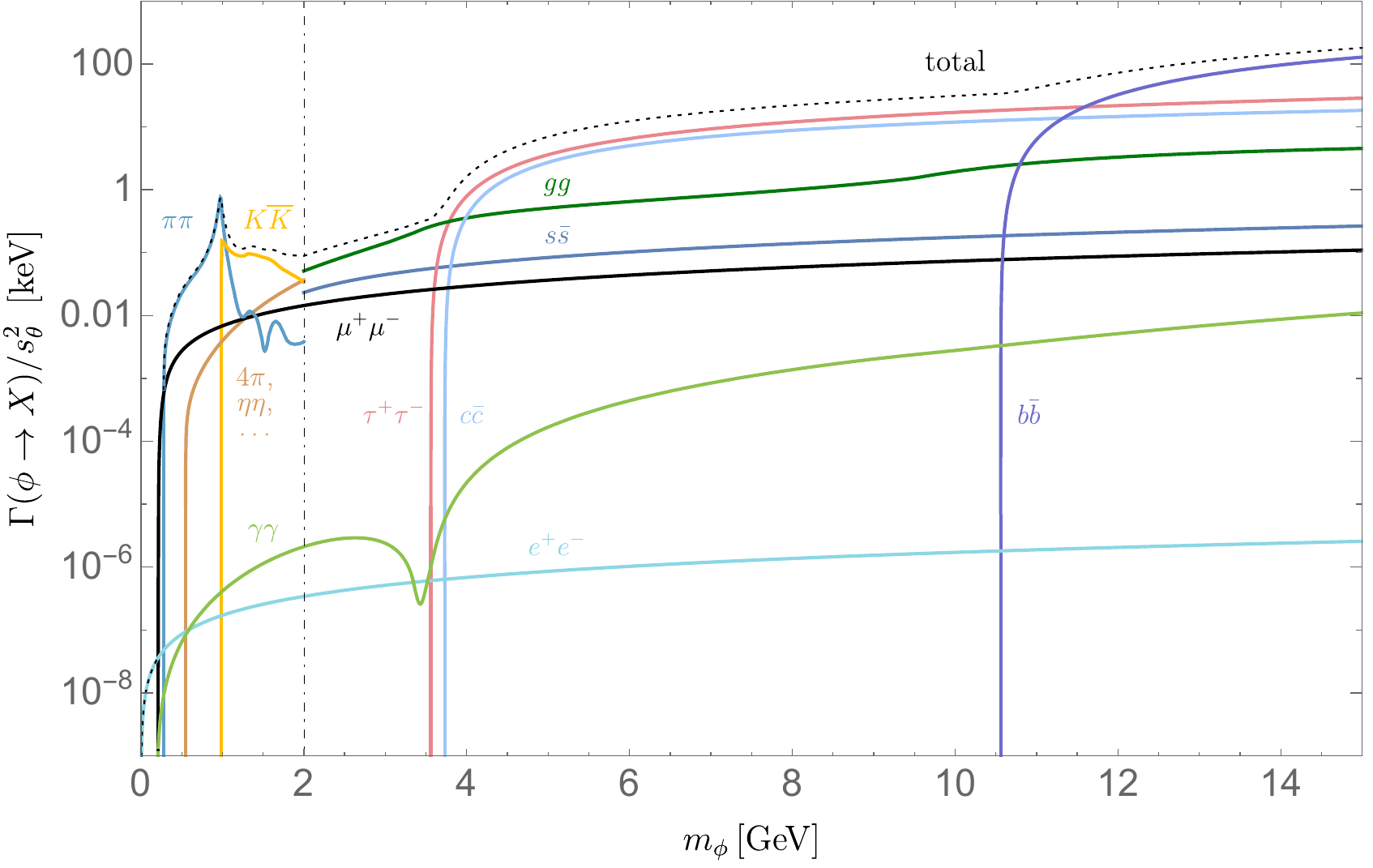}
\includegraphics[width=0.495\textwidth]{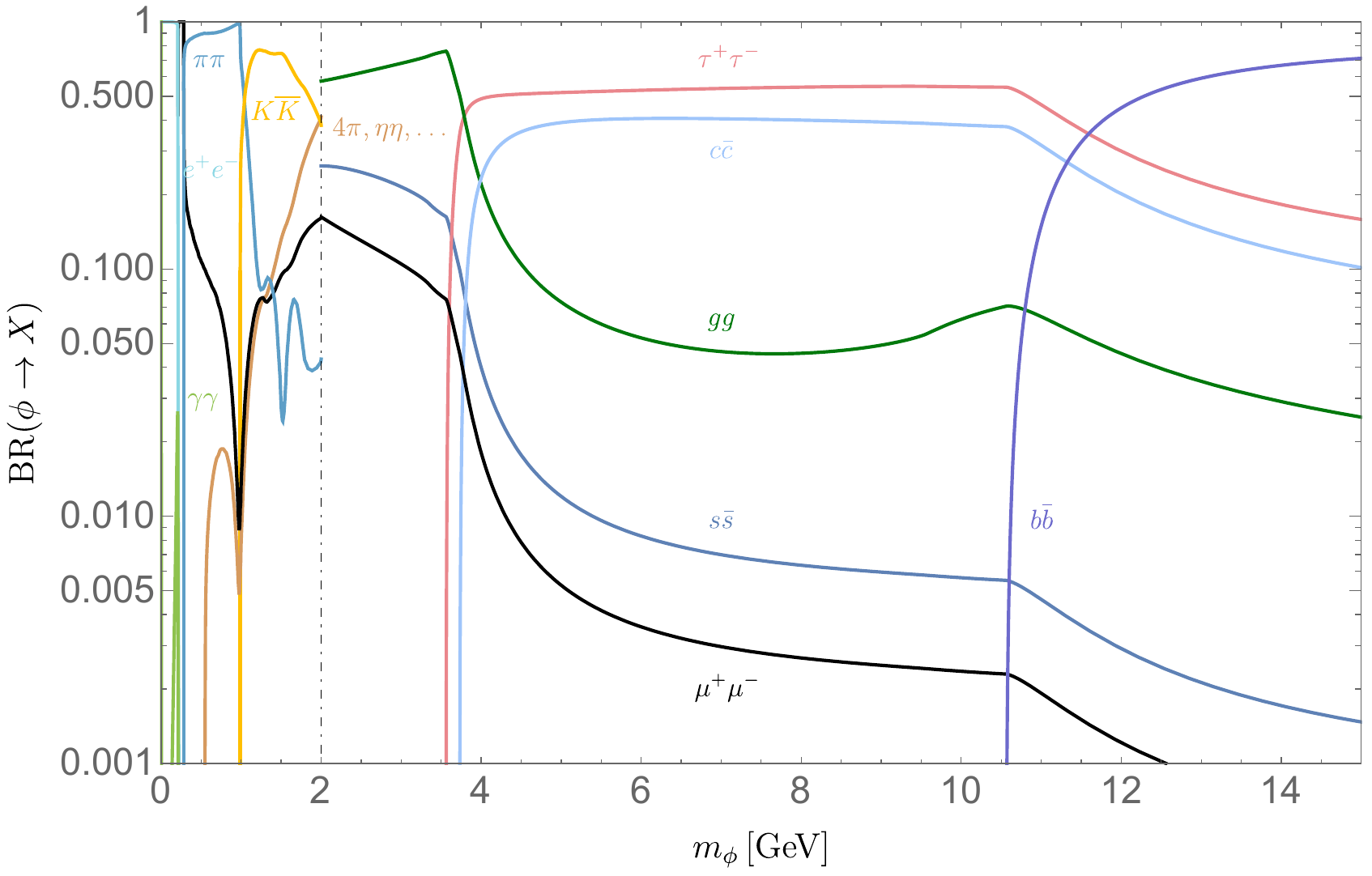}
\caption{\label{fig:PHI_widths} Decay widths {\it (left)} and branching ratios {\it (right)} of a light $CP$-even scalar coupled to the SM via Higgs mixing. The dot-dashed line at $m_\phi = 2$~GeV indicates the point where the description of hadronic decays transitions from dispersive methods to a perturbative spectator model, following Ref.~\cite{Winkler:2018qyg}.} 
\end{center}
\end{figure}

For GeV-scale dark pions, Higgs-mediated decays to exclusive hadronic SM final states become important. We do not attempt to reassess them here, but account for them following the results of Ref.~\cite{Winkler:2018qyg} (see also Ref.~\cite{Gershtein:2020mwi} for a recent reappraisal), by matching to their definition of the couplings
\begin{equation} \label{eq:Higgs_mixing}
 - s_\theta^{(a)} \frac{m_f}{v} \hat{\pi}_a \bar{f} f\,, \qquad s_\theta^{(a)} = \frac{f_{\hat{\pi}}}{2} \frac{m^2_{\hat{\pi}_a}}{\mathrm{Tr}(\bm{m}_{\psi'})} \frac{\mathrm{Tr}[i \sigma_a (\bm{B} - \bm{B}^\dagger)]}{m_h^2 - m^2_{\hat{\pi}_a}}\,.  
\end{equation}
As in Ref.~\cite{Winkler:2018qyg} we take $m_s = 95\;\mathrm{MeV}$, however we include the running of $m_{c,b}\,$ in the perturbative spectator model and consider the decay to photons, 
\begin{equation} \label{eq:CPeven_gaga}
\Gamma(\phi \to \gamma\gamma) = \frac{s_\theta^2 \alpha^2 m_\phi^3}{256 \pi^3 v^2} \Bigg| \sum_{i\, \in\, \mathrm{fermions}} \hspace{-0.3cm} 2N_c^i Q_i^2 x_i \big[1 + (1 - x_i)f(x_i)^2 \big] - \big[2 + 3 x_W + 3 x_W (2 - x_W) f(x_W)^2 \big] \Bigg|^2 ,
\end{equation}
where $x_i \equiv 4m_i^2 /m_\phi^2$ and the function $f(x)$ is defined in Eq.~\eqref{eq:f_loop}. The matching constant parametrizing $\phi \to 4\pi, \eta\eta, \ldots$~\cite{Winkler:2018qyg} is fixed to $C \approx 4.8 \times 10^{-9}$~GeV$^{-2}$. The resulting decay widths and branching ratios are shown in Fig.~\ref{fig:PHI_widths}.

Parametrically, the dark pion-Higgs mixing angle takes on the scaling,
\begin{equation}
s_\theta^{(2)} \sim 2\pi f_{\hat{\pi}}^2 \frac{Y \widetilde{Y} v}{M m_h^2} \sim 3 \times 10^{-6} \left( \frac{f_{\hat{\pi}}}{\mathrm{GeV}} \right)^2 \bigg( \frac{Y \widetilde{Y}/M}{0.03\;\mathrm{TeV}^{-1}} \bigg) ,
\end{equation}
where, as in Eq.~\eqref{eq:naive_fa}, $CP$ conservation was assumed for simplicity. Since the bound from $h\to \mathrm{invisible}$ reads roughly $Y \widetilde{Y}/M \lesssim 0.03\;\mathrm{TeV}^{-1}$ (see Eq.~\eqref{eq:h_inv}), for $f_{\hat{\pi}} \sim \mathrm{GeV}$ the $CP$-even dark pions can be viewed as Higgs-mixed scalars with mixing angles $\lesssim 10^{-6}$.

When a single dark pion decays through both the $Z$ and $h$ portals in the presence of $CP$ violation, we neglect the interference between the two amplitudes, which vanishes for $\hat{\pi}_a \to f\bar{f}$ as already noted, but can a priori be nonzero for more complex final states. In this work we focus on the mass range $m_{\hat{\pi}}\lesssim 2m_b \sim  10\;\mathrm{GeV}$, where a wider range of experiments are relevant and our results are expected to have the most impact. 

We end this section with some brief comments on the heavier dark hadrons, including non-pNGB mesons and baryons. The dark vector (and axial-vector) mesons may be relevant to intensity frontier phenomenology, where they can be produced first and subsequently decay to dark pions, if kinematically allowed. In the $N = 2$, $N_d = 3$ theory considered here, lattice QCD calculations at pion masses larger than their physical values can be repurposed~\cite{DeGrand:2019vbx} to parametrize the hidden sector, at least for moderately heavy pNGBs with $0.1 \lesssim m_{\hat{\pi}}^2/m_{\hat{V}}^2 \lesssim 0.7$, where $\hat{V}$ denotes the dark vector resonance. As for the baryons, the lightest among them is stable due to dark $U(1)_B$, but its relic density can easily be very suppressed unless a dark baryon asymmetry is present. In this work we focus on the properties and phenomenology of the dark pions, neglecting the heavier hadrons.

\section{Benchmark scenarios for dark pions}\label{sec:benchmarks}

In this section we discuss the range of possibilities for the dark pion properties, beginning with general arguments. If the theory respects the $SU(2)_V$ isospin symmetry, i.e., $\bomega, \bmass, \byuk, \byukt \propto \mathbf{1}_2$, then $\bm{A}, \widetilde{\bm{A}}, \bm{B} \propto \mathbf{1}_2$ and all dark pions are stable. It is also possible that the $SU(2)_V$ is explicitly broken to its $U(1)$ subgroup, i.e., $\bomega, \bmass, \byuk, \byukt$ and hence $\bm{A}, \widetilde{\bm{A}}, \bm{B}$ are diagonal. In this case $\pid_{\pm} \equiv ( \hat{\pi}_1 \mp i \hat{\pi}_2 )/ \sqrt{2}$ is charged under the $U(1)$ and therefore stable, while $\pid_{0} \equiv \pid_{3}$ can decay (to avoid any confusion, we remark that the subscripts do not indicate SM electric charge). 

Furthermore, specific models can give rise to distinctive patterns for the masses and couplings. For example, in a setup inspired by the tripled top~\cite{Cheng:2019yai}, but where the two hidden sectors share a common dark color gauge group, we expect $\bmass = M \bf{1}_2$, $\byuk \simeq y_t \bf{1}_2$ and $\byukt \simeq 0$, which implies $\widetilde{\bm{A}}, \bm{B} \simeq 0$. In this case the main source of isospin breaking in the hidden sector comes from the diagonal $\bmpsi \simeq \bomega$. As a result, the $U(1)$ subgroup is approximately preserved and the $\pid_{1,2}$ have much longer lifetimes than $\pid_{3}$.

The above considerations make it clear that, even for the minimal dark pion theory with $N = 2$, the parameter space is too vast to be covered systematically in this first study. Therefore we choose to discuss a few benchmark scenarios that give rise to distinct phenomenology. With these, our aim is to be illustrative rather than exhaustive, and we expect that other interesting patterns may be found in future work. We begin with a few comments on the case of stable dark pions and their possible role as dark matter, and then turn to the study of three benchmark scenarios where at least some of the pions are unstable and decay to SM particles. The key features of these three are summarized in Table~\ref{tab:mesonproperties}. For each scenario, in the phenomenological analysis we fix generic textures for the Yukawa and mass matrices, paying attention to avoiding enhanced symmetry points. This reduces the number of independent parameters to a manageable handful.

\subsection*{Scenario 0: Isospin-symmetric limit and dark pion dark matter}
As already mentioned, for $\bomega, \bmass, \byuk, \byukt \propto \mathbf{1}_2$ the dark pions form a stable triplet of $SU(2)_V$, which is a dark matter candidate.\footnote{Note that in this case the theory contains $1$ physical phase, for any $N$.} However, in this limit the dark pions do not couple to the $Z$ (see Appendix~\ref{app:dark_ChPT}), hence reducing their cosmological abundance to a viable level requires adding extra ingredients to the theory. For $N\geq 3$ the number density can be depleted via $3\to 2$ processes mediated by the Wess-Zumino-Witten action, potentially realizing Strongly Interacting Massive Particle (SIMP) dark matter~\cite{Hochberg:2014kqa}, although an additional mediation between the dark pions and the SM should still be introduced to transfer the dark matter entropy to the SM. If the mediator is a dark photon that mixes kinetically with the hypercharge~\cite{Lee:2015gsa,Hochberg:2015vrg,Katz:2020ywn}, care must be taken to check dangerous decays of singlet pions, which can be made viable through appropriate mass splittings for odd $N$~\cite{Berlin:2018tvf,Katz:2020ywn}, or prevented by imposing suitable discrete symmetries for even $N$~\cite{Berlin:2018tvf}. Such scenarios provide appealing origins for light thermal dark matter, but as they are rather tangential to the central aspects of this work, we do not discuss them further.

\renewcommand{\tabcolsep}{10pt}
\renewcommand{\arraystretch}{1.25}
\begin{table}[t]
\centering
\begin{tabular}{c|c|c|c|c|c|c}
\multirow{2}{*}{Scenario} & \multicolumn{3}{c|}{Symmetries possessed}  & \multicolumn{3}{c}{Decay portals}  \\
  & $\bm{\widetilde{Y}} = 0$ & exact $U(1)$ & exact $CP$  & {\color{blue}$\hat{\pi}_1$} & {\color{blue}$\hat{\pi}_2$} & {\color{blue}$\hat{\pi}_3$} \\ \hline
Section~\ref{sec:scenario1} & \checkmark  & \ding{55} & \ding{55} &  $Z$ & $Z$ & $Z$ \\ 
Section~\ref{sec:scenario2} & \ding{55} & \checkmark & \ding{55} & stable & stable & $Z,h$ \\ 
Section~\ref{sec:scenario3} & \ding{55} & \ding{55} & \checkmark &  $Z$ & $h$ & $Z$ \\ 
\end{tabular}
\caption{Summary of the benchmark scenarios considered in this work. A chiral symmetry can be the origin of $\bm{\widetilde{Y}} = 0$. In the second scenario, $\hat{\pi}_{\pm} = (\hat{\pi}_1 \mp i \hat{\pi}_2)/\sqrt{2}$ is stable because it is the lightest particle charged under a dark $U(1)$.}
\label{tab:mesonproperties}
\end{table}

\subsection{Scenario 1: \texorpdfstring{$\widetilde{\bm{Y}} = 0$}{p}}\label{sec:scenario1}
In this case there is no constraint from the invisible decay branching ratio of the Higgs. In general $\byuk$ contains $1$ physical phase, which can be parametrized, e.g., as
\begin{equation}
\bm{Y} =  \begin{pmatrix} y_{11} & y_{12} e^{i \alpha} \\ y_{21} & y_{22} \end{pmatrix}
\end{equation}
with real $y_{ij}$. It is convenient to perform a further redefinition which renders $\byuk^\dagger \bm{M}^{-2} \byuk$ real,\footnote{We assume $\cos\alpha\, y_{11} y_{12}/ M_1^2  + y_{21} y_{22}/M_2^2 > 0$ for definiteness.}
\begin{equation}
\psi_{R1} \to e^{i \hat{\alpha}} \psi_{R1}, \quad \hat{\alpha} = \arctan \frac{\sin \alpha \frac{y_{11} y_{12}}{M_1^2}}{\cos\alpha \frac{y_{11} y_{12}}{M_1^2} + \frac{y_{21}y_{22}}{M_2^2}} \quad \to \quad \bm{Y} =  \begin{pmatrix} e^{i \hat{\alpha}} y_{11} & y_{12} e^{i \alpha} \\  e^{i \hat{\alpha}} y_{21} & y_{22} \end{pmatrix}\, ,
\end{equation}
and the same for $\psi_{L1}$ so that the mass matrix remains real.\footnote{The low-energy quark masses are given by
\begin{equation*}
\bm{m}_\psi = \bm{\omega} \Big( \mathbf{1} - \frac{v^2}{4} \byuk^\dagger \bm{M}^{-2} \byuk \Big),
\end{equation*}
as obtained after applying the leading-order EOM to the first term on the right-hand side of Eq.~\eqref{eq:eft}. This is diagonalized to $\bm{m}_{\psi^\prime}$ by $\psi_{L,R} = U_{L,R} \psi^\prime_{L,R}\,$, but in practice we neglect the $(Yv/M)\,$-$\,$suppressed corrections.}

All three dark pions are unstable. As anticipated, $\hat{\pi}_{1}$ and $\hat{\pi}_{3}$ have unsuppressed decay to SM particles via the $Z$ portal, so their lifetimes and branching ratios can be directly obtained from Appendix~\ref{app:ALP_decays_general}. Since $\byukt = 0$, instead of the Higgs portal $\hat{\pi}_2$ decays through $CP$-violating mixing with the other pions. To estimate its lifetime, we need to take into account several corrections to the leading-order pion Lagrangian:
\begin{itemize}
\item The pion mass splitting generated by $O(p^4)$ ChPT operators with insertions of the quark mass matrix, e.g.,
\begin{equation} \label{eq:c7_operator}
\frac{c_7 \hat{B}_0^2}{(4\pi)^2} \Big( \mathrm{Tr} [\bm{m}_{\psi'} U^\dagger - U \bm{m}_{\psi'}^\dagger ] \Big)^2 \supset -\, \frac{c_7 \hat{B}_0^2}{4\pi^2 f^2_{\hat{\pi}}} (\omega_1 - \omega_2)^2 \hat{\pi}_3^2\,,
\end{equation}
where $c_7$ is a coefficient expected to be of $O(1)$ by naive dimensional analysis. For generic dark isospin breaking $|\omega_1 - \omega_2|/(\omega_1 + \omega_2) \sim O(1)$, as assumed here, Eq.~\eqref{eq:c7_operator} is the leading correction to the pion masses. Therefore, to estimate the $CP$-violating decay of $\hat{\pi}_2$ we can focus only on its mixing with $\hat{\pi}_1$.\footnote{Notice that if $\omega_1 = \omega_2$ the theory actually preserves $CP$, because the phase $\alpha$ can be removed by a $U(2)$ rotation of the $\psi$ fields. The same applies if $M_1 = M_2$.}

\item The effects of tree-level $Z$ exchange,
\begin{equation}
- \frac{f_{\hat{\pi}}^2 v^2}{32} \Big( \mathrm{Tr}[\sigma_a \byuk^\dagger \bm{M}^{-2} \byuk  ]\, \partial_\mu \hat{\pi}_a \Big)^2 \,,
\end{equation}
which correct the kinetic term of $\hat{\pi}_1$ (and $\hat{\pi}_3$), but not $\hat{\pi}_2$.

\item The one-loop contributions from box diagrams, with parametric scaling
\begin{equation}
- \frac{f_{\hat{\pi}}^2}{32\pi^2} \sum_{a, b}  \mathrm{Tr} [ \sigma_a \byuk^\dagger \bm{M}^{-2} \byuk  ]\, \mathrm{Tr} [ \sigma_b \byuk^\dagger \byuk  ] \, \partial_\mu \hat{\pi}_a \partial^\mu \hat{\pi}_b\, .
\end{equation}
These yield in particular a $CP$-violating mixing of $\hat{\pi}_1$ and $\hat{\pi}_2$, provided
\begin{equation}
\frac{1}{2} \mathrm{Tr} [ \sigma_2 \byuk^\dagger \byuk  ] = \sin\hat{\alpha}\, y_{21} y_{22} + \sin (\hat{\alpha} - \alpha) y_{11} y_{12}
\end{equation}
is nonvanishing. This is the case if $\alpha \neq 0$, all $y_{ij} \neq 0$, and $M_1 \neq M_2$ (if $M_1 = M_2$, $\byuk^\dagger \byuk$ is real because $\byuk^\dagger \bm{M}^{-2} \byuk$ is).
\end{itemize}
Once the above effects are included, the kinetic terms for $\hat{\pi}_{1,2}$ are made canonical by the rotation
\begin{equation} \label{eq:pi1pi2mixing}
\begin{pmatrix} \hat{\pi}_1 \\ \hat{\pi}_2 \end{pmatrix} \to \begin{pmatrix} c_{\theta_{12}} & s_{\theta_{12}} \\  - s_{\theta_{12}} & c_{\theta_{12}} \end{pmatrix} \begin{pmatrix} \hat{\pi}_1 \\ \hat{\pi}_2 \end{pmatrix},  \qquad \tan 2\theta_{12} = - \frac{ \mathrm{Tr}(\sigma_2 \byuk^\dagger  \byuk )}{\pi^2 v^2 \mathrm{Tr}(\sigma_1 \byuk^\dagger \bm{M}^{-2} \byuk ) + \mathrm{Tr}(\sigma_1\byuk^\dagger  \byuk )}\,.
\end{equation}

\noindent To understand quantitatively the dark pion properties we focus on the following pattern for the $\bm{Y}$ and $\bm{M}$ matrices,
\begin{equation}
\frac{y_{11}}{\sqrt{2}} = \sqrt{2}\, y_{12} = 3 y_{21} = y_{22} = y\,, \quad \alpha =\frac{ \pi}{3}\,,\qquad \frac{2}{3} M_1  = M_2 = M,
\end{equation}
which is of generic nature. From Eqs.~\eqref{eq:Leff_quarks} and~\eqref{eq:pi1pi2mixing} we find the effective decay constants of $\hat{\pi}_{1,3}\,$ and the $CP$-violating $\hat{\pi}_1$-$\hat{\pi}_2$ mixing angle, respectively,
\begin{equation} \label{eq:benchmark1_params}
f_a^{(1)} \approx 3.0\, \frac{M^2}{y^2 f_{\hat{\pi}}}\,,\qquad f_a^{(3)} \approx - 18\, \frac{M^2}{y^2 f_{\hat{\pi}}}\,,\qquad \tan 2\theta_{12} \approx \frac{0.20}{1+0.036 \left( \frac{4\pi v}{M} \right)^2}\,.
\end{equation}
The decay width of the physical $\hat{\pi}_2$ is then $\Gamma_{\hat{\pi}_2} \approx \sin^2 \theta_{12}\, \Gamma_{\hat{\pi}_1}$. Dark pion decays are mainly controlled by the three parameters $y/M, f_{\hat{\pi}},$ and $m_{\hat{\pi}}$. The mediation strength is constrained by the $Z$ invisible width: Eq.~\eqref{eq:Z_darkfermions} gives $y/M \lesssim 1.1\, \mathrm{TeV}^{-1}$ (assuming $N_d = 3$).
\begin{figure}[t]
\begin{center}
\includegraphics[width=0.49\textwidth]{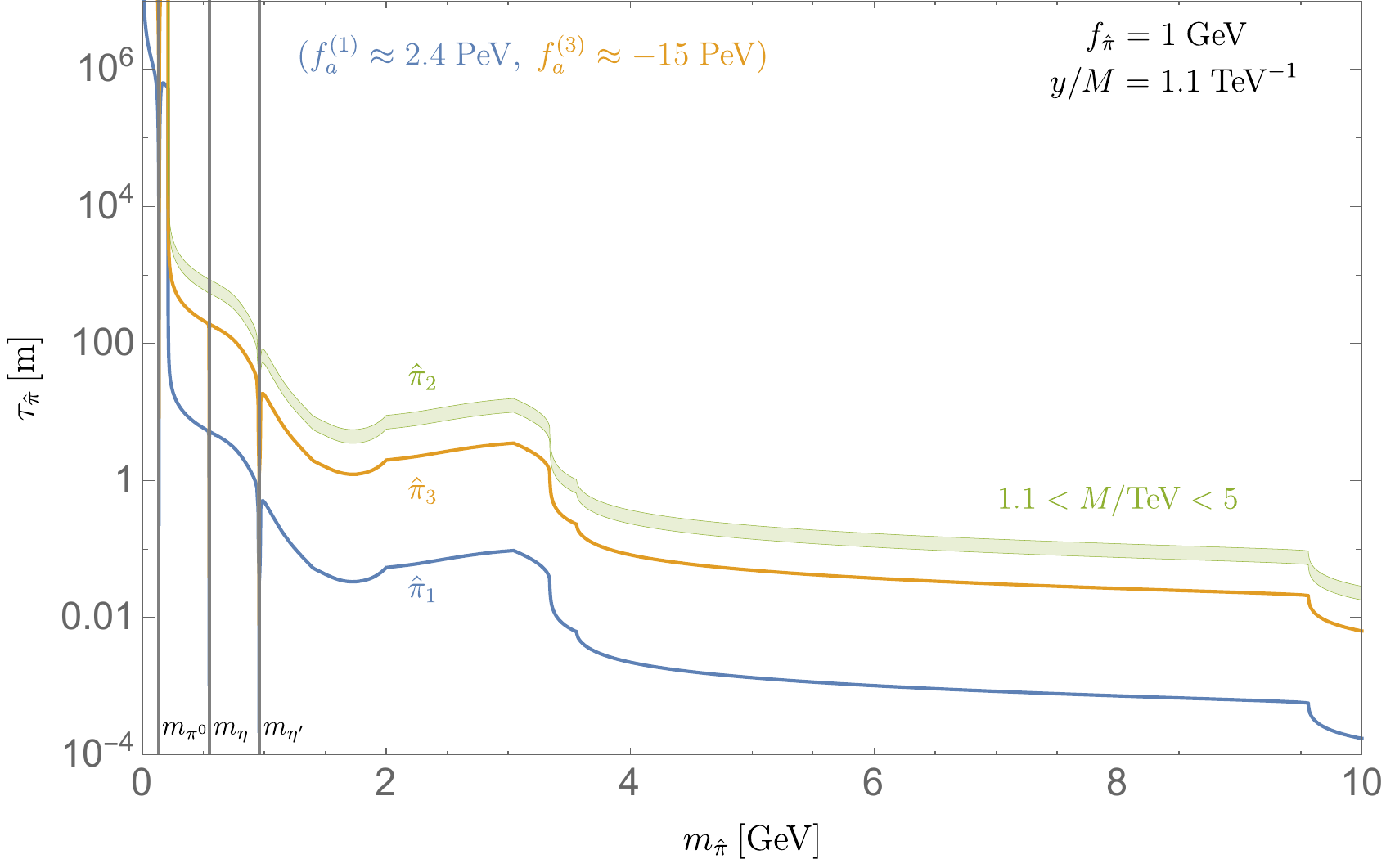}
\includegraphics[width=0.49\textwidth]{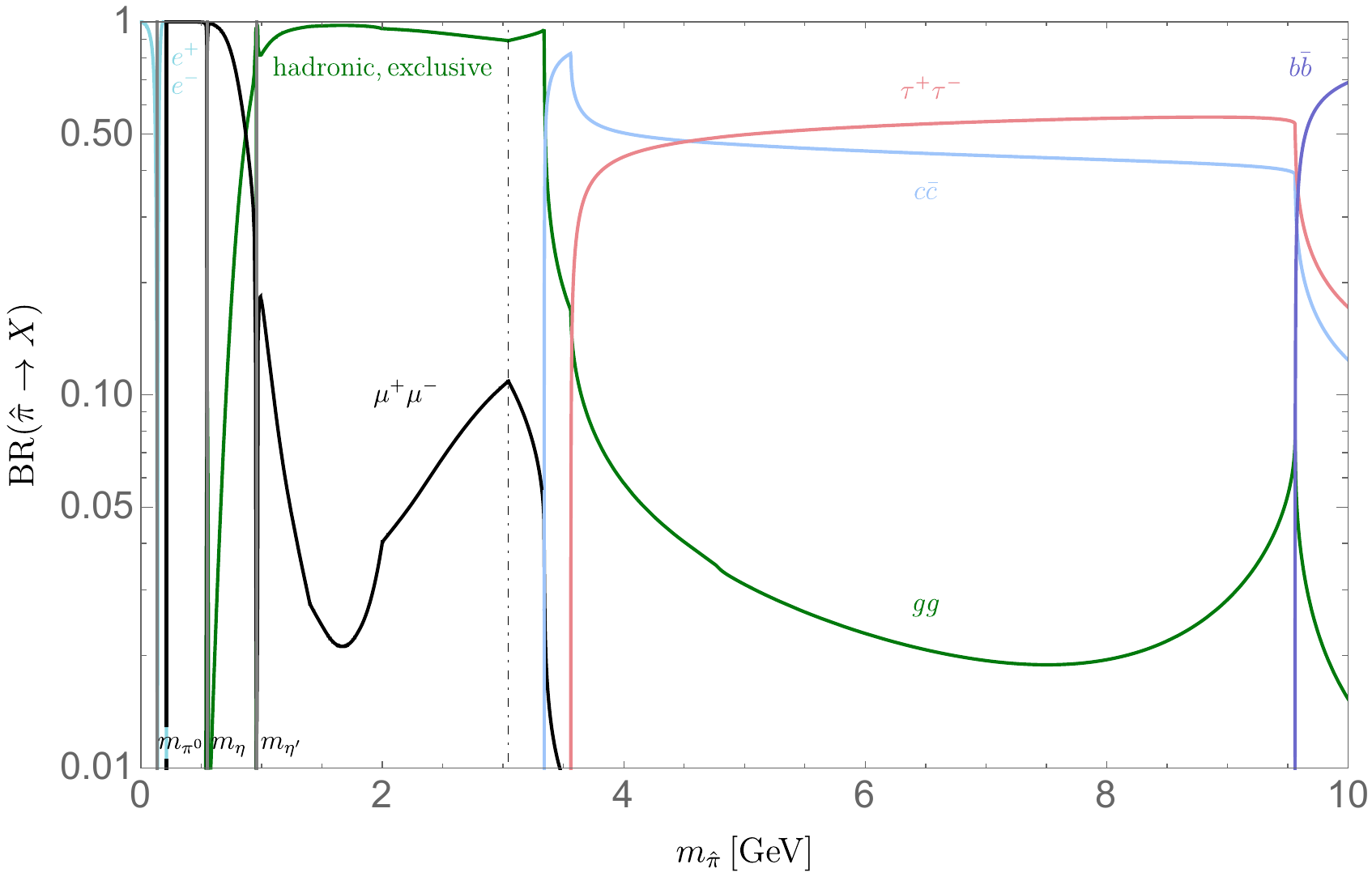}
\caption{\label{fig:Benchmark1} Lifetimes {\it (left)} and branching ratios {\it (right)} of the dark pions for $\bm{\widetilde{Y}} = 0$. In the left panel we have fixed $y/M$ to saturate the $Z \to \mathrm{invisible}$ bound and $f_{\hat{\pi}} = 1\;\mathrm{GeV}$. In the right panel only selected decay channels are shown; see Fig.~\ref{fig:ALP_widths} for the complete picture. The vertical dot-dashed line indicates the value of $m_{\hat{\pi}}$ where our description of ALP decays transitions from exclusive hadronic final states to perturbative QCD.}
\end{center}
\end{figure}

The dark pion lifetimes are shown in the left panel of Fig.~\ref{fig:Benchmark1}, choosing $y/M$ that saturates the LEP bound and fixing $f_{\hat{\pi}} = 1\;\mathrm{GeV}$; other results are obtained by rescaling $\tau \propto f_{\hat{\pi}}^{-2} (y/M)^{-4}$, see Eq.~\eqref{eq:benchmark1_params}. Remarkably, for $y \sim 1$, $M\sim 1\;\mathrm{TeV}$ and $f_{\hat{\pi}} \sim 1\;\mathrm{GeV}$, i.e. parameter choices motivated by (neutral) naturalness~\cite{Cheng:2019yai}, the lifetime of $\hat{\pi}_1$ falls between 10 meters and 1 millimeter across the mass range $2m_\mu \lesssim m_{\hat{\pi}} \lesssim 2 m_b$. Therefore, this dark pion is a {\it natural LLP target} for present and future experiments. On the other hand, $\hat{\pi}_{2}$ and $\hat{\pi}_{3}$ have much longer lifetimes. As $\theta_{12}$ depends on $M$ but not on $y$, for illustration we show the range of $\tau_{\hat{\pi}_2}$ obtained by varying $M\in [1.1,5]\;\mathrm{TeV}$, where the lower edge corresponds to the current bound on the $Q$ mass from direct searches at the LHC (see Section~\ref{sec:UV}). We stress that for the $\hat{\pi}_2$ lifetime we have performed an estimate, rather than a precise calculation, as sufficient for our purpose. The right panel of Fig.~\ref{fig:Benchmark1} shows selected branching ratios, which are the same for the three dark pions as they all decay through the $Z$ portal (if the $CP$-violating mixing with the other pions is very suppressed, $\hat{\pi}_2$ may decay through Higgs mediation via a small $\byukt \neq 0$, but we do not study that possibility here).

This benchmark scenario provides a theoretically motivated and remarkably simple target for current and future experimental probes. The constraints from and future opportunities in FCNC meson decays are discussed in Section~\ref{sec:FCNC}, whereas the prospects for discovery at the LHC via $Z$ decays to dark showers are presented in Section~\ref{sec:dark_showers}.

\subsection{Scenario 2: exact $U(1)$ }\label{sec:scenario2}

If the $U(1)$ symmetry $\{\psi, Q\} \to e^{i x \sigma_3} \{\psi,Q \}$ is preserved, the Yukawa matrices are diagonal. Parametrizing the two physical phases as $\byuk = \mathrm{diag}\,(y_1, y_2)$, $\byukt = \mathrm{diag}\,(\tilde{y}_1 e^{i \alpha_1}, \tilde{y}_2 e^{i \alpha_2})$, the EFT quark mass matrix~\eqref{eq:seesawmass} is diagonal but complex, and is transformed into a real and positive $\bm{m}_{\psi'}$ by the rephasings\footnote{We assume $\omega_i - \cos \alpha_i\, y_i \tilde{y}_i v^2/(2 M_i) > 0\,$.}
\begin{equation}
\psi_{L} \to U_L \psi_{L} \,, \quad  U_L = \mathrm{diag}\,(e^{i \hat{\alpha}_1}, \, e^{i \hat{\alpha}_2} ), \quad \hat{\alpha}_i = \arctan \frac{\sin\alpha_i \frac{y_i \tilde{y}_i v^2}{2 M_i} }{\omega_i - \cos \alpha_i \frac{y_i \tilde{y}_i v^2}{2 M_i} }\,,
\end{equation}
which also leave the (real) $Z\overline{\psi} \psi$ coupling matrix $\byukt^\dagger \bm{M}^{-2} \byukt$ unaffected. While the charged pion $\hat{\pi}_{\pm}$ is stable, $\hat{\pi}_0$ decays through the $Z$ portal and, in the presence of $CP$ violation, the Higgs portal. The cosmological history can be easily safe. The mass splitting of charged and neutral pions is controlled by the operator in Eq.~\eqref{eq:c7_operator}, which yields $m_{\hat{\pi}_0} < m_{\hat{\pi}_+}$ if $c_7 < 0$. Then $\hat{\pi}_+ \hat{\pi}_- \to \hat{\pi}_0 \hat{\pi}_0$ conversions followed by decays of $\hat{\pi}_0$ to the SM with lifetime $\tau_{\hat{\pi}_0} \ll 1\;\mathrm{s}$, as realized throughout the interesting parameter space, result in a very small $\hat{\pi}_\pm$ relic density without affecting Big Bang nucleosynthesis~\cite{Freytsis:2016dgf}.

To simplify the analysis of the parameter space we assume $\omega_i = \frac{y_i \tilde{y}_i v^2}{M_i}\cos\alpha_i$, which can be regarded as a particularly simple case of the scenario where $\omega$ and $Y \tilde{Y} v^2/M$ are of the same order. This choice leads to $\hat{\alpha}_i = \alpha_i$ and $m_{\psi'_i} = y_i \tilde{y}_i v^2/(2M_i)$.

In addition, we choose the generic patterns
\begin{equation}
 3 y_1 = y_2 = y\,, \qquad \frac{1}{3}\tilde{y}_1 =  \tilde{y}_2 = \tilde{y}\,, \quad - \frac{3}{4} \alpha_{1} = \alpha_2 = - \frac{\pi}{4}\,, \qquad  \frac{1}{2}M_1 = M_2 = M\,,
\end{equation}
giving the bounds from invisible Higgs and $Z$ decays 
\begin{equation}
\frac{y \tilde{y}}{ M } \lesssim 0.023\; \mathrm{TeV}^{-1}\,, \qquad \frac{(y \tilde{y})^{1/2}}{M} (r^2 + 6.1\, r^{-2})^{1/4} \lesssim 1.4\;\mathrm{TeV}^{-1}\,,
\end{equation}
respectively, obtained from Eqs.~\eqref{eq:h_psipsi_general} and~\eqref{eq:Z_darkfermions}. We have defined $r \equiv y/\tilde{y}$. The dark pion mass reads
\begin{equation} \label{eq:dark_pion_mass_benchmark2}
m_{\hat{\pi}} = \Big( 3 \pi f_{\hat{\pi}} \frac{y \tilde{y} v^2}{ M } \Big)^{1/2}\;\; \stackrel{h\,\to\, \mathrm{inv}}{\lesssim}\;\; 3.6\;\mathrm{GeV} \left( \frac{f_{\hat{\pi}}}{\mathrm{GeV}} \right)^{1/2} ,
\end{equation}
where we have set $\hat{B}_0 = 4\pi f_{\hat{\pi}}$.
\begin{figure}[t]
\begin{center}
\includegraphics[width=0.49\textwidth]{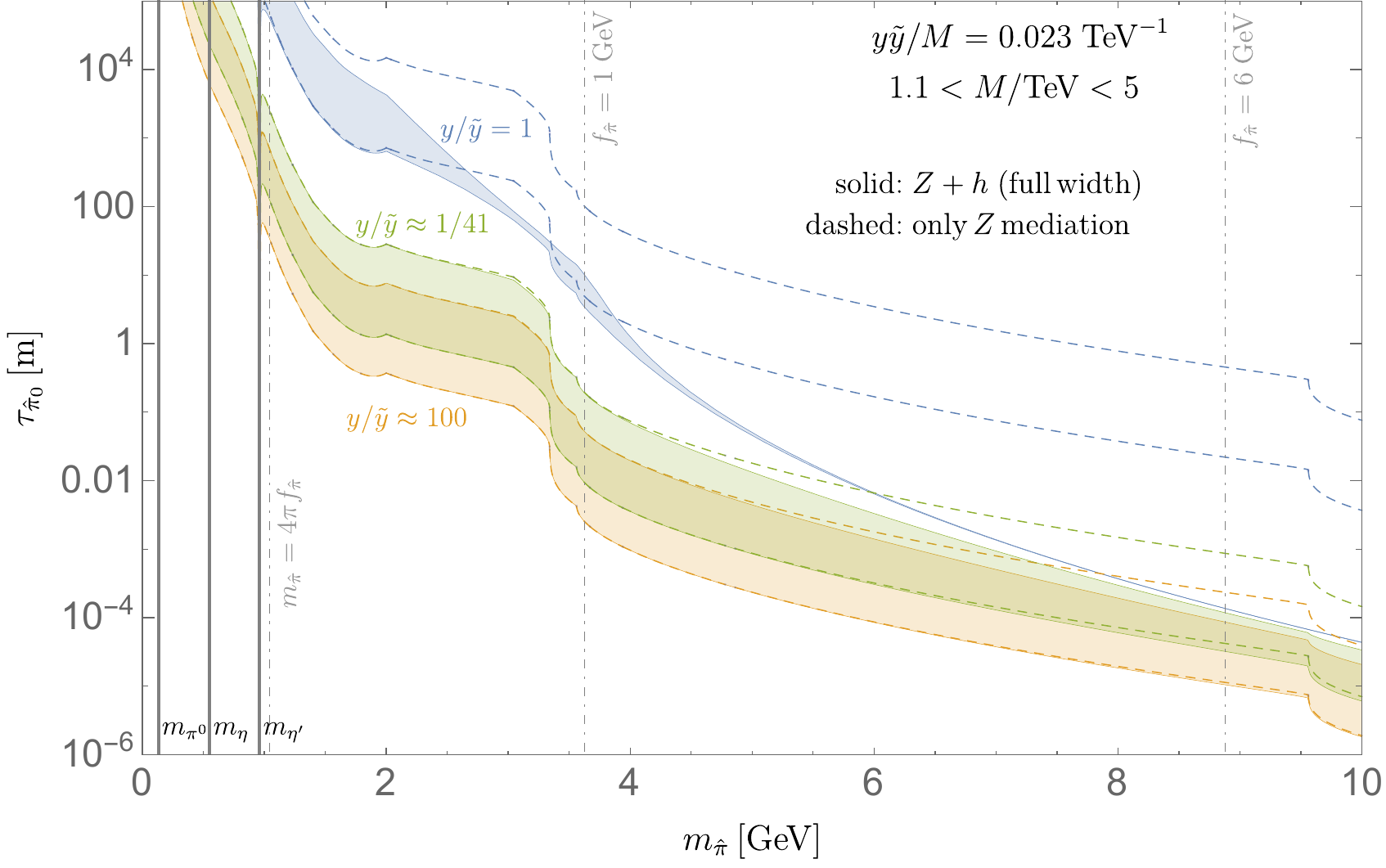}\\\vspace{2mm}
\includegraphics[width=0.49\textwidth]{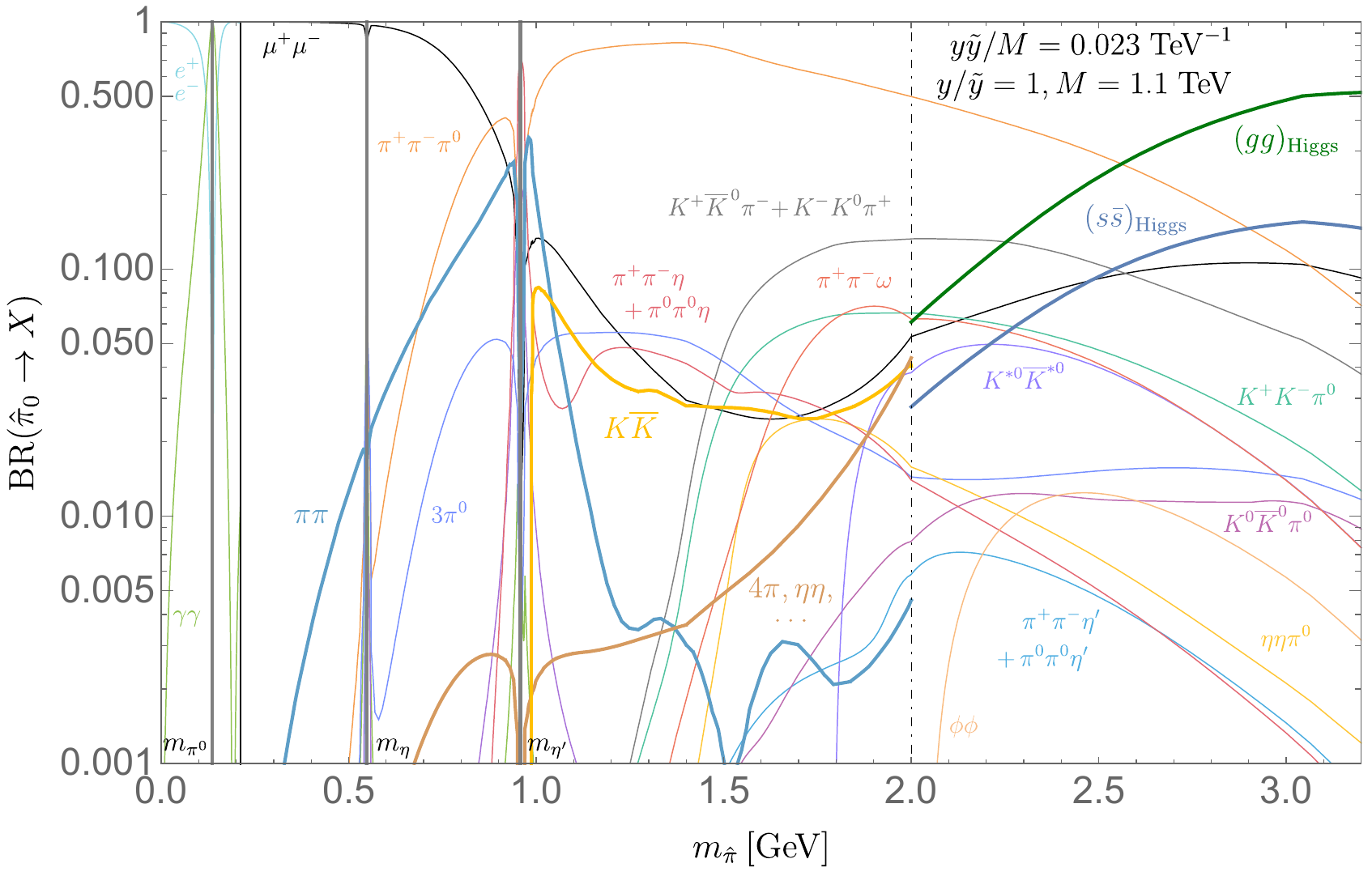}
\includegraphics[width=0.49\textwidth]{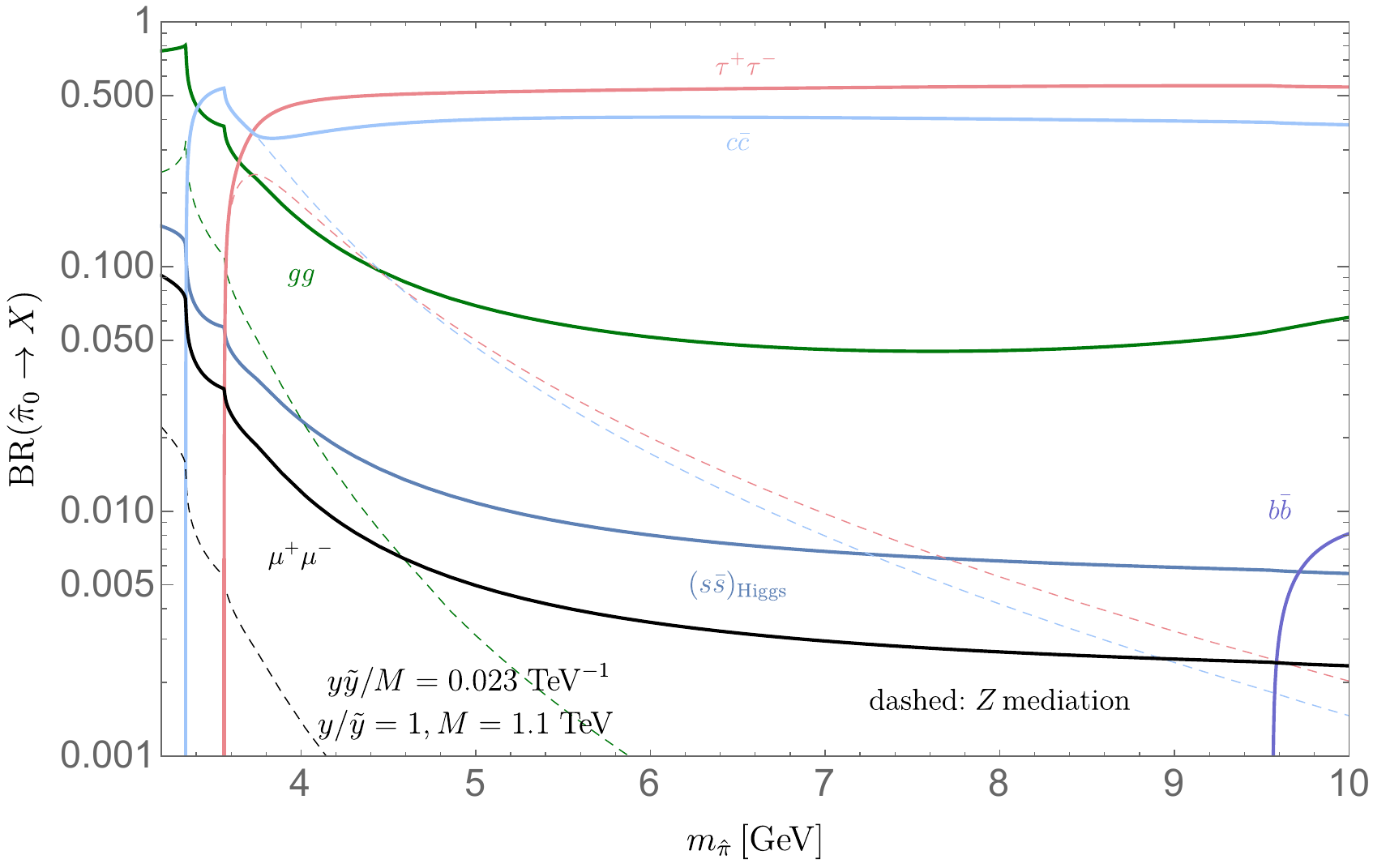}
\caption{\label{fig:Benchmark2} {\it (Top)} lifetime of $\hat{\pi}_0$ for exact dark $U(1)$. We have set $y\tilde{y}/M$ to saturate the $h\to \mathrm{invisible}$ bound, while $f_{\hat{\pi}}$ is fixed by Eq.~\eqref{eq:dark_pion_mass_benchmark2} as a function of $m_{\hat{\pi}}$ (representative values are indicated by vertical dot-dashed lines). Filled bands display the full width varying $1.1 < M/\mathrm{TeV} < 5$, for three values of $r = y/\tilde{y}$, where $r \approx 100$ and $r \approx 1/41$ are chosen to saturate the $Z\to \mathrm{invisible}$ bound for $M = 1.1$~TeV. Dashed curves indicate the same, but including only the $Z$-mediated contribution to the width. {\it (Bottom)} $\hat{\pi}_0$ branching ratios for $r = 1, M = 1.1$~TeV, and $y\tilde{y}/M$ saturating the $h\to \mathrm{invisible}$ bound. In the left panel, the hadronic channels mediated by Higgs exchange are indicated by thicker curves. For $\gamma\gamma$, $e^+ e^-$ and $\mu^+ \mu^-$ the $Z$- and $h$-exchange contributions were summed. The vertical dot-dashed line indicates $m_{\hat{\pi}} = 2$~GeV, where the description of $CP$-even decays transitions from exclusive hadronic final states to a perturbative spectator model. In the right panel, dashed curves show the contribution to the BRs of the $Z$-mediated widths.}
\end{center}
\end{figure}

Owing to the $CP$ violation, the decays of $\hat{\pi}_0$ are an intricate combination of $Z$- and $h$-mediated processes. For the former, the effective decay constant is found from Eq.~\eqref{eq:Leff_quarks}, 
\begin{equation}
f_a^{(0)} \approx -\, 4.1 \, \frac{M^2}{y \tilde{y} f_{\hat{\pi}}} ( r + 1.3\, r^{-1})^{-1}\,.
\end{equation}
For the latter, the coupling of the dark quarks to the Higgs reads $\overline{\psi}_L \bm{B} \psi_R h + \mathrm{h.c.}$ with $\bm{B} = v\, U_L^\dagger \byukt^\dagger \bm{M}^{-1} \byuk$, and we can apply Eq.~\eqref{eq:Higgs_mixing} with 
\begin{equation}
\frac{1}{2v}\mathrm{Tr}[i\sigma_3  (\bm{B} - \bm{B}^\dagger)] = \frac{y_1 \tilde{y}_1}{M_1} s_{\alpha_1 + \hat{\alpha}_1} -  \frac{y_2 \tilde{y}_2}{M_2} s_{\alpha_2 + \hat{\alpha}_2} \approx 1.4\, \frac{y\tilde{y}}{M}\; \to \; s_\theta^{(0)} \approx 5.7\, \pi f_{\hat{\pi}}^2 \frac{y \tilde{y} v}{M(m_h^2 - m^2_{\hat{\pi}})}\,.
\end{equation}
The first equality holds in general and shows that $s_\theta^{(0)}$ would vanish for $\bm{\omega} = 0$, as a consequence of $\sin(\alpha_i + \hat{\alpha}_i) = 0$. At this stage we can take $\{y\tilde{y}/M, M, r, m_{\hat{\pi}} \}$ as the four independent free parameters, with $f_{\hat{\pi}}$ fixed via Eq.~\eqref{eq:dark_pion_mass_benchmark2}. In Fig.~\ref{fig:Benchmark2}, we set $y\tilde{y}/M$ to its upper bound from $h\to \mathrm{invisible}$ and explore the remaining three-dimensional parameter space. As expected, the $\hat{\pi}_0$ lifetime depends strongly on $r$: if $M \sim O(\mathrm{TeV})$, for $r \gg 1$ or $\ll 1$ the $Z$ portal dominates (with branching ratios that are well described by Fig.~\ref{fig:ALP_widths}). Note that the dependence of the dark pion lifetime on its mass is very different from scenario 1, as can be observed by comparing with Fig.~\ref{fig:Benchmark1}. The reason is that, while in scenario 1 $f_{\hat{\pi}}$ is independent from $m_{\hat{\pi}}$, here Eq.~\eqref{eq:dark_pion_mass_benchmark2} dictates the scaling $f_{\hat{\pi}} \propto m_{\hat{\pi}}^2$, resulting in a much shorter lifetime as the dark pion mass increases. 

Conversely, for $r \sim 1$ the Higgs portal plays an important role, dominating the total width for $m_{\hat{\pi}} \gtrsim 2\,$-$\,3$~GeV. For this reason, in the bottom panels of Fig.~\ref{fig:Benchmark2} we show the branching ratios at $r = 1$, which best illustrate the complexity of the decay pattern. If $m_{\hat{\pi}} \lesssim 2$~GeV the branching ratio to the $CP$-even $K\overline{K}$ final state is of several percent, which could rise up to $\sim 15\%$ in the region below the $c\bar{c}$ threshold, although the description adopted here~\cite{Winkler:2018qyg}, based on the $s\bar{s}$ final state in a perturbative spectator model, does not permit a more accurate prediction. For $m_{\hat{\pi}}$ above the $c\bar{c}$ threshold Higgs exchange completely dominates the width, yielding the interesting prediction that a heavier (and therefore shorter-lived) $\hat{\pi}_0$ mainly decays to $CP$-even final states if the dark Yukawa interactions contain sizable $CP$ violation.

\subsection{Scenario 3: exact $CP$} \label{sec:scenario3}
\begin{figure}[t]
\begin{center}
\includegraphics[width=0.49\textwidth]{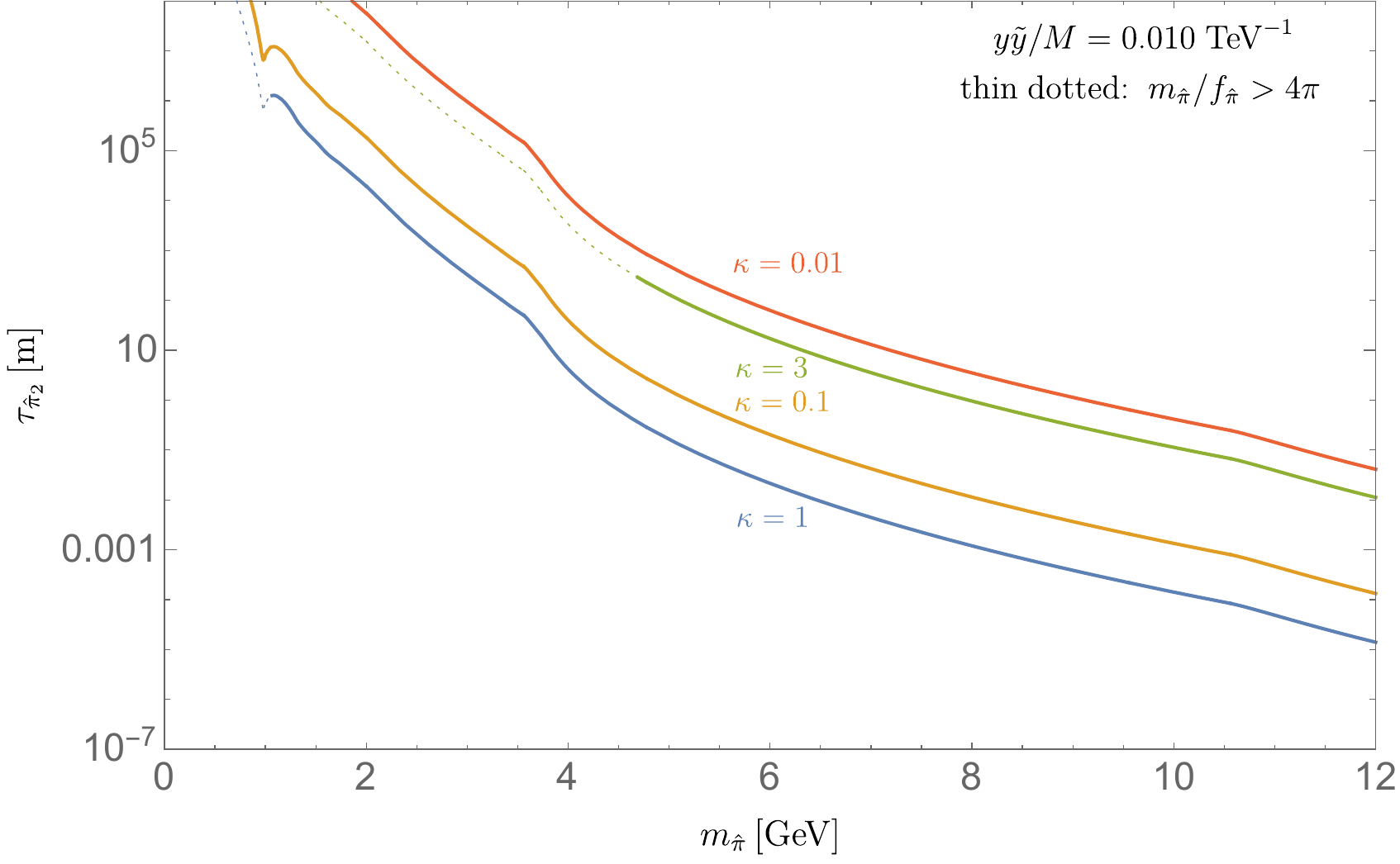}
\includegraphics[width=0.49\textwidth]{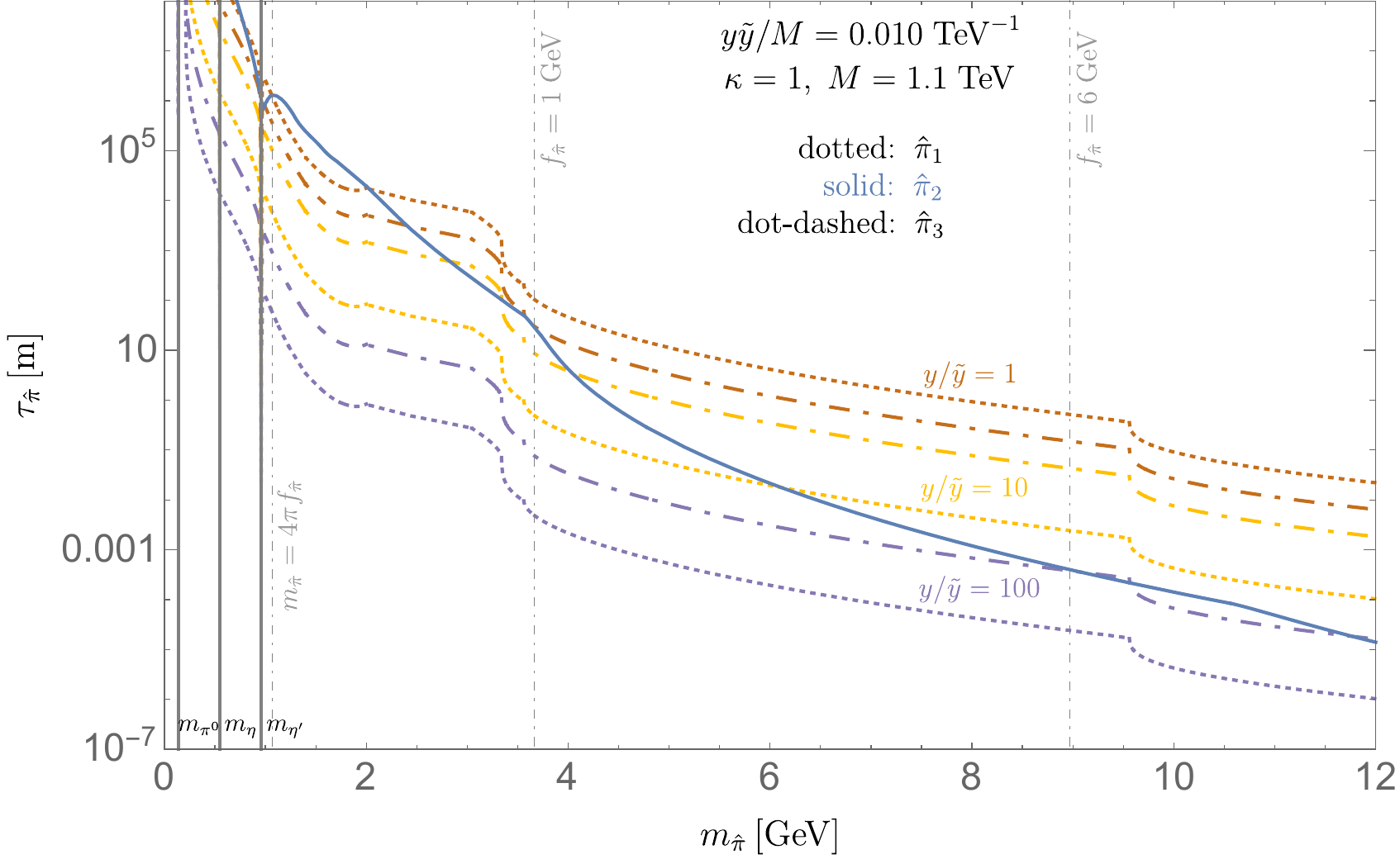}\\ \vspace{2.5mm}
\includegraphics[width=0.49\textwidth]{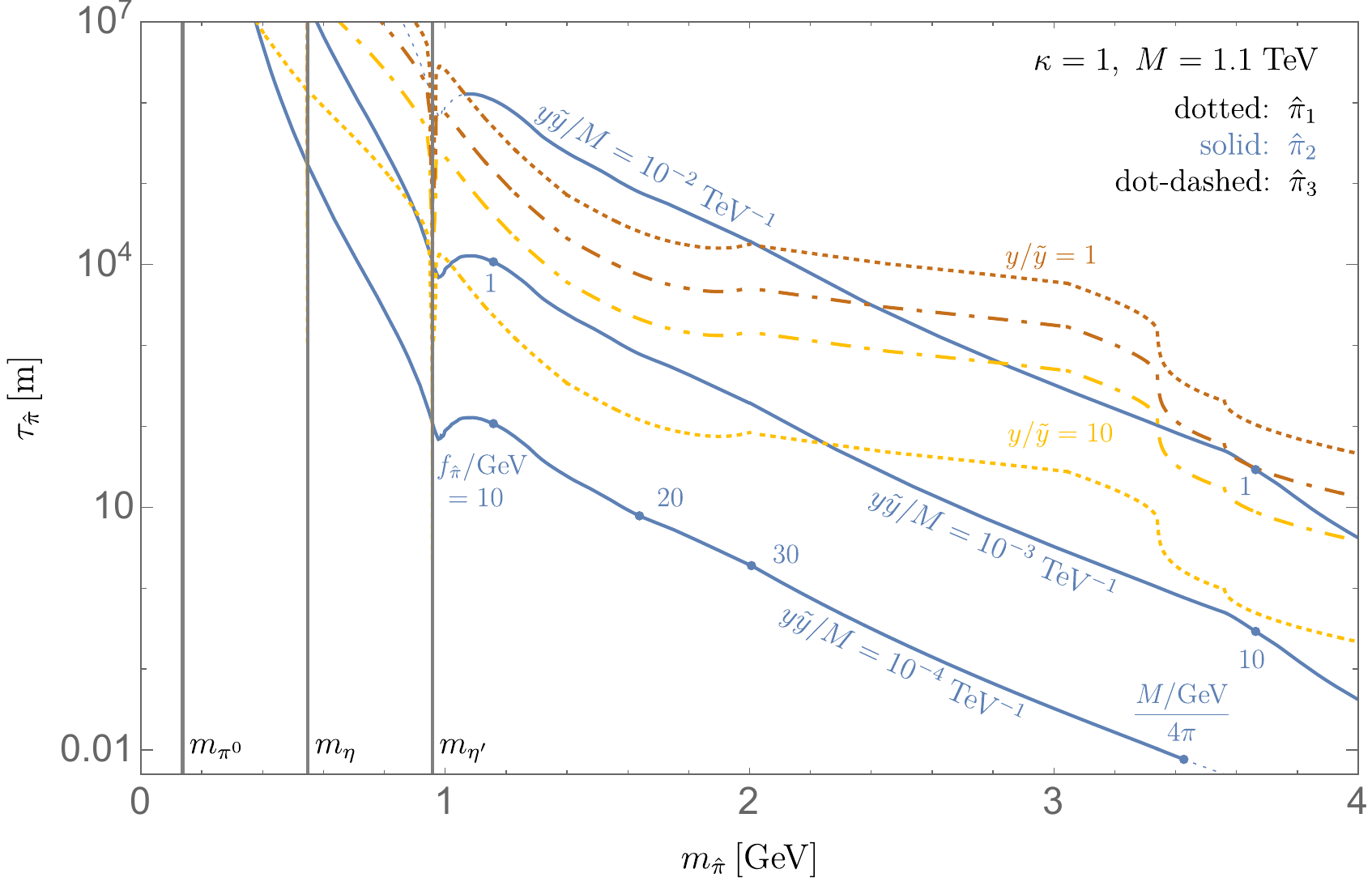}\vspace{2mm}
\caption{\label{fig:Benchmark3} Lifetimes of the dark pions for a $CP$-conserving dark sector. {\it (Top left)} lifetime of $\hat{\pi}_2$ for representative values of $\kappa$, setting $y\tilde{y}/M$ to saturate the $h\to \mathrm{inv}$ bound. Note that for a given $m_{\hat{\pi}}$, to each curve corresponds a different value of $f_{\hat{\pi}}$ via Eq.~\eqref{eq:dark_pion_mass_benchmark3}; the thin dotted portions have $m_{\hat{\pi}} > 4\pi f_{\hat{\pi}}$ and are therefore not physical. {\it (Top right)} lifetimes of the three dark pions for $\kappa = 1$, $M$ saturating the constraint from direct $Q$ searches, $y\tilde{y}/M$ saturating the $h\to \mathrm{inv}$ bound, and illustrative choices of $y/\tilde{y}$ consistent with the $Z\to \mathrm{inv}$ bound. {\it (Bottom)} zooming in on light dark pions with $m_{\hat{\pi}} \lesssim 2\,m_{c,\tau}$ and showing the effect of decreasing $y\tilde{y}/M$. Due to a compensating increase of $f_{\hat{\pi}}$ via Eq.~\eqref{eq:dark_pion_mass_benchmark3}, the $\hat{\pi}_{1,3}$ lifetimes are not affected, whereas the one of $\hat{\pi}_2$ is {\it shortened}. Dots mark representative values of $f_{\hat{\pi}}$ in GeV.} 
\end{center}
\end{figure}

The third and last scenario we consider is one where $CP$ is exactly preserved by the dark Yukawa interactions. As in scenarios 1 and 2, to reduce the number of independent parameters the Yukawa and mass matrices are set to definite patterns. These are chosen to be of generic nature, avoiding points of enhanced symmetry. We take
\begin{equation}
\bm{Y} = \begin{pmatrix} \sqrt{2} & 1/\sqrt{2} \\ 1/3 & 1 \end{pmatrix} y\,, \qquad \bm{\widetilde{Y}} = \begin{pmatrix} 4/3 & 1/5 \\ 1 & 1 \end{pmatrix} \tilde{y}\,, \qquad \bm{M} = \begin{pmatrix} 3/2 & \\ & 1 \end{pmatrix} M\,,
\end{equation}
leading to the $h,Z \to \mathrm{invisible}$ bounds,
\begin{equation}
\frac{y \tilde{y}}{ M } \lesssim 0.010\; \mathrm{TeV}^{-1}\,, \qquad \frac{(y \tilde{y})^{1/2}}{M} (r^2 + 1.8\, r^{-2})^{1/4} \lesssim 1.0\;\mathrm{TeV}^{-1}\,.
\end{equation}
In addition we take $\{ \omega_1, \omega_2 \} = \kappa \{1, 2 \} y \tilde{y} v^2/M$, with $\kappa$ being a dimensionless free parameter, so that after mass diagonalization $\mathrm{Tr}(\bm{m}_{\psi'}) = c(\kappa) y \tilde{y} v^2/M$ where $c$ is a dimensionless function. The pion mass is from Eq.~\eqref{eq:pion_mass_chpt}
\begin{equation} \label{eq:dark_pion_mass_benchmark3}
m_{\hat{\pi}} = \left( 4\pi c(\kappa) f_{\hat{\pi}}  \frac{y \tilde{y} v^2}{M} \right)^{1/2}.
\end{equation}
As a consequence of $CP$ invariance, $\hat{\pi}_2$ decays only through Higgs exchange, with lifetime dictated by $\mathrm{Tr}[i\sigma_2 (\bm{B} - \bm{B}^\dagger)] = d(\kappa) y \tilde{y} v/M$, with $d$ being a dimensionless function. Setting $\hat{B}_0 = 4\pi f_{\hat{\pi}}$ leads from Eq.~\eqref{eq:Higgs_mixing} to
\begin{equation} \label{eq:dark_pion_lifetime_benchmark3}
s_\theta^{(2)} = 2\pi d(\kappa) f_{\hat{\pi}}^2 \frac{y \tilde{y} v}{M(m_h^2 - m^2_{\hat{\pi}})}\,.
\end{equation}
In the top left panel of Fig.~\ref{fig:Benchmark3} we show the $\hat{\pi}_2$ lifetime as a function of $m_{\hat{\pi}}$, for several values of $\kappa$. $\tau_{\hat{\pi}_2}$ becomes very long for $\kappa \ll 1$, because in the limit $\kappa \to 0$ we have $\bm{B} \propto \bm{m}_{\psi'}$ which is diagonal, hence $d(\kappa) \to 0\,$. $\tau_{\hat{\pi}_2}$ also increases for $\kappa \gg 1$, due to the larger $c(\kappa)$ which for fixed $m_{\hat{\pi}}$ requires a smaller value of $f_{\hat{\pi}}$, thereby suppressing $s_\theta^{(2)}$. The shortest lifetime for a given $m_{\hat{\pi}}$ is thus obtained for $\kappa \sim 1$, i.e., when the mass scales $\omega$ and $Y\widetilde{Y} v^2/M$ are close. 

The $CP$-odd dark pions $\hat{\pi}_{1,3}$ decay only via the $Z$ portal, with decay constants that depend strongly on $r = y/\tilde{y}\,$,
\begin{equation}
f_a^{(b)} = p^{(b)}(\kappa) \frac{M^2}{y \tilde{y} f_{\hat{\pi}}} \big(r + q^{(b)}(\kappa) r^{-1} \big)^{-1}, \qquad\quad b = 1, 3\,,
\end{equation}
where $p^{(b)},q^{(b)}$ are dimensionless functions. It is instructive to compare the lifetimes of all three pions. To do so we focus on $\kappa = 1$, showing in the top right panel of Fig.~\ref{fig:Benchmark3} the lifetimes for illustrative values of $r$.\footnote{For $\kappa = 1$, the dimensionless functions take the values $c \approx 1.7$, $d \approx -\, 3.8$, $p^{(1)} \approx 2.6,\,q^{(1)} \approx -\,0.32$, $p^{(3)} \approx 10,\,q^{(3)} \approx - \,6.0$.} At small masses the $CP$-even pion $\hat{\pi}_2$ has the longest lifetime irrespective of $r$, but for $m_{\hat{\pi}} \gtrsim 4\,(6)\;\mathrm{GeV}$ it becomes the shortest-lived for $r \sim 1\,(10)$. Recalling the branching ratio patterns shown in Figs.~\ref{fig:ALP_widths} and \ref{fig:PHI_widths} for $CP$-odd and -even pions, we conclude that the expected signatures from dark shower events display a striking dependence on $m_{\hat{\pi}}$. For simplicity, here we have considered $r \geq 1$; in the opposite regime $r \leq 1$ the behavior is very similar, but with the roles of $\hat{\pi}_1$ and $\hat{\pi}_3$ reversed: in particular, for $r \ll 1$ it is $\hat{\pi}_3$ that has the shortest lifetime.

Thus far, we have fixed $y\tilde{y}/M$ to the upper bound from $h\to \mathrm{invisible}$. If this parameter is decreased by a factor $n > 1$, $f_{\hat{\pi}}$ must be correspondingly increased by $n$ in order to keep the same dark pion mass, as dictated by Eq.~\eqref{eq:dark_pion_mass_benchmark3}. These two effects exactly compensate (for fixed $M$) in the decay constants $f_a^{(1,3)}$, leaving $\tau_{\hat{\pi}_{1,3}}$ unvaried, whereas from Eq.~\eqref{eq:dark_pion_lifetime_benchmark3} we read that the net effect on $s_\theta^{(2)}$ is an $n$-fold increase, and therefore the $\hat{\pi}_2$ lifetime becomes $n^2$ times {\it shorter}. We illustrate this somewhat counter-intuitive effect in the bottom panel of Fig.~\ref{fig:Benchmark3}, which shows that even for $m_{\hat{\pi}} \lesssim 2\,m_{c,\tau}$ the $\hat{\pi}_2$ lifetime can be as short as $O(1\,$-$\,10)$~m, provided $f_{\hat{\pi}} \sim 10\;\mathrm{GeV}$. Furthermore, $\hat{\pi}_2$ can easily have the smallest lifetime among the dark pions. These results are especially interesting in view of a proposed LHCb search for LLPs decaying to $K^+ K^-\,$\cite{CidVidal:2019urm}, which may have sensitivity to our $\hat{\pi}_2$ since in this mass region its BR to $K\overline{K}$ is sizable, see Fig.~\ref{fig:PHI_widths}. As for the largest plausible value of $f_{\hat{\pi}}$, the neutral naturalness framework suggests $\Lambda \sim 4\pi f_{\hat{\pi}} \lesssim 100\;\mathrm{GeV}$, corresponding to $f_{\hat{\pi}} \lesssim 10\;\mathrm{GeV}$, while the ultimate limit is $\Lambda \ll M$, otherwise, the $Q$'s cannot be treated as heavy dark quarks anymore, the global symmetry pattern is modified and the EFT breaks down.

\section{FCNC meson decays}\label{sec:FCNC}
Light dark pions may be produced in FCNC meson decays if kinematically allowed. To describe these decay rates, we calculate the four-fermion effective operators of the form $\bar{d}_{L\alpha} d_{L\beta} \overline{\psi}' \psi'$ with $\alpha < \beta$. In our theory they arise through two classes of one-loop diagrams: $Z$ exchange with insertion of the $\bar{d}_{L\alpha} d_{L\beta}Z$ coupling, and box diagrams containing $W$ and $Q_u$ internal lines. The amplitudes can be fully obtained from the classic results for $d\bar{s}\to \nu \bar{\nu}$ in Ref.~\cite{Inami:1980fz}, leading to
\begin{align}
&\,\mathcal{L}_{\rm eff} = \frac{G_F}{\sqrt{2}} \frac{g^2}{4\pi^2} \bar{d}_{L\alpha} \gamma_\mu d_{L\beta} \sum_{q = c,t} V_{q\alpha}^\ast V_{q\beta} \\ &\,\sum_{i, k, j = 1,2}  \overline{\psi}'_{i} \gamma^\mu \frac{v^2}{2} \Bigg( \frac{ (U_L^\dagger \bm{\widetilde{Y}}^\dagger)_{ik} (\bm{\widetilde{Y}} U_L)_{kj} }{M_k^2} P_L + \frac{ (U_R^\dagger \bm{Y}^\dagger)_{ik} (\bm{Y} U_R)_{kj} }{M_k^2} P_R \Bigg) \psi^{\prime}_{j} \overline{D}(x_q, x_u =0; y_k) + \mathrm{h.c.} \nonumber
\end{align}
where $x_q \equiv m_q^2/m_W^2$ and $y_k \equiv M_k^2/m_W^2$ (recall that the mass of $Q_{u}^k$ is simply $M_k$). For our purposes we can safely take the large-$y_k$ limit of $\overline{D}$,
\begin{equation}
\overline{D}(x_q, x_u =0; y_k \to \infty) \simeq \frac{x_q}{8(x_q - 1)^2} \Big[ - x_q^2 \big( \log \frac{y_k}{x_q} - 2 \big) + x_q \big( 2 \log \frac{y_k}{x_q} - 7 \big) - \log \frac{y_k}{x_q} + 3 \log x_q + 5 \Big].
\end{equation}
If the $k$-dependence of the $\overline{D}$ function can be neglected (e.g., for $\bm{M} = M \bm{1}_2$), we arrive at a simple expression for the relevant effective Hamiltonian,
\begin{equation}
\mathcal{H}_{\rm eff} \supset \frac{G_F}{\sqrt{2}} \frac{g^2}{64\pi^2} \bar{d}_{L\alpha} \gamma_\mu d_{L\beta} V_{t\alpha}^\ast V_{t\beta} \sum_{b = 1}^3 \mathrm{Tr}\,[\sigma_b(\bm{A} - \bm{\widetilde{A}})] j^\mu_{5b} \Big[\frac{m_t^2}{m_W^2}\Big( \log \frac{M^2}{m_t^2} - 2 \Big) + 3 \Big] + \mathrm{h.c.},
\end{equation}
where only the first few terms of the dominant top loop were retained. The meson decay amplitude is then, assuming factorization of the hadronic matrix elements into a SM factor and a hidden factor,
\begin{equation}
\langle \hat{\pi}_a X | \mathcal{H}_{\rm eff} | B \rangle = \langle \hat{\pi}_a | \langle X | \mathcal{H}_{\rm eff} | B \rangle | 0 \rangle = \frac{i \hspace{0.2mm}g^2}{64\pi^2} V_{ts}^\ast V_{tb} \langle X | \bar{s}_L \gamma_\mu b_L | B \rangle \frac{p_{\hat{\pi}}^\mu}{f_{a}^{(a)}} \Big[\frac{m_t^2}{m_W^2}\Big( \log \frac{M^2}{m_t^2} - 2 \Big) + 3 \Big],
\end{equation}
where we have focused on $B \to X \hat{\pi}_a$ decays with $X = K, K^\ast$, and applied Eq.~\eqref{eq:f_def}. For the decay widths we find
\begin{align}
\Gamma(B \to K \hat{\pi}_a) =\;& \frac{m_B^3}{64\pi} f_0(m^2_{\hat{\pi}})^2  \left| \frac{g^2 V_{ts}^\ast V_{tb}}{64\pi^2 f_a^{(a)}} \Big[\frac{m_t^2}{m_W^2} \Big( \log \frac{M^2}{m_t^2} - 2 \Big) + 3 \Big]  \right|^2 \Big( 1 - \frac{m_K^2}{m_B^2}\Big)^2 \lambda_{K \hat{\pi}}^{1/2}\;, \nonumber \\
\frac{\Gamma(B \to K^\ast \hat{\pi}_a)}{\Gamma(B \to K \hat{\pi}_a)} =\;& \frac{A_0(m^2_{\hat{\pi}})^2}{f_0(m^2_{\hat{\pi}})^2}\frac{ \lambda_{K^\ast \hat{\pi}}^{3/2} }{ \Big( 1 - \frac{m_K^2}{m_B^2}\Big)^2 \lambda_{K \hat{\pi}}^{1/2} }\;, \qquad \qquad\qquad (CP\;\mathrm{odd})
\label{eq:FCNC_odd}\end{align}
where $\lambda_{X \hat{\pi}} = \big( 1 - \frac{(m_X + m_{\hat{\pi}})^2}{m_B^2} \big) \big( 1 - \frac{(m_X - m_{\hat{\pi}})^2}{m_B^2} \big)$. The log-enhanced contribution to $\Gamma(B \to K \hat{\pi}_a)$ is in agreement with what one finds~\cite{Gavela:2019wzg} from Eq.~\eqref{eq:FCNC_LL}, but the finite terms have an important quantitative impact: for $M = 1\;\mathrm{TeV}$, retaining only the logarithmic piece overestimates the rate by a factor $\approx 3$. The definitions and numerical values of the form factors $f_0, A_0$ are taken from the light-cone QCD sum rules analysis of Ref.~\cite{Gubernari:2018wyi}, with $f_0^{B\to K}(0) \approx 0.27$ and $A_0^{B\to K}(0) \approx 0.31$. An expression analogous to the first line in Eq.~\eqref{eq:FCNC_odd} applies to $K\to \pi \hat{\pi}_a$, with the appropriate replacements of masses, CKM elements, and the form factors available from lattice QCD with $f_0^{K\to \pi} (0) \approx 0.97$~\cite{Carrasco:2016kpy}.\footnote{Tree-level contributions to $K\to \pi a$ have also been considered~\cite{Guerrera:2021yss}, but are negligible here since the ALP couples to fermions with universal strength (in absolute value).}  

FCNC decays can also produce the $CP$-even dark pions, through Higgs mixing. The corresponding amplitudes are proportional to the Higgs penguin, resulting in~\cite{Batell:2009jf,Boiarska:2019jym}
\begin{equation}
\mathcal{L}_{\rm eff} \simeq \frac{3\sqrt{2} G_F}{16\pi^2} \frac{m_{d_\beta}}{v} \bar{d}_{L\alpha} d_{R\beta} s_\theta^{(a)} \hat{\pi}_a \sum_{q = u,c,t} m_q^2 V_{q\alpha}^\ast V_{q\beta} + \mathrm{h.c.},
\end{equation}
and
\begin{align} \label{eq:FCNC_even}
\Gamma(B \to K \hat{\pi}_a) =\;& \frac{m_B^3}{64\pi} f_0(m^2_{\hat{\pi}})^2  \left| \frac{3 V_{ts}^\ast V_{tb}}{16\pi^2} \frac{s_\theta^{(a)}}{v} \frac{m_t^2}{v^2} \right|^2 \Big( 1 - \frac{m_K^2}{m_B^2}\Big)^2 \lambda_{K \hat{\pi}}^{1/2}\;, \nonumber \\
\frac{\Gamma(B \to K^\ast \hat{\pi}_a)}{\Gamma(B \to K \hat{\pi}_a)} =\;& \frac{A_0(m^2_{\hat{\pi}})^2}{f_0(m^2_{\hat{\pi}})^2}\frac{ \lambda_{K^\ast \hat{\pi}}^{3/2} }{ \Big( 1 - \frac{m_K^2}{m_B^2}\Big)^2 \lambda_{K \hat{\pi}}^{1/2} }\; ,\qquad\qquad (CP\;\mathrm{even})
\end{align}
for the decay widths. Evaluating Eqs.~\eqref{eq:FCNC_odd} and~\eqref{eq:FCNC_even} we find
\begin{align} \label{eq:B_BR_theory}
\mathrm{BR}(B^{\{+, 0 \}} \to \{ K^+ \hat{\pi}_{b}, K^{\ast 0} \hat{\pi}_{b}\}) \,&\,\approx \{0.92,\, 1.1\} \times 10^{-8} \Bigg( \frac{10^3\;\mathrm{TeV}}{f_a^{(b)}} \Bigg)^2 \{ \lambda_{K \hat{\pi}}^{1/2} , \lambda_{K^{\ast } \hat{\pi}}^{3/2} \} \,, (CP\;\mathrm{odd})  \\
\mathrm{BR}(B^{\{+, 0 \}} \to \{ K^+ \hat{\pi}_{b}, K^{\ast 0} \hat{\pi}_{b}\}) \,&\,\approx \{2.6,\, 3.3\} \times 10^{-12} \Bigg( \frac{s_\theta^{(b)}}{3\times 10^{-6}} \Bigg)^2 \{ \lambda_{K \hat{\pi}}^{1/2} , \lambda_{K^{\ast } \hat{\pi}}^{3/2} \} \,, (CP\;\mathrm{even}) \nonumber
\end{align}
where in the $CP$-odd case we have set $M = 1\;\mathrm{TeV}$ in the logarithm. 

\subsection{Constraints and projected sensitivity}\label{sec:FCNC_constraints}
We now highlight a few implications for our parameter space, focusing mainly on $m_{\hat{\pi}} > 2 m_\mu$. The theoretical predictions in Eq.~\eqref{eq:B_BR_theory} can be compared with the current BaBar~\cite{BaBar:2013npw} and Belle~\cite{Belle:2017oht} $90\%$ CL bounds on invisible decays,
\begin{equation}
\mathrm{BR}(B^+ \to K^+ \nu\bar{\nu}) < 1.6 \times 10^{-5}\;, \qquad \mathrm{BR}(B^0 \to K^{\ast 0} \nu \bar{\nu}) < 1.8 \times 10^{-5}\,.
\end{equation}
For $CP$-odd scalars, branching ratios at the $10^{-5}$ level require $f_a^{(b)} < 100\;\mathrm{TeV}$, but in this regime the dark pion lifetimes become sufficiently short to ensure that decays to SM particles occur inside the detector (see Fig.~\ref{fig:ALP_widths} or~\ref{fig:f2_comparison}), thus violating the search assumptions. 

\begin{figure}[t]
\begin{center}
\includegraphics[width=0.495\textwidth]{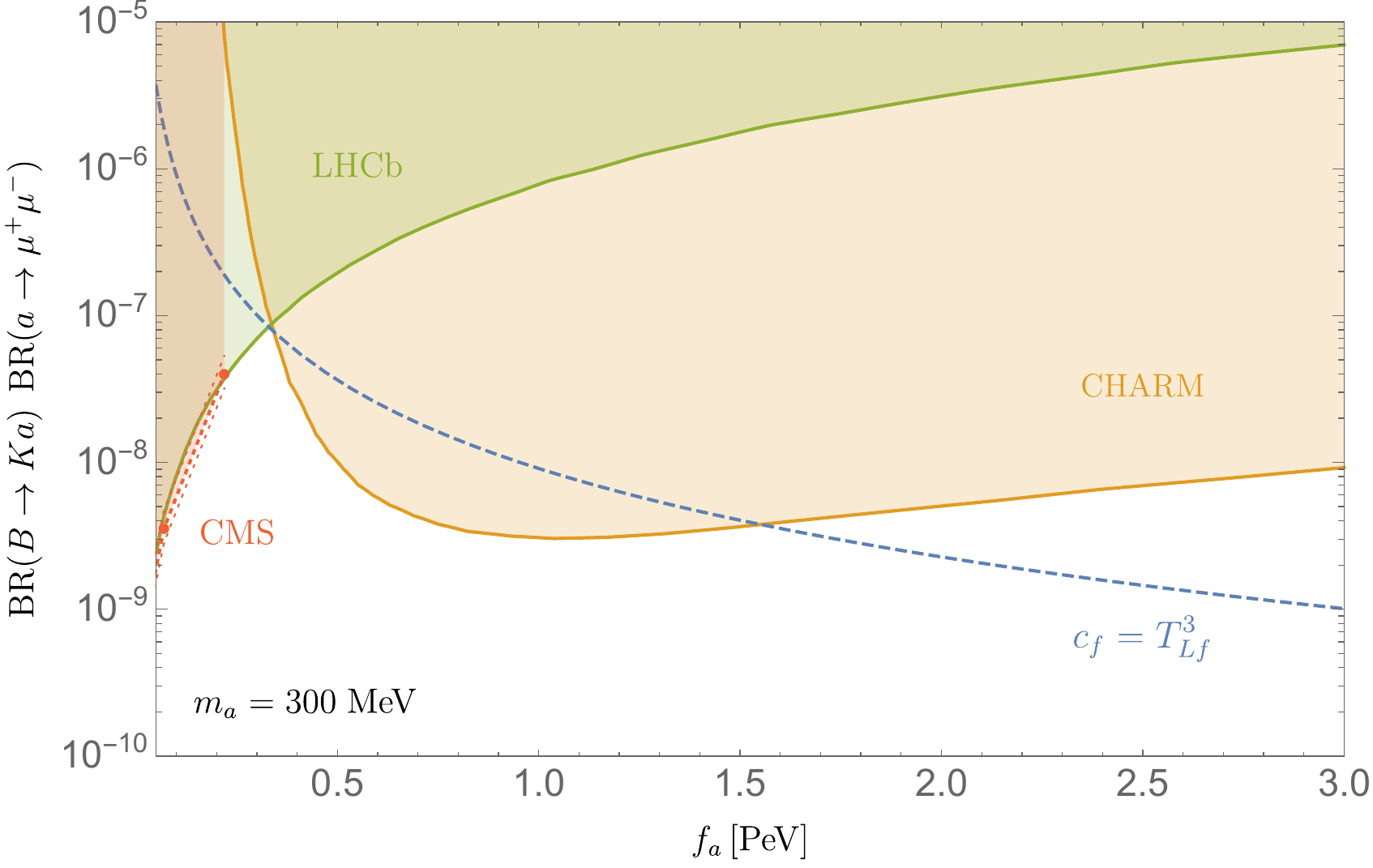}
\includegraphics[width=0.495\textwidth]{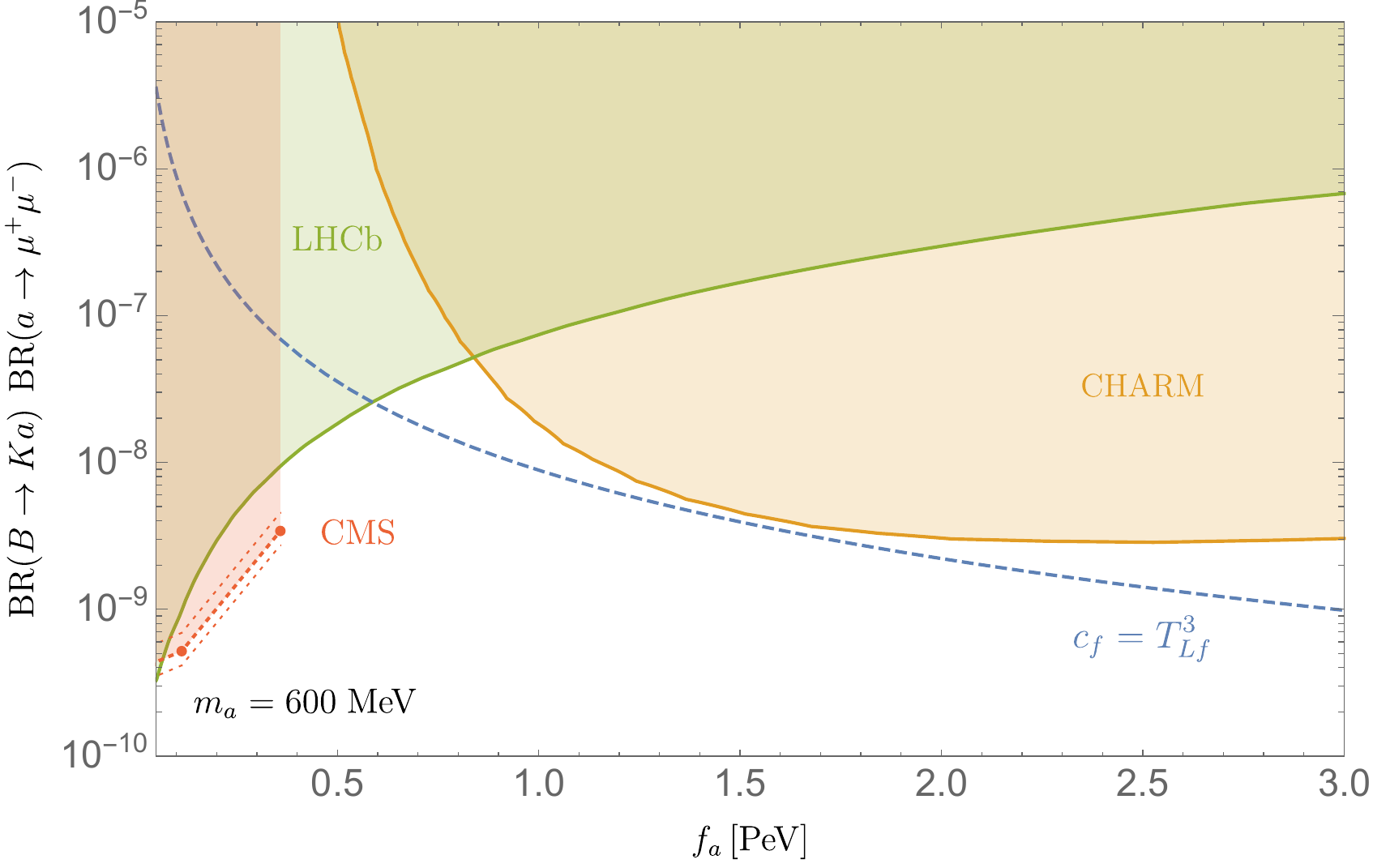}\vspace{2mm}
\includegraphics[width=0.495\textwidth]{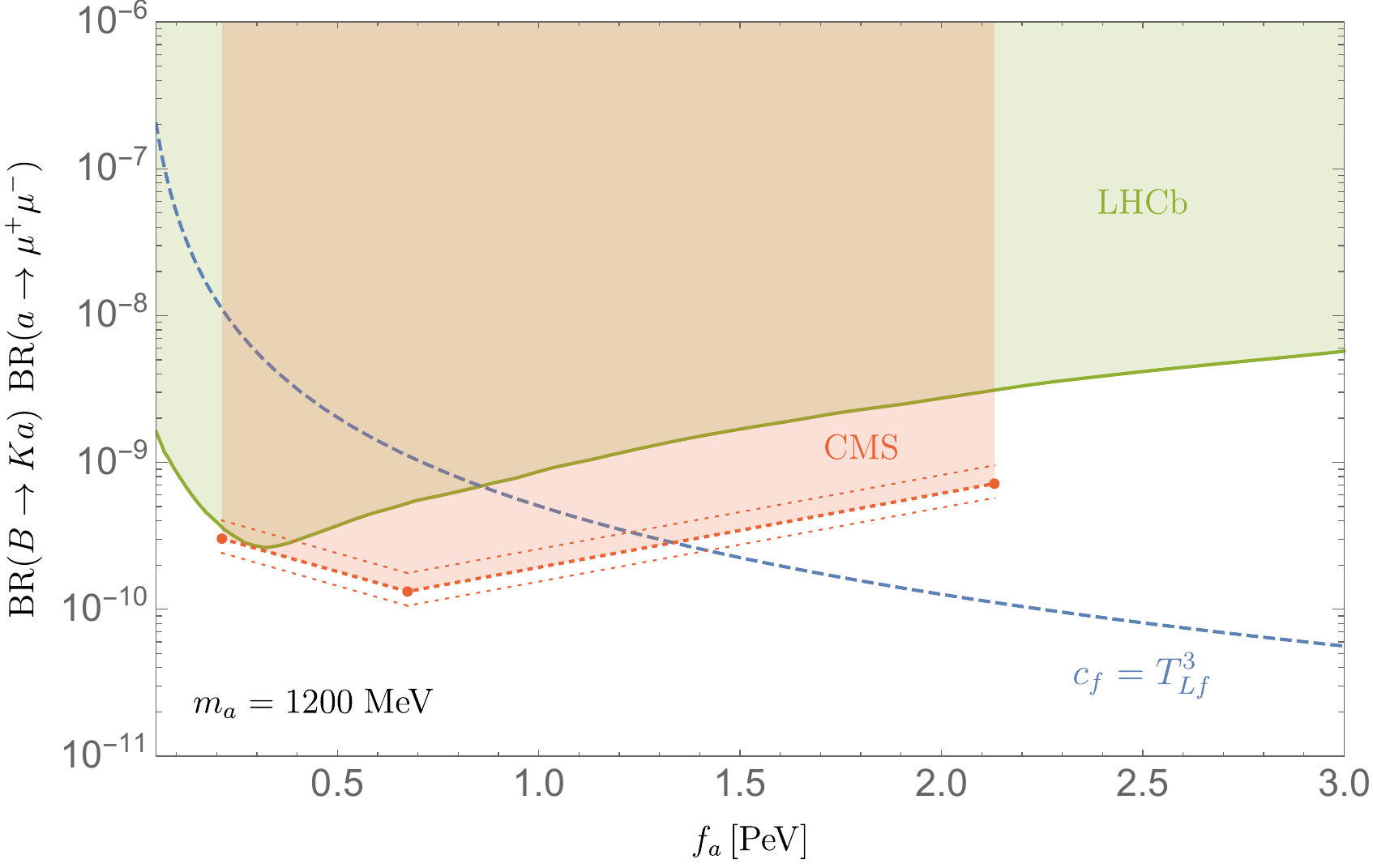}
\includegraphics[width=0.495\textwidth]{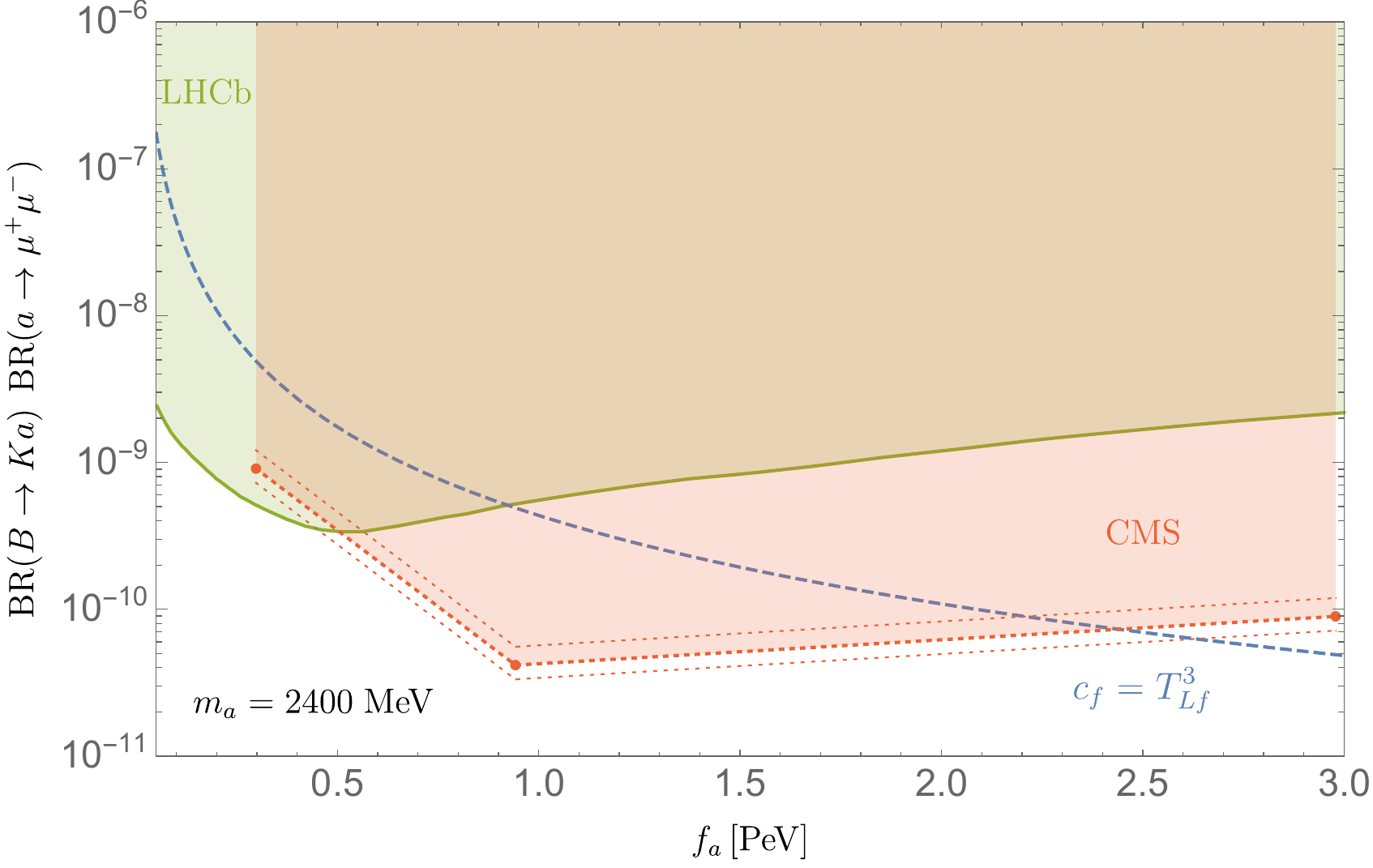}
\end{center}
\caption{$B\to K (a\to \mu^+ \mu^-)$ branching ratio of an ALP coupled to weak isospin (dashed blue curve), compared to regions excluded by LHCb (\cite{Dobrich:2018jyi}, green), CHARM (\cite{Dobrich:2018jyi}, orange), and CMS (\cite{CMS:2021ogd}, red), for four representative values of $m_a$. For CMS we display the uncertainty stemming from the relation between inclusive and exclusive branching ratios.}
\label{fig:LHCb_CHARM_CMS}
\end{figure}

Therefore more relevant are searches for $B \to K^{(\ast)} (\chi \to \mu\mu)$ with long-lived (scalar or pseudoscalar) $\chi$ at LHCb~\cite{LHCb:2016awg,LHCb:2015nkv}, as well as the re-interpretation in terms of these decays~\cite{Dobrich:2018jyi} of results from the CHARM beam dump experiment~\cite{CHARM:1985anb}. In addition, CMS has recently presented a novel search based on data scouting~\cite{CMS:2021ogd}, setting limits on the inclusive branching ratio for $B\to X_s (\chi \to \mu\mu)$~\cite{Evans:2020aqs}. In our setup this may be related to the exclusive branching ratios via
\begin{equation} \label{eq:incl_excl_Bdecays}
\mathrm{BR}(B \to X_s a) = (4 \pm 1) \times \big[ \mathrm{BR}(B\to K a) +  \mathrm{BR}(B\to K^\ast a) \big]\,,
\end{equation}
as estimated from the observed values of $\mathrm{BR}(b \to s \ell \ell)$ and $\mathrm{BR}(B \to K^{(\ast)} \ell\ell)$~\cite{Zyla:2020zbs}. The sizable uncertainty reflects the still-unsettled experimental status of these measurements. The relation~\eqref{eq:incl_excl_Bdecays} enables a direct comparison of the CMS and LHCb/CHARM bounds. In Fig.~\ref{fig:LHCb_CHARM_CMS} we show such comparison for four representative ALP masses in the range $m_{a} \lesssim 2m_c\,$, where searches for $a\to \mu^+ \mu^-$ are relevant, as seen from the branching ratios in Fig.~\ref{fig:ALP_widths}. The LHCb and CHARM constraints are taken from Ref.~\cite{Dobrich:2018jyi}, whereas we apply here for the first time the CMS bound~\cite{CMS:2021ogd} with the help of Eqs.~\eqref{eq:incl_excl_Bdecays} and~\eqref{eq:FCNC_odd}. For each value of $m_a$, CMS provides limits for $\tau = 1, 10, 100\;\mathrm{mm}$, corresponding to the red points in the $(f_a, \mathrm{BR})$ plane of Fig.~\ref{fig:LHCb_CHARM_CMS}; we simply interpolate between those points and include the uncertainty band arising from the relation between inclusive and exclusive branching ratios.\footnote{In the top right panel of Fig.~\ref{fig:LHCb_CHARM_CMS} we actually use the CMS bound for $m_a = 610\;\mathrm{MeV}$, as $600\;\mathrm{MeV}$ is masked in the analysis~\cite{CMS:2021ogd}. We neglect the impact of this small difference.}

Figure~\ref{fig:LHCb_CHARM_CMS} shows a clear pattern: for low (high) mass the strongest constraint comes from CHARM (CMS), while in the intermediate region LHCb has the best sensitivity. To better estimate the bounds as functions of the dark pion mass, we combine the above results at fixed $m_a$ with the findings of the Expression of Interest for CODEX-b~\cite{Aielli:2019ivi}, where constraints on $f_a$ for an ALP coupled {\it universally} to SM fermions were reported following the analysis of Ref.~\cite{Dobrich:2018jyi}, but with updated lifetime and branching ratio calculations employing data-driven methods~\cite{Aloni:2018vki}. The main differences between our setup (where the ALP couples to weak isospin) and the universal coupling scenario~\cite{Aielli:2019ivi,Batell:2009jf,Dobrich:2018jyi} are the ALP total width and the treatment of finite terms in the $B\to K a$ calculation. For the former, a detailed comparison in Fig.~\ref{fig:f2_comparison} (right panel) shows qualitative agreement, although important quantitative differences are present; for the latter, in Refs.~\cite{Batell:2009jf,Dobrich:2018jyi} only the leading-log term was retained and the cutoff was set to $1\;\mathrm{TeV}$, which combined with slightly different values for the form factors gives a rate $\sim 4$ times larger than here. In light of these considerations we apply the $f_a$ bounds for universal couplings~\cite{Aielli:2019ivi} to our setup, after weakening them by a factor $\sim 2$ to account for the smaller production rate. Where relevant, the resulting estimates agree with Fig.~\ref{fig:LHCb_CHARM_CMS}.  

For $2m_\mu \lesssim m_a \lesssim 0.6\;\mathrm{GeV}$, the re-interpretation~\cite{Aielli:2019ivi,Dobrich:2018jyi} of CHARM results gives the strongest constraint.\footnote{In the universal coupling scenario~\cite{Aielli:2019ivi} the CHARM bound was found to extend up to $m_a \sim 1\;\mathrm{GeV}$, but a direct comparison shows that in our setup it is limited to $\sim 600\;\mathrm{MeV}$, see Fig.~\ref{fig:LHCb_CHARM_CMS}.} In this region we estimate $f_a \gtrsim 1.3$\,-$1.9\;\mathrm{PeV}$,\footnote{Here we quote the lower limit on $f_a$ from CHARM, but note that a small ``wedge'' of allowed $f_a$ may remain between the LHCb, CHARM and CMS exclusions for $0.3 \lesssim m_a/\mathrm{GeV} \lesssim 0.6$, see the top panels of Fig.~\ref{fig:LHCb_CHARM_CMS}.} translating in benchmark scenario 1 to
\begin{equation}
\left( \frac{y/M}{1.1\; \mathrm{TeV}^{-1}} \right)^2 \left( \frac{f_{\hat{\pi}}}{\mathrm{GeV}} \right) \lesssim 2 \qquad (\mathrm{scenario}\;1,\;\, 0.21 \lesssim m_{\hat{\pi}}/\mathrm{GeV} \lesssim 0.6)
\end{equation}
when applied to $\hat{\pi}_1$ using Eq.~\eqref{eq:benchmark1_params}, and taking conservatively the weakest bound in the given mass range. Thus, for $f_{\hat{\pi}} \gtrsim 2\;\mathrm{GeV}$ the CHARM sensitivity surpasses $Z\to \mathrm{invisible}$. Considering inclusive decays would likely strengthen the CHARM bounds compared to those used here, which were derived from $B\to K^{(\ast)} a$ only~\cite{Dobrich:2018jyi}. 

For $m_a \in [0.6,1.1]$~GeV, the limits set by LHCb on $\mathrm{BR}(B^+ \to K^+ \chi)\, \mathrm{BR}(\chi \to \mu\mu)$ for $\chi$ lifetimes in the range $0.1\,$-$\,10^3$ ps~\cite{LHCb:2016awg} are the strongest. We estimate $f_a \gtrsim 0.6$\,-$\,0.8\;\mathrm{PeV}$, which reads
\begin{equation}\label{eq:FCNC_bound_intermediate}
\left( \frac{y/M}{1.1\; \mathrm{TeV}^{-1}} \right)^2 \left( \frac{f_{\hat{\pi}}}{\mathrm{GeV}} \right) \lesssim 4\,, \qquad (\mathrm{scenario}\;1,\;\, 0.6 \lesssim m_{\hat{\pi}}/\mathrm{GeV} \lesssim 1.1)
\end{equation}
when expressed in terms of the underlying model parameters. 

Above $m_a \sim 1.1\;\mathrm{GeV}$ the CMS scouting search~\cite{CMS:2021ogd} provides the best sensitivity to date, $f_a \gtrsim 1.3$\,-$\,2.8\;\mathrm{PeV}$, yielding
\begin{equation}
\left( \frac{y/M}{1.1\; \mathrm{TeV}^{-1}} \right)^2 \left( \frac{f_{\hat{\pi}}}{\mathrm{GeV}} \right) \lesssim 2\,. \qquad (\mathrm{scenario}\;1,\;\, 1.1 \lesssim m_{\hat{\pi}}/\mathrm{GeV} \lesssim 2.8)
\end{equation}
Here again we have been conservative, adopting the weakest bound in this mass range; close to the upper end, the constraint can actually be about twice as strong, as seen in the bottom right panel of Fig.~\ref{fig:LHCb_CHARM_CMS}. It should be emphasized that a theoretical uncertainty affects this bound, stemming from Eq.~\eqref{eq:incl_excl_Bdecays}.

Looking ahead, several proposed LLP experiments at the LHC have the potential to improve the sensitivity on ALPs coupled to fermions in the mass range $2m_\mu \lesssim m_a \lesssim 2m_c$, including CODEX-b~\cite{Aielli:2019ivi}, FASER~2~\cite{Ariga:2018uku}, and MATHUSLA~\cite{Curtin:2018mvb}. Importantly, in contrast to current bounds that rely on $a\to \mu^+ \mu^-$, these experiments would be sensitive to any decays to $\geq 2$ charged tracks and therefore to $a\to \pi^+ \pi^- \pi^0$, which in our model dominates between $1$ and $3\;\mathrm{GeV}$ (see Fig.~\ref{fig:ALP_widths}). As already discussed above for LHCb and CHARM constraints, we can roughly estimate the projected sensitivities from the results for universal ALP-fermion couplings in Ref.~\cite{Aielli:2019ivi}. Caveats concern the total ALP width, as shown in Fig.~\ref{fig:f2_comparison}, and the production rate, which is assumed to arise dominantly from FCNC $B$ meson decays. For example, in our setup mixing with $\pi_0, \eta, \eta'$ may enhance the production, owing to the non-trivial $U(3)$ transformation properties of the ALP. With these disclaimers, we obtain for $m_a = 1\;\mathrm{GeV}$ the projections $f_a \gtrsim 10\;\mathrm{PeV}$ at FASER~2, $f_a \gtrsim 20\;\mathrm{PeV}$ at CODEX-b, and $f_a \gtrsim 80\;\mathrm{PeV}$ at MATHUSLA200. In addition, for SHiP with $10^{20}$ protons on target we find $f_a \gtrsim 14\;\mathrm{PeV}$~\cite{Dobrich:2018jyi}, based on the $a\to \mu^+ \mu^-$ signature. 

The decays $B\to K\, a,\, a \to \mathrm{hadrons}$ with $m_a$ in the GeV range have also been studied as probes of a heavy QCD axion, where the dominant coupling to the SM is $aG\widetilde{G}$. Both prompt $a\to \pi^+ \pi^- \pi^0,\, \eta \pi^+ \pi^-,\, K K \pi,\, \phi \phi$~\cite{Chakraborty:2021wda} and displaced $a \to \pi^+ \pi^- \pi^0$~\cite{Bertholet:2021hjl} have been considered and projections for Belle II obtained. Our branching ratio calculations in Appendix~\ref{app:ALP_decays_general} can serve as the basis to extend those results to the class of models where the ALP couples dominantly to SM fermions.

For smaller masses, kaon decays provide relevant constraints through $K\to \pi\, +\,$invisible final states: comparing the theory prediction
\begin{equation}
\mathrm{BR}(K^+ \to \pi^+ \hat{\pi}^{(b)}) \approx 3.9 \times 10^{-11} \Bigg( \frac{10^3\;\mathrm{TeV}}{f_a^{(b)}} \Bigg)^2 \lambda_{\pi \hat{\pi}}^{1/2}\;,
\end{equation}
with the strongest NA62 bound $\mathrm{BR}(K^+ \to \pi^+ X) < 5 \times 10^{-11}$ ($90\%$ CL)~\cite{NA62:2020xlg}, valid for $m_X \in [160, 250]$~MeV and $\tau_X \gtrsim 10$~ns, we learn that PeV decay constants are currently being tested. In fact, for benchmark scenario 1 we obtain
\begin{equation}
\mathrm{BR}(K^+ \to \pi^+ \hat{\pi}_1) \approx 5.7 \times 10^{-11} \left( \frac{f_{\hat{\pi}}}{3\;\mathrm{GeV}} \right)^2 \left( \frac{y/M}{1.1\; \mathrm{TeV}^{-1}} \right)^4 \lambda_{\pi \hat{\pi}}^{1/2}\,, \qquad (\mathrm{scenario}\;1)
\end{equation}
hence for $f_{\hat{\pi}} \gtrsim 3$~GeV NA62 surpasses $Z$ invisible decays, at least for $160\;\mathrm{MeV} < m_{\hat{\pi}} < 2m_\mu$, where the lifetime is extremely long. For $m_{\hat{\pi}} > 2m_\mu$ such values of $f_{\hat{\pi}}$ correspond to a shorter lifetime of $O$(ns), causing a deterioration of the bound~\cite{NA62:2020xlg}.\footnote{The KOTO experiment has produced a bound on $K_L \to \pi^0 X$, reaching $\mathrm{BR} \sim 2 \times 10^{-9}$ ($90\%$~CL) for $m_X \lesssim 150$~MeV~\cite{KOTO:2018dsc}. From the relation with $K^+ \to \pi^+ X$~\cite{Grossman:1997sk}, we estimate that its impact on our parameter space is weaker compared to NA62.} For even larger $f_{\hat{\pi}} \gtrsim 10\;\mathrm{GeV}$ the NA48/2 limits~\cite{NA482:2016sfh} become relevant, $\mathrm{BR}(K^+ \to \pi^+ \chi)\mathrm{BR}(\chi \to \mu\mu) \lesssim 10^{- 10}\,$-$\,10^{-9}$, valid for $\tau_\chi \lesssim 100\;\mathrm{ps}$. For $m_{\hat{\pi}} < m_{\pi_0}$, the NA62 bounds are very similar to those from E949~\cite{BNL-E949:2009dza}.

Finally, for decays to $CP$-even dark pions we find from Eq.~\eqref{eq:FCNC_even}
\begin{equation}
\mathrm{BR}(K^+ \to \pi^+ \hat{\pi}^{(b)}) \approx 1.3 \times 10^{-11} \bigg( \frac{s_\theta^{(b)}}{10^{-4}} \bigg)^2 \lambda_{\pi \hat{\pi}}^{1/2}\;,
\end{equation}
hence NA62~\cite{NA62:2020xlg} is probing mixing angles of $O(10^{-4})$. Comparing this with the expectation in benchmark scenario 3,
\begin{equation}
s_\theta^{(2)} \sim -\,  10^{-7} \left( \frac{f_{\hat{\pi}}}{5\;\mathrm{GeV}} \right)^2 \left( \frac{y \tilde{y}/M}{10^{-5} \;\mathrm{TeV}^{-1}} \right)\,, \qquad (\mathrm{scenario}\;3)
\end{equation}
where we have set $m_{\hat{\pi}} \sim 250$~MeV and $\kappa = 1$ (recall that the product $f_{\hat{\pi}} \times y\tilde{y}/M$ is then fixed by Eq.~\eqref{eq:dark_pion_mass_benchmark3}), suggests that FCNC meson decays to $CP$-even dark pions are out of experimental reach, unless one is willing to consider an {\it extreme} hierarchy between $f_{\hat{\pi}}$ and $m_{\hat{\pi}}$, with the former exceeding the TeV.

\section{$Z\,$-$\,$initiated, muon-rich dark showers at the LHC}\label{sec:dark_showers}

In the previous section we have discussed processes at energies well below the weak scale, where the dark pion properties can be fully described through the low-energy parameter combinations $f_a$ and $s_\theta$, for $CP$-odd and -even states respectively. Here we take a step up in energy and consider production of dark pions via $Z$ and Higgs decays to dark partons, followed by showering and hadronization. As we are going to show, these processes access new directions in parameter space compared to FCNC meson decays. 

The LHC inclusive production cross sections for $Z$ and Higgs bosons are (see, e.g., Ref.~\cite{Cheng:2019yai})
\begin{equation}\label{eq:Z_h_xsecs}
\sigma(pp\to Z) \approx 54.5\;(58.9)\;\mathrm{nb}, \qquad\quad \sigma(pp\to h) \approx 48.6\;(54.7)\;\mathrm{pb},
\end{equation}
at $13\,(14)$~TeV. The coupling structure of our model implies that $Z$ decays dominate in scenarios with $Y$ or $\widetilde{Y} \sim 0$, whereas $h$ decays are most important if $Y \sim \widetilde{Y}$, as quantified by the branching ratios to dark quarks in Eqs.~\eqref{eq:Z_BR} and~\eqref{eq:hffwidth}. Here we focus on $Z$ decays to the dark sector, which are largely unexplored but hold a strong LHC discovery potential, as the forthcoming discussion illustrates.
\enlargethispage{20pt}

The $Z\to \psi' \overline{\psi}'$ decay results in two dark jets, dominantly composed of dark pions with high multiplicity. GeV-scale dark pions eventually decay to a variety of SM final states, as seen in Figs.~\ref{fig:ALP_widths} and~\ref{fig:PHI_widths}. For $m_{\hat{\pi}} \lesssim 2m_c$, the FCNC meson decays discussed in Section~\ref{sec:FCNC_constraints} set a lower bound $f_a \gtrsim O(\mathrm{PeV})$, implying in turn a lower bound on the dark pion lifetimes. Concretely, in scenario 1 with $m_{\hat{\pi}} \sim 1 \;\mathrm{GeV}$ we obtain from Eq.~\eqref{eq:FCNC_bound_intermediate} a constraint $\tau_{\hat{\pi}_1} > O(1\,$-$10)\;\mathrm{cm}$, which sets the target for dark shower searches in this mass range. Differently from scenarios with $t$-channel mediation such as emerging jets~\cite{Schwaller:2015gea,Sirunyan:2018njd}, here the signal is not automatically accompanied by hard SM jet activity, hence the trigger strategy is a central issue. For this reason in the first exploration we focus on $\hat{\pi} \to \mu^+ \mu^-$ decays, which result in striking displaced vertices (DVs) at the LHC and a narrow resonance peak that can be exploited to suppress the combinatorial and misidentification backgrounds~\cite{Pierce:2017taw}.\footnote{Hadronic $\pid$ decays are alternative opportunities, especially when the final states are fully charged: for example, $\pid\to K^{\ast 0} \overline{K}^{\ast 0} \to (K^+ \pi^-) (K^- \pi^+)$ through the $Z$ portal, or $\pid \to K^+ K^-$ through the $h$ portal. The phenomenology of these hadronic final states within dark showers deserves future study.}

The sensitivity of LHCb to dark shower signals is well established~\cite{Pierce:2017taw,Cheng:2019yai} (see also a recent overview~\cite{Borsato:2021aum}) and the most recent search for dimuon resonances~\cite{Aaij:2020ikh} has already provided a HV interpretation. Building on these results, in Section~\ref{sec:LHCb} we perform a detailed recast to set current bounds and estimate the future reach of LHCb on our $Z$-initiated, muon-rich dark shower signals. By contrast, for ATLAS and CMS we limit ourselves to some qualitative comments in Section~\ref{sec:ATLAS_CMS}, whereas a detailed study is deferred to a separate publication due to its more complex nature~\cite{Future:showers} (see also Refs.~\cite{Cohen:2015toa,Cohen:2017pzm,Cohen:2020afv,Knapen:2021eip,Linthorne:2021oiz} for discussions of other dark shower signals).

\subsection{LHCb sensitivity}\label{sec:LHCb}

We base our reinterpretation on the latest LHCb search for displaced dimuons~\cite{Aaij:2020ikh}. We generate $pp \to Z \to \psi'\overline{\psi}'$ at 13~TeV using the HV module of Pythia8~\cite{Carloni:2010tw,Carloni:2011kk,Sjostrand:2014zea}, with the production cross section in Eq.~\eqref{eq:Z_h_xsecs} as normalization. To set the dark pion parameters we focus on benchmark scenario 1 (Section~\ref{sec:scenario1}), where all three dark pions decay through the $Z$ portal, considering two mass points with the following characteristics:
\begin{align}
m_{\hat{\pi}} \,=&\; 650\;\mathrm{MeV}\,,\qquad \mathrm{BR}(\hat{\pi} \to \mu\mu) \approx 0.96\,, \qquad \langle N_{\hat{\pi}} \rangle = 7\,, \nonumber \\
m_{\hat{\pi}} \,=&\; 1\;\mathrm{GeV}\,,\qquad\quad\,\mathrm{BR}(\hat{\pi} \to \mu\mu) \approx 0.18\,, \qquad \langle N_{\hat{\pi}} \rangle = 5\,.
\end{align}
Here $\langle N_{\hat{\pi}} \rangle$ is the average number of dark pions per dark jet. As three different lifetimes cannot be accommodated by the HV module, we neglect the longest-lived $\hat{\pi}_2$ (which is also subject to larger uncertainties) and fix the ratio $\tau_{\hat{\pi}_3} /\tau_{\hat{\pi}_1} \approx 37$ as expected from Eq.~\eqref{eq:benchmark1_params}. This leaves $\tau_{\hat{\pi}_1}$ and $\mathrm{BR}(Z\to \psi' \overline{\psi}')$ as free parameters of our analysis.

\begin{figure}[t]
\begin{center}
\includegraphics[width=0.6\textwidth]{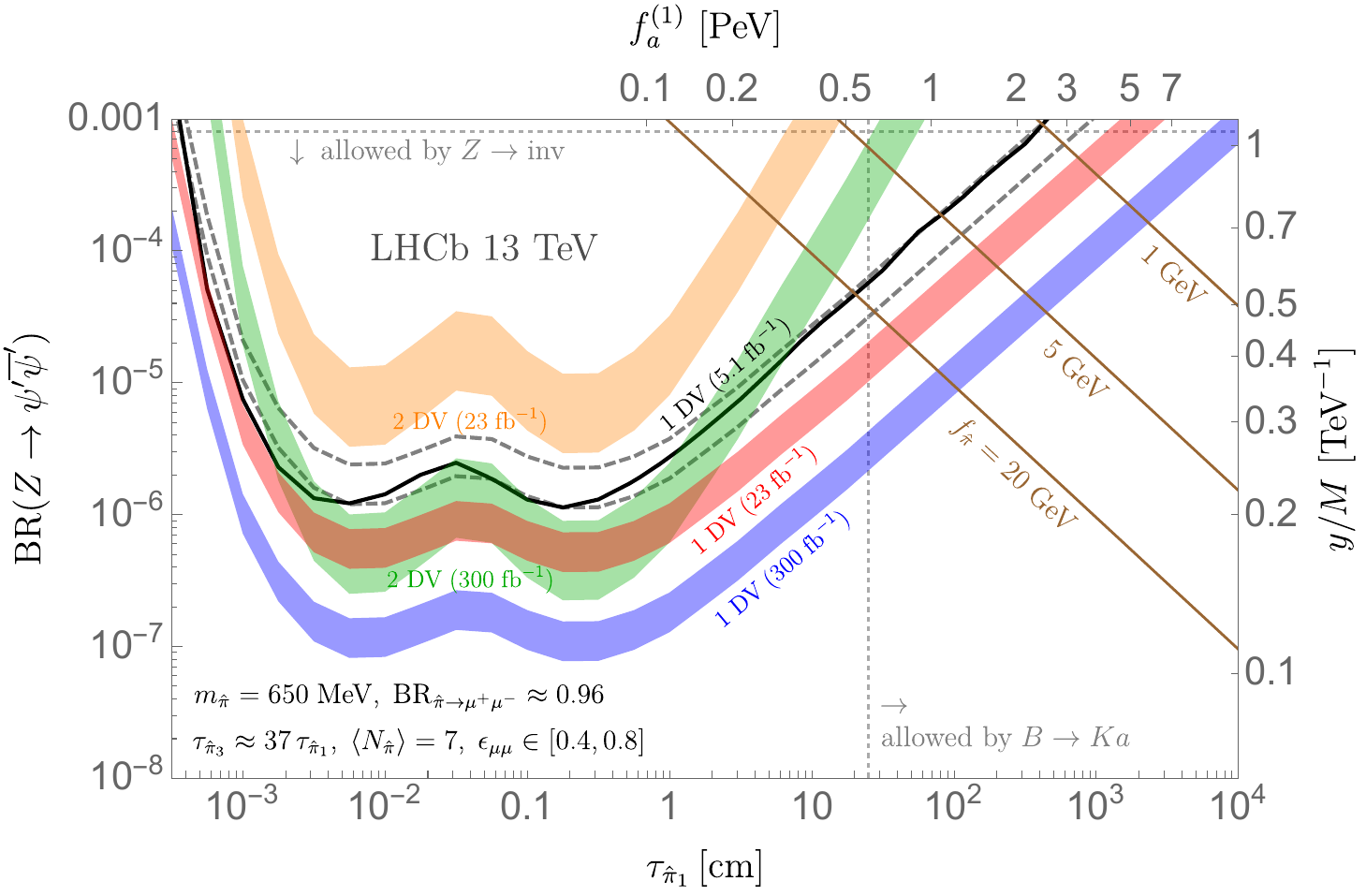}\vspace{2mm}

\includegraphics[width=0.6\textwidth]{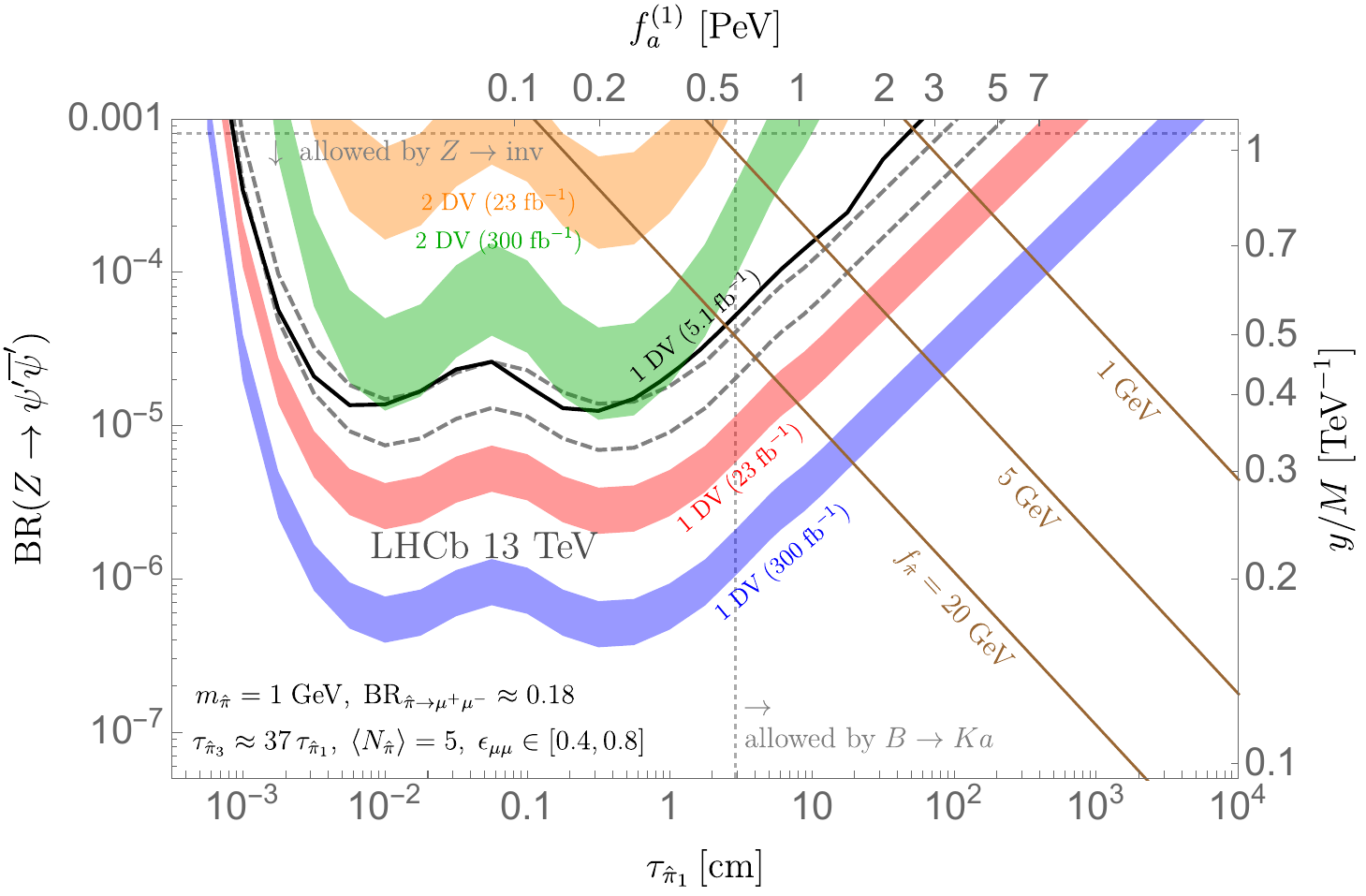}\vspace{-2mm}
\end{center}
\caption{Projection of the $90\%$~CL LHCb sensitivity~\cite{Aaij:2020ikh} to $Z$-initiated, muon-rich dark showers for $m_{\hat{\pi}} = 650\;\mathrm{MeV}$ {\it (top)} and $m_{\hat{\pi}} = 1\;\mathrm{GeV}$ {\it (bottom)}. The two minima of the exclusion contours correspond to optimal sensitivity to decays of two dark pion species with different lifetimes, $\hat{\pi}_1$ and $\hat{\pi}_3$, while decays of the longest-lived $\hat{\pi}_2$ are neglected. The current exclusion is shown by the black curve, while the widths of all other bands are obtained by varying the single-DV efficiency $\epsilon_{\mu\mu}\in[0.4,0.8]$. Brown lines indicate the relation between BR($Z\to\psi'\overline{\psi}'$) and $\tau_{\pid_1}$ obtained from benchmark scenario 1, for representative choices of $f_{\pid}$.}
\label{fig:LHCblimit}
\end{figure}

To derive the current constraint, we apply at truth level the displaced search cuts listed in Table~1 of Ref.~\cite{Aaij:2020ikh} and compare to the cross section limits for promptly-produced $X\to \mu^+ \mu^-$~\cite{Aaij:2020ikh} (this is the appropriate choice, as dark parton shower and hadronization are prompt in our model, and we require the reconstructed $X$ to come from the primary vertex). We find the $p_T^X \in [5,10]$~GeV bin dominates the sensitivity, resulting in the solid black exclusion curves in Fig.~\ref{fig:LHCblimit}. The right minimum of the exclusion contours corresponds to optimal sensitivity to the $\hat{\pi}_1$ signal with $\tau_{\hat{\pi}_1} \sim \mathrm{few}\; \mathrm{mm}$, whereas the left minimum corresponds to optimal sensitivity to decays of $\hat{\pi}_3$, with $\tau_{\hat{\pi}_3} \approx 37 \tau_{\hat{\pi}_1} \sim \mathrm{few}\; \mathrm{mm}$. 

To estimate the future reach, we follow a slightly different strategy: we calculate the signal rate after cuts and parametrize remaining detector effects through a DV efficiency $\epsilon_{\mu\mu}$ that is varied in the range $[0.4,0.8]$. This is compared to the background rate extracted from Fig.~2 in Ref.~\cite{Aaij:2020ikh}, which is found to be $\approx 1.6\,(\approx 0.7)$ events per $5.1$~fb$^{-1}$ for the $m_{\hat{\pi}} = 650\;\mathrm{MeV}\,(1\;\mathrm{GeV})$ hypothesis, by averaging over the $m_{\mu^+ \mu^-} \in [600, 700]$~MeV~($[0.9, 1.1]$~GeV) window and considering a bump-search interval $|m_{\mu^+ \mu^-} - 650\;\mathrm{MeV}\,(1\;\mathrm{GeV})| < 2\sigma$ with $\sigma$ being the experimental resolution. When applied to the current luminosity, this procedure gives the dashed gray bands in Fig.~\ref{fig:LHCblimit}. The reasonable agreement with the actual LHCb constraint gives us confidence in the method, which is then applied to Run 3 ($23\;\mathrm{fb}^{-1}$) and High-Luminosity LHC (HL-LHC, $300\;\mathrm{fb}^{-1}$) scenarios to obtain the red and blue bands. For $m_{\hat{\pi}} = 650\;\mathrm{MeV}$, LHCb will probe $Z$ branching ratios down to $\sim 10^{-7}$ in the high-luminosity phase, with further improvements possible either through optimization to the dark shower signal or future detector upgrades. The reach for $m_{\hat{\pi}} = 1\;\mathrm{GeV}$ is somewhat weaker, due to the lower dark pion multiplicity and smaller dimuon branching ratio.
\enlargethispage{10pt}

The brown lines in Fig.~\ref{fig:LHCblimit} show the relation between BR($Z\to\psi'\overline{\psi}'$) and $\tau_{\pid_1}$ that is realized in benchmark scenario 1, as a function of $f_{\hat{\pi}}$. The dependence on the underlying parameters should be contrasted with complementary bounds from other processes, namely $Z\to \mathrm{invisible}$, which probes $y/M$, and $B$ decays, sensitive to $f_a \propto M^2/(y^2 f_{\hat{\pi}})$. We learn that for $1 \lesssim f_{\hat{\pi}}/\mathrm{GeV} \lesssim 20$ the LHCb dark shower search has already probed new parameter space, highlighting the strongly complementary role of this type of analysis with current and upcoming data.

In addition to the single-DV analysis we consider requiring 2 DVs per event, assuming zero background in this case. The corresponding exclusions, shown by the orange and green bands in Fig.~\ref{fig:LHCblimit}, turn out to be weaker than the single-DV ones. This is explained by the fact that the background is already very suppressed for 1 DV, hence removing it completely results in a limited gain, and by the additional efficiency cost.

The potential sensitivity of LHCb to heavier pseudoscalars, with masses above a few GeV, has also been discussed in several final states~\cite{BuarqueFranzosi:2021kky}.

\subsection{ATLAS and CMS prospects}\label{sec:ATLAS_CMS}

In the light of the results shown in Fig.~\ref{fig:LHCblimit}, and in particular the correlation observed in our framework between $\mathrm{BR}(Z\to \psi' \overline{\psi}')$ and the dark pion lifetimes, a priori ATLAS and CMS may lead to dramatic improvements in the region $\tau_{\hat{\pi}} \sim 0.1\,$-$\,1\;\mathrm{m}$, thanks to their larger volumes (and integrated luminosities). However, owing to the soft nature of the signals considered here, progress requires targeted experimental strategies that enable efficient triggering on low-$p_T$ displaced muons.

A major step in this direction has recently been achieved by CMS with the search for dimuon DVs~\cite{CMS:2021ogd} in data collected with scouting triggers, which permit the unprecedented exploration of very low muon transverse momenta and thus DV masses, down to the $m_{\mu\mu} \sim 2 m_\mu$ threshold. This approach is well suited to test the $\hat{\pi} \to \mu \mu$ signals discussed here, as demonstrated by the new constraints on the parameter space we have derived in Section~\ref{sec:FCNC} from the CMS $B\to X_s (\chi \to \mu \mu)$ results~\cite{CMS:2021ogd}. Thus a recast to the dark shower signal is warranted, which will be presented elsewhere~\cite{Future:showers}. We note that the CMS analysis imposes a cut $l_{xy} < 11$~cm on the transverse displacement of the dimuon DVs, due to the definition of the scouting trigger stream which requires hits in at least two pixel layers. Looking ahead to future upgrades, CMS-specific triggers targeting LLP dimuon signals have also been proposed~\cite{Gershtein:2019dhy}.

At ATLAS, a search for two ``dark photon jets''~\cite{ATLAS:2019tkk} targeted final states related to those of interest here: Higgs decays to two jet-like structures, each composed of an invisible particle and two GeV-mass LLPs decaying to $\mu^+ \mu^-$. Events were selected by means of a trigger requiring $\geq 3$ L1 muons with $p_T > 6\;\mathrm{GeV}$, then confirmed at HLT using only muon spectrometer information. It results in optimal sensitivity for $O(\mathrm{cm})$ lifetimes. Compared to the signal model used by ATLAS, our $Z$-initiated dark shower has larger multiplicity, lower transverse momenta, and for $m_{\hat{\pi}} \lesssim 1$~GeV larger branching fraction to muons. 

Heavier LLPs have been searched for in a number of analyses by ATLAS and CMS, mainly focusing on rare Higgs decays to the hidden sector, see e.g., Refs.~\cite{ATLAS-CONF-2021-032,CMS:2021juv} for very recent results.

\section{Probing the ultraviolet completion}\label{sec:UV}

Finally, we take another step up in energy and discuss the expected LHC signals from direct production of the heavy fermions $Q$. Since these carry SM EW charges, they undergo Drell-Yan (DY) pair production such as, for instance, $u\bar{d} \to W^{+\ast} \to Q_u \overline{Q}_d$. The decay patterns can be read from the Yukawa interactions by means of the Goldstone equivalence theorem: $Q_u \to W^+ \psi$, whereas $Q_d$ decays to $Z \psi$ and $h \psi$ with $\approx 1/2$ branching ratios for $M \gg m_{Z,h}\,$. Flavor indices have been suppressed for simplicity. The $\psi \overline{\psi}$ pair in the final state give rise to two dark jets, which characterize the signal. 

Assuming the dark pions are sufficiently long-lived to escape the detector we obtain $WZ/Wh\,$+$\,$MET, a typical signature of EWinos in supersymmetry. Similar considerations apply to the production of the electrically neutral pairs $Q_u \overline{Q}_u, Q_d \overline{Q}_d\,$. Consequently, bounds on $M$ can be directly set by applying the results of searches for Higgsinos, which are assumed to decay directly to the lightest supersymmetric particle (LSP), taken to be the bino-like neutralino.  Our signal matches this topology in the limit of very light neutralino LSP. The strongest sensitivity has been achieved, remarkably, in the all-hadronic$\,+\,$MET search by ATLAS~\cite{ATLAS:2021yqv}, which outperforms analogous searches for $3\ell\,$+$\,$MET and $\ell b\bar{b}\,$+$\,$MET. For degenerate Higgsinos $\tilde{\chi}$ and massless bino LSP, a bound $m_{\tilde{\chi}} > 900$~GeV ($95\%$ CL) was obtained. Our signal cross section reads at partonic level
\begin{equation}
\hat{\sigma}(u \bar{d} \to Q_u \overline{Q}_d) = \frac{N_d}{N_c} \frac{\pi \alpha_W^2}{6\hat{s}}  \frac{\hat{s}^2}{(\hat{s} - m_W^2)^2} \left( 1 - \frac{4 M^2}{\hat{s}} \right)^{1/2} \left( 1 + \frac{2M^2}{\hat{s}} \right),
\end{equation}
which is $N_d$ times larger than for the Higgsinos, if SM QCD and dark QCD corrections are neglected. After convoluting with the parton luminosities and summing over all charge combinations, the $13$~TeV cross section is $\sigma(pp \to Q\overline{Q}, M = 900\;\mathrm{GeV}) \approx N_d \,1.6\;\mathrm{fb}$, where we used MSTW2008 NLO parton distribution functions~\cite{Martin:2009iq} and the factorization scale was set to $\sqrt{\hat{s}}/2$. To obtain an approximate but reliable exclusion on the $Q$'s, we solve
\begin{equation}
\sigma(pp \to Q\overline{Q}, M) = \sigma(pp \to \tilde{\chi}\tilde{\chi}, m_{\tilde{\chi}} =  900\;\mathrm{GeV})\,,
\end{equation}
for the mass $M$ of the lightest $Q_i$, obtaining for $N_d = 3$
\begin{equation}
M \gtrsim 1.1\;\mathrm{TeV}\,. \qquad (\mathrm{direct \;searches})
\end{equation}
We have assumed the $Q_i$ are not mass-degenerate, which applies to all benchmark models considered in Section~\ref{sec:benchmarks} (for two degenerate $Q_i$, the constraint strengthens to $1.2$~TeV). Given the current Higgsino expected bound~\cite{ATLAS:2021yqv}, we rescale the cross section by $\sqrt{L/L'}$ with $L, L' = 139, 3000\;\mathrm{fb}^{-1}$ and derive $M \gtrsim 1.3$~TeV as our estimate of the ($13\;\mathrm{TeV}$) HL-LHC sensitivity. 

If the dark pions are heavy enough to decay inside the detector, the phenomenology becomes similar to the emerging jets scenario~\cite{Schwaller:2015gea,Mies:2020mzw}, albeit with EW rather than QCD production of the mediators. Evaluating the impact of the existing CMS search~\cite{Sirunyan:2018njd} on our signals is beyond the scope of this work, and left as an interesting avenue for future studies.

Beside DY production we consider single $Q$ production mediated by off-shell Higgs, $gg\to h^\ast \to Q_d \overline{\psi}$. This yields $Z/h + \psi\overline{\psi}$ final states, leading to mono-$Z/h$ signatures if the dark pions escape undetected. The partonic cross section is found to be, neglecting the $\psi$ mass,
\begin{equation}
\hat{\sigma}(gg \to Q_d \overline{\psi}) = \frac{N_d \alpha_s^2 }{1024 \pi^3 (N_c^2-1)} \frac{ |Y|^2 + |\widetilde{Y}|^2 }{v^2} \Big| x \big[1 + (1-x)f(x)^2 \big] \Big|^2 \left( \frac{1 - M^2/ \hat{s} }{1 - m_h^2/\hat{s}} \right)^2\,,
\end{equation}
where $x(\hat{s}) = 4m_t^2 / \hat{s}$, $f(x)$ is defined in Eq.~\eqref{eq:f_loop}, and again we have neglected flavor indices. By folding in the $gg$ parton luminosity we obtain the $13$~TeV cross sections
\begin{equation}
\sigma \Big(pp \to Q_d \overline{\psi} + \overline{Q}_d \psi, M = \{0.5, 1\}\;\mathrm{TeV} \Big) \approx\, N_d \left(|Y|^2 + |\widetilde{Y}|^2\right)\; \{0.70 \;\mathrm{fb}, 9.7\;\mathrm{ab}\}\,,
\end{equation}
where the renormalization and factorization scales were set to $M$. These results show that single production cannot compete in rate with DY, though the sensitivity to the Yukawa couplings makes it a complementary probe of the UV completion.

\section{Conclusions}\label{sec:conclusions}
 
In this paper we have formulated a theory and initiated the study of dark pions, coupled to the SM via irrelevant $Z$ and Higgs portals. The corresponding operators are obtained by integrating out TeV-scale EW-doublet fermion mediators. This setup has strong UV motivations, appearing in various modern approaches to the hierarchy problem, such as neutral naturalness models and the relaxion scenario. It provides a concrete framework where the GeV-scale phenomenology of the dark pions, the EW-scale decays of $Z$ and $h$ bosons to the hidden sector, and the TeV-scale signals of the mediators are all coherently linked. 

The decays of $CP$-odd and $CP$-even dark pions proceed via tree-level mixing with the $Z$ and $h$, respectively, providing explicit realizations of light composite ALPs and scalars coupled feebly to the SM. For $CP$-odd dark pions, we have provided a new comprehensive calculation of the decay widths to exclusive hadronic SM final states, obtained by applying data-driven methods. The results are valid for any ALP with arbitrary flavor-diagonal couplings to SM fermions, and can therefore be widely used to study other models.

The dark pion phenomenology depends on the symmetries possessed by the model, including $CP$, dark isospin, and chiral symmetries. To illustrate the range of possibilities we have analyzed in detail three benchmark scenarios. We find that for masses and couplings of the mediators that can be related to the hierarchy problem while satisfying experimental constraints, and for dark pion decay constants around the GeV scale, dark pions with $2m_\mu \lesssim m_{\hat{\pi}} \lesssim 2m_b$ have lifetimes varying from a millimeter to 10 meters. Intriguingly, this is the most interesting range for LLP searches at the LHC (and beyond), making the dark pions a natural target. We have begun the exploration of the signatures with two applications, meson FCNC decays and $Z$-initiated dark shower searches, focusing primarily on the mass region $2m_\mu \lesssim m_{\hat{\pi}} \lesssim 2m_c$ where the striking dark pion decay to dimuons has a significant branching ratio.

Searches for flavor-changing $b\to s\hspace{0.2mm} a$ decays, with long-lived $a \to \mu^+\mu^-$, set important bounds on the effective decay constant of the $CP$-odd dark pions, $f_a \gtrsim \mathrm{PeV}$. In addition to well-known constraints from CHARM and LHCb, we have derived new ones from a recent CMS search leveraging the data scouting technique. Each of these experiments turns out to have the strongest sensitivity in a different $m_{\hat{\pi}}$ window. Proposed LLP detectors at the LHC interaction points, including FASER 2, CODEX-b, and MATHUSLA, have the potential to extend the sensitivity on $f_a$ by $1\,$-$\,2$ orders of magnitude. For $m_{\hat{\pi}} < 2 m_\mu$, there are lower bounds $f_a \gtrsim \mathrm{PeV}$ from $K\to \pi \,+\,$invisible searches at E949 and NA62. On the other hand, the $CP$-even dark pions remain out of reach due to their very small mixing with the Higgs.

Dark shower searches at the LHC access the additional structure that partially completes the theory at the EW scale. They probe decays of on-shell $Z$ and $h$ bosons to dark jets composed mainly of long-lived dark pions. $Z$ decays to the dark sector, in particular, have been largely overlooked so far, but here we have shown that they probe new directions in the parameter space, supplying orthogonal information to meson FCNC decays. We have performed a thorough recast of the most recent LHCb search for displaced dimuons. The resulting constraints demonstrate that the sensitivity to $Z$-initiated dark showers has already reached new parameter space, surpassing competing bounds from meson FCNC and $Z\to\mathrm{invisible}$ decays. ATLAS and CMS have strong potential to extend the reach to longer dark pion lifetimes, which are well motivated in our framework, by exploiting larger decay volumes and luminosities. Dedicated experimental strategies are increasingly being implemented, such as data scouting/trigger-level analysis, and a detailed assessment of their impact on our framework will appear elsewhere.

As for the direct LHC reach on the EW-charged mediators, a straightforward reinterpretation of Higgsino searches in all hadronic + MET final states gives the constraint $M \gtrsim 1.1\;\mathrm{TeV}$. The improvement expected in the high-luminosity phase is mild, leaving open the possibility that a dark pion discovery may take place at the LHC, while the direct production of the mediators would need to wait for a future collider.

Looking ahead, many paths deserve further exploration. Hadronic decays of GeV-scale dark pions are shown to be important by our results, warranting new studies both for FCNC meson decays and dark shower searches at the LHC. Notable modes include: $\hat{\pi} \to \pi^+ \pi^- \pi^0$, which we find to dominate the width of light $CP$-odd dark pions, $\hat{\pi} \to K^{\ast 0} \overline{K}^{\ast 0} \to (K^+ \pi^-)(K^- \pi^+)$ and $\hat{\pi} \to K^+ K^-$, which can be fully reconstructed and have sizable branching ratios in some parameter regions, and several others discussed in Sections~\ref{sec:theory_hadrons} and~\ref{sec:benchmarks}. The sensitivity of Belle II to such modes requires detailed studies, as well. In addition, we have not touched upon the heavier mass range $m_{\hat{\pi}} \gtrsim 2m_c$, where hadronic decays dominate and lifetimes become significantly shorter. In particular, it would be interesting to understand if in this region there are any constraints on the EW pair-production of the heavy dark quark mediators from the existing CMS search for emerging jets.

The dark pion phenomenology at fixed-target experiments also remains to be investigated. We note that dark hadrons heavier than the dark pions may be relevant there, due to different production mechanisms which could be exploited to test specific regions of parameter space. For instance, bremsstrahlung production of dark vector mesons can be strongly enhanced if their mass is around 1 GeV, due to mixing with SM vector meson resonances.

Finally, the sensitivity of future colliders to the scenario presented here warrants further studies. In particular, an $e^+ e^-$ machine like FCC-ee would offer extraordinary possibilities to probe decays to the hidden sector at a Tera-$Z$ phase, as it has already been demonstrated for one-flavor dark QCD models. We believe the present work sets a solid foundation to tackle all the above aspects, while providing several new results of general applicability in the study of light, feebly coupled hidden sectors.

\acknowledgments
We have benefited for communications with Daniel Aloni about Ref.~\cite{Aloni:2018vki} and with Martino Borsato concerning Ref.~\cite{Aaij:2020ikh}. We thank Chris Verhaaren for collaboration at an early stage of this project, and Simon Knapen, Michele Redi, Diego Redigolo, and Yuhsin Tsai for discussions. HC thanks for hospitality the National Center for Theoretical Sciences in Taiwan, where part of the work was done. HC is supported by the US Department of Energy grant No.~DE-SC0009999. LL was partly supported by the General Research Fund (GRF) under Grant No.~16312716, which was issued by the Research Grants Council of Hong Kong S.A.R.. LL was also partly supported by NASA grant No.~80NSSC18K1010. ES thanks the organizers of the MIAPP ``Novel Hidden Sectors'' and Portoro\v z ``Physics of the flavorful Universe'' workshops for kind invitations to present preliminary results of this work.

\appendix 
 \section{Decays of a light ALP coupled to Standard Model fermions}\label{app:ALP_decays_general}
The starting point is the Lagrangian
\begin{equation} \label{eq:ALP_Lagr}
\mathcal{L}_a = \frac{1}{2}(\partial_\mu a)^2 - \frac{1}{2} m^2_a a^2 -  \frac{\partial_\mu a}{f_a} \sum_f c_f \bar{f} \gamma^\mu \gamma_5 f\;,
\end{equation}
with $f \in \{q, \ell , \nu \}$ for quarks, charged leptons and neutrinos. The width for decay to a pair of charged leptons is
\begin{equation}
\Gamma(a \to \ell^+ \ell^-) = \frac{c_\ell^2}{2\pi f_a^2} m_a m_\ell^2 \Big( 1 - \frac{4m_\ell^2}{m_a^2} \Big)^{1/2}.
\end{equation}
If the ALP is much heavier than the SM QCD scale, $\Lambda_{\rm SM} \ll m_{a}$, its hadronic decays can be analyzed perturbatively. The width for decay to two gluons is~\cite{Bauer:2017ris,Bauer:2020jbp}
\begin{equation} \label{eq:pihat_gg}
\Gamma(a \to gg) = \frac{128 \pi \alpha_s^2 (m_{a}) }{f_a^2}\,m_{a}^3\,  \bigg| \sum_{q \, =\, u,d,s,c,b,t} \frac{c_{q} B_1 (4m_q^2/m_{a}^2)}{32 \pi^2} \bigg|^2 \Big[ 1 + \Big(\frac{97}{4} - \frac{7 n_q}{6}\Big) \frac{\alpha_s (m_{a})}{\pi} \Big],
\end{equation}
where $n_q$ counts the quarks lighter than $m_{a}$, while the loop function is\footnote{The structure of $B_1$ can be understood upon integrating by parts the interaction with quarks in Eq.~\eqref{eq:ALP_Lagr} and using the expression of the divergence of the axial current,
\begin{equation*}
\partial_\mu (\bar{q} \gamma_\mu \gamma_5 q) = 2 i m_q \bar{q} \gamma_5 q - \frac{g_s^2}{16\pi^2} G_{\mu\nu}^a \widetilde{G}^{\mu\nu\,a} \quad \to \quad \frac{a}{f_a} \sum_{q}  c_q \big( 2 i m_q \bar{q} \gamma_5 q - \frac{g_s^2}{16\pi^2} G_{\mu\nu}^a \widetilde{G}^{\mu\nu\,a} \big).
\end{equation*}
}
\begin{equation} \label{eq:f_loop}
B_1(x) = 1 - x f (x)^2\,, \qquad f(x) =\begin{cases}
               \arcsin \frac{1}{\sqrt{x}}\quad\qquad\quad\quad\;\; x \geq 1\,,\\
               \frac{\pi}{2} + \frac{i}{2} \log \frac{1 + \sqrt{1 - x}}{1 - \sqrt{1 - x}}\quad\quad\, x < 1\,.
            \end{cases}
\end{equation}
We have $B_1(x) \approx 1\,(-\frac{1}{3x})$ for $x \ll 1\;(\gg 1)$, implying that light quarks contribute $\approx c_q/(32\pi^2)$ to the sum in Eq.~\eqref{eq:pihat_gg} whereas heavy quarks rapidly decouple. For decay to heavy quarks $Q = c, b$,
\begin{equation}
\Gamma(a \to Q \overline{Q}) = \frac{N_c c_Q^2}{2\pi f_a^2} m_a \overline{m}^2_Q(m_a) \Big( 1 - \frac{4m_Q^2}{m_a^2} \Big)^{1/2} ,
\end{equation}
where $\overline{m}_Q$ is the running quark mass in the $\overline{\mathrm{MS}}$ scheme. We use two-loop running for both $\alpha_s$ and $\overline{m}_{c,b}\,$, and set $m_c = 1.67$~GeV, $m_b = 4.78$~GeV.

For $m_a \lesssim \Lambda_{\rm SM}$ we must consider decays to exclusive hadronic final states instead. To do so we match Eq.~\eqref{eq:ALP_Lagr} to the low-energy effective Lagrangian~\cite{Georgi:1986df,Aloni:2018vki,Bauer:2021wjo},
\begin{equation} \label{eq:L_eff}
\mathcal{L}_{\rm eff} = \frac{f_\pi^2}{4} \mathrm{Tr} [(D_\mu \Sigma)^\dagger (D^\mu \Sigma) ] + \frac{B_0 f_\pi^2}{2} \mathrm{Tr}[M^\dagger \Sigma + \Sigma^\dagger M] + \frac{1}{2}(\partial_\mu a)^2 - \frac{1}{2} m^2_a a^2 - \frac{1}{2}m_0^2 \eta_0^2 + \mathcal{L}_{V}\,,
\end{equation}
where $\Sigma \to L \Sigma R^\dagger$ under $SU(3)_L \times SU(3)_R$ and the covariant derivative is $D_\mu \Sigma = \partial_\mu \Sigma - i e A_\mu [\bm{Q}, \Sigma] - i \partial_\mu a  \{ \bm{c_q} , \Sigma \} / f_a$, with $\bm{Q} = \mathrm{diag}\,(2, -1, -1)/3$ and $\bm{c_q} = \mathrm{diag}\,(c_u, c_d, c_s)$. The pseudoscalar matrix is written as 
\begin{equation}
\Sigma = \exp(2 i \bm{P}/f_\pi),\qquad \bm{P} = \frac{1}{\sqrt{2}} \begin{pmatrix} \frac{\pi_0}{\sqrt{2}} + \frac{\eta}{\sqrt{3}} + \frac{\eta'}{\sqrt{6}} & \pi^+ & K^+ & \\ \pi^- & - \frac{\pi_0}{\sqrt{2}} + \frac{\eta}{\sqrt{3}} + \frac{\eta'}{\sqrt{6}} & K^0 \\ K^- & \overline{K}^{\,0} & - \frac{\eta}{\sqrt{3}} + \frac{2\eta'}{\sqrt{6}} \end{pmatrix},
\end{equation}
where $f_\pi \approx 93$~MeV. The hard $U(1)_A$ breaking due to the anomaly is parametrized by $m_0^2$ and the physical $\eta,\eta'$ are related to the octet and singlet fields by
\begin{equation}
\begin{pmatrix} \eta \\ \eta' \end{pmatrix} = \begin{pmatrix} \cos\theta_{\eta\eta'} & - \sin\theta_{\eta\eta'} \\ \sin\theta_{\eta\eta'} & \cos\theta_{\eta\eta'} \end{pmatrix}  \begin{pmatrix} \eta_8 \\ \eta_0 \end{pmatrix}, \quad \sin\theta_{\eta\eta'} = - \frac{1}{3}\,,\quad \cos\theta_{\eta\eta'} = \frac{2\sqrt{2}}{3}\,.
\end{equation}
This approximate value of the mixing angle is sufficiently accurate for our purpose, while simplifying analytical expressions~\cite{Aloni:2018vki}. The relevant pieces of the Lagrangian describing the vector resonances are~\cite{Fujiwara:1984mp}
\begin{align} \label{eq:VMD}
\mathcal{L}_{V} =&\, g_{VVP} \mathrm{Tr} \big( \bm{P} \bm{V}_{\mu\nu} \bm{\widetilde{V}}^{\mu\nu} \big) +  \frac{i \hspace{0.1mm}N_c e}{6\pi^2 f_\pi^3} \epsilon^{\mu\nu\rho\sigma} A_\mu \mathrm{Tr} \big( \bm{Q} \partial_\nu \bm{P} \partial_\rho \bm{P} \partial_\sigma \bm{P} \big) \nonumber \\ 
+&\, 2 f_\pi^2 \mathrm{Tr} \left| g \bm{V}_\mu - e A_\mu \bm{Q} - \frac{i}{2 f_\pi^2} [\bm{P}, \partial_\mu \bm{P} ] \right|^2 + \ldots \,,
\end{align}
where $\widetilde{V}^{\mu\nu} = \frac{1}{2} \epsilon^{\mu\nu\rho\sigma}V_{\rho\sigma}$ (with $\epsilon^{0123} = 1$) and $g_{VVP} = - N_c g^2 / (8\pi^2 f_\pi)$ is determined by the anomaly. The coupling $g$ is fixed by the Kawarabayashi-Suzuki-Riazuddin-Fayyazuddin (KSRF) relation~\cite{Kawarabayashi:1966kd,Riazuddin:1966sw} to $g = g_{V \pi\pi} = m_{V}/(\sqrt{2}f_\pi) \approx 6.0$, where for $m_V$ we have taken the $\rho$ mass.\footnote{This $g$ should not be confused with the $SU(2)_L$ gauge coupling, which never appears in this appendix.} The vector meson matrix reads
\begin{equation} \label{eq:V_def}
\bm{V} = \frac{1}{\sqrt{2}} \begin{pmatrix} \frac{\rho_0 + \omega}{\sqrt{2}} & \rho^+  & K^{\ast +} \\ \rho^- & \frac{-\rho_0 + \omega}{\sqrt{2}} & K^{\ast 0} \\ K^{\ast -} & \overline{K}^{\ast 0} & \phi  \end{pmatrix}.
\end{equation}
It is important to note that Eq.~\eqref{eq:VMD} realizes vector meson dominance (VMD) for $\pi_0 \to \gamma\gamma$ but retains an (anomalous) $\gamma P^3$ contact interaction, with coefficient equal to $-1/2$ of the one given by the WZW action. This choice was shown to provide a better fit to data compared to ``complete VMD''~\cite{Fujiwara:1984mp}, and will impact the calculation of $a\to \pi^+ \pi^- \gamma$.

Equation~\eqref{eq:L_eff} contains a piece $i \frac{f_\pi^2}{2 f_a} \partial_\mu a \mathrm{Tr} [ \bm{c_q} (\Sigma^\dagger D^\mu \Sigma - \Sigma (D^\mu \Sigma)^\dagger)]$ that mixes kinetically the ALP and pseudoscalar mesons, $\mathcal{L}_{\rm eff} \supset - \frac{f_\pi}{f_a} \partial_\mu a \sum_{P = \pi_0, \eta, \eta'}  K_{aP}  \partial^\mu P$ with
\begin{equation}
K_{a\pi_0} = c_u - c_d\,, \quad K_{a \eta} = \sqrt{\frac{2}{3}} (c_u + c_d - c_s)\,,\quad K_{a\eta'} = \frac{1}{\sqrt{3}} (c_u + c_d + 2 c_s).
\end{equation}
This is diagonalized at $O(f_\pi/f_a)$ by the transformations~\cite{Aloni:2018vki}
\begin{equation}
a \to a - \frac{f_\pi}{f_a} \sum_{P = \pi_0, \eta, \eta'} \frac{m_P^2}{m_a^2} \langle \bm{a} \bm{P} \rangle  P\,, \qquad P \to P + \frac{f_\pi}{f_a} \langle \bm{a} \bm{P} \rangle\, a \,, \qquad \langle \bm{a} \bm{P} \rangle = \frac{m_a^2 K_{aP}}{m_a^2 - m_P^2}\,,
\end{equation}
where isospin breaking due to $m_u \neq m_d$ was neglected and we defined $\langle \ldots \rangle \equiv 2\, \mathrm{Tr}(\ldots)$. We then assign to the ALP the $U(3)$ representation $\bm{a} = \frac{1}{\sqrt{6}}\, \mathrm{diag}\,(\mathcal{C}_u, \mathcal{C}_d, \mathcal{C}_s)$, with
\begin{equation}
\mathcal{C}_u = \sqrt{\tfrac{3}{2}} \langle \bm{a} \bm{\pi_0} \rangle + \langle \bm{a} \bm{\eta} \rangle + \tfrac{1}{\sqrt{2}}  \langle \bm{a} \bm{\eta'} \rangle \,, \quad \mathcal{C}_d =  -  \sqrt{\tfrac{3}{2}} \langle \bm{a} \bm{\pi_0} \rangle + \langle \bm{a} \bm{\eta} \rangle + \tfrac{1}{\sqrt{2}}  \langle \bm{a} \bm{\eta'} \rangle\,, \quad \mathcal{C}_s =  - \langle \bm{a} \bm{\eta} \rangle + \sqrt{2}  \langle \bm{a} \bm{\eta'} \rangle \,,
\end{equation}
which are taken to be valid up to $m_a \approx 3\;\mathrm{GeV}$. Above this mass we switch to the perturbative description. The model studied in this paper has $c_f = T^3_{Lf}$ i.e. $c_u = - c_d = -c_s = 1/2$, giving $K_{a\pi_0} = 1$, $K_{a\eta} = 1/\sqrt{6}$ and $K_{a\eta'} = - 1/\sqrt{3}$. However, we stress that our results are general and also apply to other models with different patterns of ALP-fermion couplings, for example those in Refs.~\cite{Renner:2018fhh,Carmona:2021seb}. We are now in the position to calculate the decays of low-mass ALPs to exclusive final states. 

\subsection{$a\to \gamma\gamma$}
We begin with the decay to two photons~\cite{Aloni:2018vki,Bauer:2017ris},
\begin{equation} \label{eq:width_gaga}
\Gamma(a \to \gamma\gamma) = \frac{\alpha^2 m_{a}^3}{(4\pi)^3 f_a^2} \left| C_\gamma^{\rm VMD} + C_\gamma^{{\rm pQCD},uds} + C_\gamma^{{\rm pQCD},cbt} + C_\gamma^{{\rm leptons}} \right|^2 ,
\end{equation}
where $C_\gamma$ is defined by the effective operator $\frac{C_\gamma \alpha}{8\pi f_a} a\, \epsilon^{\mu\nu\rho\sigma} F_{\mu\nu} F_{\rho\sigma}$. The individual contributions are
\begin{align}
C_{\gamma}^{\rm VMD} &= -  N_c \big[ \langle \bm{a} \bm{\rho_0} \bm{\rho_0} \rangle + \tfrac{1}{9} \langle \bm{a} \bm{\omega}\bm{\omega} \rangle + \tfrac{2}{9} \langle \bm{a} \bm{\phi} \bm{\phi} \rangle + \tfrac{2}{3} \langle \bm{a} \bm{\rho_0} \bm{\omega} \rangle \big]  \Theta (m_a^\ast - m_a) \mathcal{F}(m_a),\\
C_\gamma^{{\rm pQCD},uds}&= - \sum_{q = u,d,s} 2 N_c Q_q^2 c_q \Theta ( m_{a} - m_a^\ast), \\
C_\gamma^{{\rm pQCD},cbt}&= - \sum_{q = c,b,t} 2 N_c Q_q^2 c_q B_1 (4m_q^2/m_{a}^2) \Theta ( m_{a} - m_a^\ast), \\
C_\gamma^{{\rm leptons}}&= - \sum_{\ell = e,\mu,\tau} 2 Q_\ell^2 c_\ell B_1 (4m_\ell^2/m_{a}^2),
\end{align}
where the $\bm{\rho_0}, \bm{\omega}, \bm{\phi}$ matrices are implicitly defined by Eq.~\eqref{eq:V_def} and $m_a^\ast$ is the scale where the VMD and pQCD terms are matched, which equals $\approx 2.9$~GeV for our benchmark model. The form factor $\mathcal{F} \equiv \mathcal{F}_4$ accounts for the suppression of the $VVP$ interaction at high mass, extracted in Ref.~\cite{Aloni:2018vki} by comparison to $e^+ e^-$ data,
\begin{equation} \label{eq:form_factors}
\mathcal{F}_n(m_a) =\begin{cases}
              1\quad\qquad\quad\quad\qquad\qquad\quad\qquad\;\;\, m_a < 1.4\;\mathrm{GeV},\\
               1 + [(\frac{1.4}{2})^n - 1] \frac{ (m_a - 1.4\;\mathrm{GeV}) }{(2 - 1.4)\,\mathrm{GeV}}           \quad\;\, 1.4 \;\mathrm{GeV} \leq m_a \leq 2\;\mathrm{GeV},\\
             \Big(\frac{1.4\;\mathrm{GeV}}{m_a}\Big)^n\qquad\;\,\qquad \qquad\quad\quad\, m_a > 2\;\mathrm{GeV}.
            \end{cases}
\end{equation}
A basic cross-check of Eq.~\eqref{eq:width_gaga} is that, setting $\bm{a}\to \bm{\pi_0}$ and $f_a \to f_\pi$, it reproduces the classic result for $\Gamma(\pi_0 \to \gamma\gamma)$, which in the VMD picture is mediated by $\langle \bm{\pi_0} \bm{\rho_0} \bm{\omega} \rangle = 1/2$. In addition, the predicted widths for $\eta, \eta' \to \gamma\gamma$ match the experimental values within $20\%$.

\subsection{$a\to \pi^+ \pi^- \gamma$}
The amplitude is described by $5$ diagrams: two with $\rho_0$ exchange, two with $\rho_\pm$ exchange, and one contact interaction. For the spin-summed squared matrix element we find
\begin{align} \label{eq:pipigamma_M2}
&\overline{|\mathcal{M}|^2} = \frac{1}{4} \, [m_{12}^2(m_{13}^2 - m_\pi^2)(m_{23}^2 - m_\pi^2) - m_\pi^2(m_a^2 - m_{12}^2)^2 ]\, \Bigg| \Big\{ \hspace{-1mm} -g_{VVP} \frac{2e}{3}  \frac{f_\pi}{f_a} \nonumber\\ 
&\times \Big[ \big(\sqrt{6} \langle \bm{a} \bm{\eta} \rangle + \sqrt{3} \langle \bm{a} \bm{\eta'} \rangle + \langle \bm{a} \bm{\pi_0} \rangle \big) \mathrm{BW}_{\rho_0} (m_{12}^2) - \langle \bm{a} \bm{\pi_0} \rangle \big(  \mathrm{BW}_{\rho_\pm} (m_{13}^2) +  \mathrm{BW}_{\rho_\pm} (m_{23}^2) \big) \Big] \nonumber \\ 
&- \frac{N_c \hspace{0.1mm}e}{6\pi^2 f_\pi^2 f_a } \big( \tfrac{1}{2\sqrt{6}}  \langle \bm{a} \bm{\eta} \rangle + \tfrac{1}{4\sqrt{3}}  \langle \bm{a} \bm{\eta'} \rangle +  \tfrac{1}{4}  \langle \bm{a} \bm{\pi_0} \rangle \big)   \Big\} \mathcal{F}(m_a) \Bigg|^2 ,
\end{align}
where 
\begin{equation}
\mathrm{BW}_{x} (m_{ij}^2) \equiv (m_x^2 - m_{ij}^2 - i m_x \Gamma_x)^{-1}
\end{equation}
and, adopting a convention we follow consistently, the final-state particles were ordered according to how we define the decay (i.e. in $a\to \pi^+ \pi^- \gamma$, $1$ denotes the $\pi^+$ and so on). In addition, the four-momenta satisfy $p_a = \sum_{i\, \in\, {\rm final}} p_i$. The width is
\begin{equation}
\Gamma(a \to \pi^+ \pi^- \gamma) = \frac{1}{2 S m_a} \int \overline{|\mathcal{M}|^2} d\Phi_3 = \frac{1}{32 m_a^3 (2\pi)^3} \int_{4m_\pi^2}^{m_a^2} dm_{12}^2 \int_{m_-^2}^{m_+^2} dm_{23}^2\, \overline{|\mathcal{M}|^2}
\end{equation}
with $2m^2_{\pm} = \pm (m_a^2 - m_{12}^2)(1 - \tfrac{4m_\pi^2}{m_{12}^2})^{1/2} + m_a^2 + 2 m_\pi^2 - m_{12}^2\,$ and the symmetry factor $S = 1$. We cross-check this result by applying it to the $\eta'$: setting $a\to \eta'$, $\langle \bm{\eta'} \bm{P} \rangle = \delta_{\eta' P}$ and $f_a \to f_\pi$ gives $\Gamma(\eta' \to \pi^+ \pi^- \gamma) \approx 56$~keV, in excellent agreement with the PDG value of $55$~keV. The same procedure applied to the $\eta$ yields $\Gamma(\eta \to \pi^+ \pi^- \gamma) \approx 90$~eV, to be compared with the experimental value of $55$~eV. Our $\eta$ prediction would get significantly closer to the observed rate if we used a more precise value of $\theta_{\eta \eta'}$ and accounted for the $SU(3)$-breaking differences among the pseudoscalar decay constants~\cite{Picciotto:1991ae}, which however go beyond the scope of this work. Nonetheless, we remark that the $\gamma P^3$ contact term in the vector meson Lagrangian~\eqref{eq:VMD} is important to improve agreement with data: omitting this term (as done, e.g., in Ref.~\cite{Aloni:2018vki}) we obtain $154$~eV~($63$~keV) for $\eta^{(\prime)}\to \pi^+ \pi^- \gamma$, so the $\eta$ partial width is off by a factor $\approx 3$ relative to the observed value.\footnote{In Ref.~\cite{Aloni:2018vki} only the diagram containing an $a\rho_0 \rho_0$ vertex was included, which corresponds to retaining only the piece proportional to
\begin{equation*}
\big(\sqrt{6} \langle \bm{a} \bm{\eta} \rangle + \sqrt{3} \langle \bm{a} \bm{\eta'} \rangle \big) \mathrm{BW}_{\rho_0} (m_{12}^2) = 6 \langle \bm{a} \bm{\rho_0} \bm{\rho_0} \rangle \mathrm{BW}_{\rho_0} (m_{12}^2)
\end{equation*}
in the second line of Eq.~\eqref{eq:pipigamma_M2}. Upon integrating this partial amplitude over $dm_{23}^2$ we find agreement with Ref.~\cite{Aloni:2018vki}.}

\subsection{$a\to \pi^+ \pi^- \pi^0$}
We include $5$ contributions to the amplitude, $\mathcal{M} = \mathcal{M}_{\rm ChPT} + \mathcal{M}_{\rm VMD} + \mathcal{M}_{\sigma} + \mathcal{M}_{f_0} + \mathcal{M}_{f_2}$. The chiral Lagrangian gives
\begin{equation}
\mathcal{M}_{\rm ChPT} = \frac{\sqrt{k}}{3 f_a f_\pi} \Big[ \langle \bm{a} \bm{\pi_0} \rangle (3m_{12}^2 - m_a^2 - 2 m_\pi^2) \Big] \Theta (m_{\eta'} - m_a).
\end{equation}
In the numerics we actually replace the quantity in square parentheses with its expression including isospin breaking up to $O(\delta_I)$, where $\delta_I \equiv (m_d - m_u)/(m_d + m_u)$, as provided in Eq.~(S32) of Ref.~\cite{Aloni:2018vki}. A $k$ factor equal to $2.7$ is included, derived from comparison with $\eta^{(\prime)} \to 3\pi$ data~\cite{Aloni:2018vki}. The vector meson Lagrangian gives
\begin{align}
&\mathcal{M}_{\rm VMD} =  \frac{\langle \bm{a} \bm{\pi_0} \rangle}{f_a} \Big\{ g^2 f_\pi \big[ (2m_{12}^2 + m_{23}^2 - m_a^2 - 3m_\pi^2) \mathrm{BW}_{\rho}(m_{23}^2) \\ +\,&  (2m_{12}^2 + m_{13}^2 - m_a^2 - 3m_\pi^2) \mathrm{BW}_{\rho}(m_{13}^2) \big] \mathcal{F}_V (m_a) - \frac{1}{2f_\pi} (3m_{12}^2 - m_a^2 - 3m_\pi^2)  \Theta (m_{\eta'} - m_a) \Big\} , \nonumber
\end{align}
where $\mathcal{F}_V \equiv \mathcal{F}_3$. The first two pieces arise from $\rho_\pm$ exchange diagrams, while the third one originates from the $\partial^2 P^4/f_\pi^2$ interactions in Eq.~\eqref{eq:VMD} and is essential to ensure that $\mathcal{M}_{\rm VMD}$ vanishes at low energy, as can be explicitly verified by taking $\mathrm{BW}_\rho \simeq m_\rho^{-2}$ and applying the KSRF relation. Exchange of the $\sigma$ scalar meson yields
\begin{equation} \label{eq:pippimpi0_sigma}
\mathcal{M}_\sigma = - 2 \gamma_{\sigma\pi\pi}^2 \frac{f_\pi}{f_a} \langle \bm{a} \bm{\pi_0} \rangle p_a \cdot p_3\, p_1 \cdot p_2\, \mathrm{BW}_{\sigma} (m_{12}^2) \Theta(4m_K^2 - m_{12}^2) \mathcal{F}(m_a)\, ,
\end{equation}
where $\gamma_{\sigma \pi\pi} = 7.27\;\mathrm{GeV}^{-1}$, as well as all the couplings of the scalar nonet mesons that appear in the following, are taken from the fit to data performed in Ref.~\cite{Fariborz:1999gr} without assuming $U(3)$ symmetry (we use the second set of couplings given in Ref.~\cite{Fariborz:1999gr}). To avoid issues with unitarity we turn off the $\sigma$ contribution at the di-kaon threshold~\cite{Aloni:2018vki}. The amplitude for $f_0$ exchange is given by Eq.~\eqref{eq:pippimpi0_sigma} with $\sigma \to f_0$ everywhere, $\gamma_{f_0 \pi\pi} = 1.47\;\mathrm{GeV}^{-1}$, and removing the cutoff at $m_{12}^2 = 4m_K^2$. 

The tensor meson $f_2$, denoted by $\phi_{\mu\nu}$, is assumed to couple to the energy-momentum tensor as~\cite{Suzuki:1993zs,Han:1998sg,Katz:2005ir}
\begin{equation}\label{eq:f2_coupling}
\mathcal{L}_{f_2} = - g_{f_2 \pi\pi} \frac{f_\pi^2}{4} \langle (\partial^\mu \Sigma^\dagger \partial^\nu \Sigma - \tfrac{1}{2}g^{\mu\nu} \partial^\alpha \Sigma^\dagger \partial_\alpha \Sigma) \bm{f_2} \rangle \phi_{\mu\nu}\,.
\end{equation}
Choosing the $U(3)$ representation $\bm{f_2} = \mathrm{diag}\,(1,1,0)/2$ allows to reproduce approximately the branching ratios into $\pi\pi, \eta\eta, KK$,\footnote{The agreement can be mildly improved by taking $\bm{f_2} = \mathrm{diag}\,(\sqrt{1-s^2},\sqrt{1-s^2},\sqrt{2}\,s)/2$ with $s \sim 0.1$, but for simplicity we stick to $s = 0$.} and the overall coupling strength is fixed via
\begin{equation} 
\Gamma(f_2 \to \pi^+ \pi^-) = \frac{g_{f_2 \pi\pi}^2 m_{f_2}^3}{960\pi} \Big(1 - \frac{4m_\pi^2}{m_{f_2}^2} \Big)^{5/2}  = \frac{2}{3}\, \Gamma_{f_2}^{\rm exp} \,\mathrm{BR}(f_2 \to \pi^+ \pi^- + \pi^0 \pi^0)_{\rm exp}
\end{equation}
to $g_{f_2 \pi\pi} = 13.1\;\mathrm{GeV}^{-1}$, using the PDG values~\cite{Zyla:2020zbs} for the $f_2$ total width and $\pi\pi$ branching ratio. Then we obtain the amplitude for $a\to \pi^+ \pi^- \pi^0$
\begin{equation} \label{eq:f2_pi+pi-pi0}
\mathcal{M}_{f_2} = -\, g_{f_2 \pi\pi}^2 \frac{f_\pi}{f_a} \langle \bm{a} \bm{\pi_0} \rangle \mathrm{BW}_{f_2} (m_{12}^2) \,\widehat{\mathcal{M}}_{f_2} \Theta \big[ m_{12}^2 - (m_{f_2} - \Gamma_{f_2})^2 \big] \mathcal{F}(m_a)
\end{equation}
with
\begin{equation}   \label{eq:f2_pi+pi-pi0_reduced}
\widehat{\mathcal{M}}_{f_2} = (p_a^\mu p_3^\nu - \tfrac{1}{2}g^{\mu\nu} p_a \cdot p_3)(p_1^\rho p_2^\sigma - \tfrac{1}{2}g^{\rho\sigma} p_1 \cdot p_2) B_{\mu\nu,\, \rho\sigma}(p_1 + p_2)\,,
\end{equation}
where the definition of $B_{\mu\nu,\, \rho\sigma}(k)$ is~\cite{Han:1998sg}
\begin{equation}\label{eq:B_tensor}
B_{\mu\nu,\, \rho\sigma}(k) = \big( g_{\mu\rho} - \tfrac{k_\mu k_\rho}{m_{f_2}^2} \big) \big( g_{\nu\sigma} - \tfrac{k_\nu k_\sigma}{m_{f_2}^2} \big) + \big( g_{\mu\sigma} - \tfrac{k_\mu k_\sigma}{m_{f_2}^2} \big)\big( g_{\nu\rho} - \tfrac{k_\nu k_\rho}{m_{f_2}^2} \big) - \tfrac{2}{3}  \big( g_{\mu\nu} - \tfrac{k_\mu k_\nu}{m_{f_2}^2} \big)\big( g_{\rho\sigma} - \tfrac{k_\rho k_\sigma}{m_{f_2}^2} \big)\,.
\end{equation}
The low-energy expansion of this amplitude is
\begin{equation} \label{eq:Mf2_low_energy}
\widehat{\mathcal{M}}_{f_2} = \frac{1}{12} \big[ m_{12}^2 m_{13}^2 + m_{12}^2 m_{23}^2 + 6 m_{13}^2 m_{23}^2 - 8  (m_a^2 + m_\pi^2)m_\pi^2 \big] + O(1/m_{f_2}^2).
\end{equation}
However, it is well known that the $O(p^4)$ low-energy constants in ChPT do not include any sizable contributions from tensor mesons; in fact, it was shown~\cite{Ecker:2007us} that imposing QCD short-distance constraints removes, or suppresses strongly, terms such as Eq.~\eqref{eq:Mf2_low_energy}. For this reason we turn on the $f_2$ exchange amplitudes like Eq.~\eqref{eq:f2_pi+pi-pi0} only for $m_{ij}^2 > (m_{f_2} - \Gamma_{f_2})^2$, thus retaining most of the resonance peak but discarding unphysical low-energy pieces. Combining all terms, the width for $a\to \pi^+ \pi^- \pi^0$ is
\begin{equation}\label{eq:width_3body_scalars}
\Gamma(a \to \pi^+ \pi^- \pi^0) = \frac{1}{2 S m_a} \int |\mathcal{M}|^2 d\Phi_3 
\end{equation}
with $S = 1$.

\subsection{$a\to 3\pi^0$}
This amplitude is not mediated by vector mesons, owing to the absence of a $\rho_0 \pi_0 \pi_0$ coupling in $\mathcal{L}_V$, hence $\mathcal{M} = \mathcal{M}_{\rm ChPT} + \mathcal{M}_{\sigma} + \mathcal{M}_{f_0} + \mathcal{M}_{f_2}$. The chiral Lagrangian contributes only through the pNGB mass term,
\begin{equation}
\mathcal{M}_{\rm ChPT} = \sqrt{k}\, \frac{m_\pi^2}{ f_a f_\pi} \Big[ \langle \bm{a} \bm{\pi_0} \rangle  \Big] \Theta (m_{\eta'} - m_a).
\end{equation}
In the numerical evaluation the quantity in square parentheses is replaced with its expression up to $O(\delta_I)$, found in Eq.~(S31) of Ref.~\cite{Aloni:2018vki}. For $\sigma$ exchange,
\begin{equation} \label{eq:3pi0_sigma}
\mathcal{M}_{\sigma} = - 2 \gamma_{\sigma \pi\pi}^2 \frac{f_\pi}{f_a} \langle \bm{a} \bm{\pi_0} \rangle \big[ p_a \cdot p_3\, p_1 \cdot p_2\, \mathrm{BW}_\sigma (m_{12}^2) \Theta(4m_K^2 - m_{12}^2) + \{1 \leftrightarrow 3\} + \{2 \leftrightarrow 3\} \big] \mathcal{F}(m_a).
\end{equation}
The amplitude for $f_0$ exchange is obtained from Eq.~\eqref{eq:3pi0_sigma} by replacing $\sigma \to f_0$ everywhere and removing the cutoffs on the $\sigma$ propagators. The tensor meson $f_2$ contributes
\begin{equation}
\mathcal{M}_{f_2} = - g_{f_2 \pi\pi}^2 \frac{f_\pi}{f_a} \langle \bm{a} \bm{\pi_0} \rangle \Big\{ \mathrm{BW}_{f_2} (m_{12}^2) \,\widehat{\mathcal{M}}_{f_2} \Theta \big[ m_{12}^2 - (m_{f_2} - \Gamma_{f_2})^2 \big] + \{1 \leftrightarrow 3\} + \{2 \leftrightarrow 3\} \Big\}\mathcal{F}(m_a).
\end{equation}
The width is given by Eq.~\eqref{eq:width_3body_scalars} with $S = 3!$. The $f_2$ amplitude is responsible for the increase of the width above $m_a = 2\;\mathrm{GeV}$ seen in Fig.~\ref{fig:ALP_widths}.

\subsection{$a \to \pi^0 \pi^0 \eta, \pi^+ \pi^- \eta$}
For $a \to \pi^0 \pi^0 \eta$ we consider five contributions, $\mathcal{M} = \mathcal{M}_{\rm ChPT} + \mathcal{M}_\sigma + \mathcal{M}_{f_0} + \mathcal{M}_{a_0} + \mathcal{M}_{f_2}$. The chiral Lagrangian contributes via the pNGB mass term,
\begin{equation}
\mathcal{M}_{\rm ChPT} = \frac{2m_\pi^2}{3 f_\pi f_a} \big( \langle \bm{a} \bm{\eta} \rangle + \tfrac{1}{\sqrt{2}} \langle \bm{a} \bm{\eta'} \rangle \big) \Theta (m_{\eta'} - m_a),
\end{equation}
while the $\sigma$ contribution is
\begin{equation} \label{eq:pipieta_sigma}
\mathcal{M}_\sigma = - \sqrt{2} \frac{f_\pi}{f_a} \gamma_{\sigma \pi\pi} \big(2 \gamma_{\sigma\eta\eta} \langle \bm{a} \bm{\eta} \rangle + \gamma_{\sigma\eta\eta'} \langle \bm{a} \bm{\eta'} \rangle \big) p_a \cdot p_3\, p_1 \cdot p_2 \mathrm{BW}_\sigma (m_{12}^2) \Theta(4m_K^2 - m_{12}^2) \mathcal{F}(m_a)\,,
\end{equation}
and $\mathcal{M}_{f_0}$ is obtained from Eq.~\eqref{eq:pipieta_sigma} by replacing $\sigma\to f_0$ and removing the cutoff at the di-kaon mass. Exchange of the $a_0$ scalar triplet gives
\begin{equation}\label{eq:pipieta_a0}
\mathcal{M}_{a_0} = - \gamma_{a_0 \pi \eta} \frac{f_\pi}{f_a}  ( \gamma_{a_0 \pi \eta} \langle \bm{a} \bm{\eta} \rangle + \gamma_{a_0 \pi \eta'} \langle \bm{a} \bm{\eta'} \rangle ) p_a \cdot p_2  \,p_1 \cdot p_3\, \mathrm{BW}_{a_0}(m_{13}^2)\mathcal{F}(m_a) + \{ 1 \leftrightarrow 2\} \,,
\end{equation}
and finally,
\begin{equation}
\mathcal{M}_{f_2} = - \frac{2}{3} g_{f_2 \pi\pi}^2 \frac{f_\pi}{f_a} \big( \langle \bm{a} \bm{\eta} \rangle + \tfrac{1}{\sqrt{2}} \langle \bm{a} \bm{\eta'} \rangle \big)  \mathrm{BW}_{f_2}(m_{12}^2) \widehat{\mathcal{M}}_{f_2} \Theta \big[ m_{12}^2 - (m_{f_2} - \Gamma_{f_2})^2 \big] \mathcal{F}(m_a).
\end{equation}
The width is obtained from Eq.~\eqref{eq:width_3body_scalars} with $S = 2!$. For the analogous final state involving charged pions we find $\Gamma(a\to \pi^+ \pi^- \eta) = 2\, \Gamma(a\to \pi^0 \pi^0 \eta)$. As a check of these results, we apply them to $\eta' \to \pi\pi \eta$ by setting $\bm{a} \to \bm{\eta'}$ and $f_a \to f_\pi$. The $f_2$ contribution is not included since $m_{\eta'} <  m_{f_2} - \Gamma_{f_2}$, and we obtain $\Gamma(\eta'\to \pi^0 \pi^0 \eta + \pi^+ \pi^- \eta) \approx 129$~keV in very good agreement with the PDG value of $122$~keV, as it should be since the remaining four amplitudes were fit to data including $\eta'\to \eta \pi\pi$ in Ref.~\cite{Fariborz:1999gr}.

\subsection{$a \to \pi^0 \pi^0 \eta', \pi^+ \pi^- \eta'$}
This channel is similar to the previous one, with a few notable differences. In accordance with our choice to cut off ChPT terms at $m_a = m_{\eta'}$, this contribution is not included, hence for $a\to \pi^0 \pi^0\eta'$ we have $\mathcal{M} =  \mathcal{M}_\sigma + \mathcal{M}_{f_0} + \mathcal{M}_{a_0} + \mathcal{M}_{f_2}$. The $\sigma, f_0$ and $a_0$ terms are simply obtained from Eqs.~\eqref{eq:pipieta_sigma} and~\eqref{eq:pipieta_a0} by exchanging everywhere $\eta \leftrightarrow \eta'$, whereas the $f_2$ amplitude reads
\begin{equation}
\mathcal{M}_{f_2} = - \frac{1}{3} g_{f_2 \pi\pi}^2 \frac{f_\pi}{f_a} \big( \langle \bm{a} \bm{\eta'} \rangle + \sqrt{2} \langle \bm{a} \bm{\eta} \rangle \big) \mathrm{BW}_{f_2}(m_{12}^2)  \widehat{\mathcal{M}}_{f_2} \Theta \big[ m_{12}^2 - (m_{f_2} - \Gamma_{f_2})^2 \big] \mathcal{F}(m_a).
\end{equation}
Again we find $\Gamma(a\to \pi^+ \pi^- \eta') = 2\, \Gamma(a\to \pi^0 \pi^0 \eta')$.

\subsection{$a\to \eta \eta \pi^0$}
Here $\mathcal{M} = \mathcal{M}_{f_0} + \mathcal{M}_{a_0} + \mathcal{M}_{f_2}$, with
\begin{align}
 \mathcal{M}_{f_0} =&\, - 2 \sqrt{2} \gamma_{f_0 \pi\pi} \gamma_{f_0 \eta\eta} \frac{f_\pi}{f_a} \langle \bm{a} \bm{\pi_0} \rangle p_a \cdot p_3\,p_1 \cdot p_2\,\mathrm{BW}_{f_0}(m_{12}^2) \mathcal{F}(m_a), \\
  \mathcal{M}_{a_0} =&\, - \gamma_{a_0 \pi\eta}^2 \frac{f_\pi}{f_a} \langle \bm{a} \bm{\pi_0} \rangle p_a \cdot p_1\,p_2 \cdot p_3\,\mathrm{BW}_{a_0}(m_{23}^2) \mathcal{F}(m_a) + \{1 \leftrightarrow 2\}, \\
   \mathcal{M}_{f_2} =&\, - \frac{2}{3} g_{f_2 \pi\pi}^2  \frac{f_\pi}{f_a} \langle \bm{a} \bm{\pi_0} \rangle \mathrm{BW}_{f_2}(m_{12}^2)\widehat{\mathcal{M}}_{f_2} \mathcal{F}(m_a).
\end{align}
We do not include a $\sigma$ contribution analogous to $\mathcal{M}_{f_0}$, consistently with the prescription of cutting off the $\sigma$ propagator at $4m_K^2$, and the usual $\Theta$ function in $\mathcal{M}_{f_2}$ is trivial since $2m_\eta > m_{f_2} - \Gamma_{f_2}$. The symmetry factor to be used in Eq.~\eqref{eq:width_3body_scalars} is $S = 2!$.

\subsection{$a\to K^0 \overline{K}^0 \pi^0$}
The amplitude reads $\mathcal{M} = \mathcal{M}_{\rm VMD} + \mathcal{M}_{a_0} + \mathcal{M}_{S(K\pi)} + \mathcal{M}_{f_2}$. Exchange of the $K^\ast$ vector gives
\begin{align}\label{eq:Kstar_K0K0barpi0}
 \mathcal{M}_{\rm VMD} =&\, - \frac{g^2}{2}\frac{f_\pi}{f_a} \big( \tfrac{2}{\sqrt{6}} \langle \bm{a} \bm{\eta} \rangle - \tfrac{1}{2\sqrt{3}} \langle \bm{a} \bm{\eta'} \rangle - \tfrac{1}{2} \langle \bm{a} \bm{\pi_0} \rangle \big) \\ 
 \times&\,\Big[ m_{12}^2 - m_{13}^2  - \frac{(m_a^2 - m_K^2)(m_K^2 - m_\pi^2)}{m_{K^\ast}^2} \Big] \mathrm{BW}_{K^\ast} (m_{23}^2) \mathcal{F}_V(m_a) + \{1 \leftrightarrow 2\}, \nonumber
 \end{align}
whereas for $a_0$ we find
\begin{equation}
\mathcal{M}_{a_0} = \frac{\gamma_{a_0 KK}}{\sqrt{2}}\frac{f_\pi}{f_a} (\gamma_{a_0 \pi\eta} \langle \bm{a} \bm{\eta} \rangle + \gamma_{a_0 \pi\eta'} \langle \bm{a} \bm{\eta'} \rangle ) \,p_a \cdot p_3\, p_1 \cdot p_2\, \mathrm{BW}_{a_0}(m_{12}^2) \mathcal{F}(m_a).
\end{equation}
In addition, we include the $S$-wave $K\pi$ amplitude measured by BaBar in $\eta_c$ decays~\cite{Lees:2015zzr} as
\begin{align} \label{eq:S_Kpi}
\mathcal{M}_{S(K\pi)} = \frac{\gamma_{\kappa K \pi}}{\sqrt{2}} \frac{f_\pi}{f_a} \,& \Big(\hspace{-1mm} - \frac{\gamma_{\kappa K\pi}}{\sqrt{2}} \langle \bm{a} \bm{\pi_0} \rangle + \gamma_{\kappa K \eta} \langle \bm{a} \bm{\eta} \rangle + \gamma_{\kappa K \eta'} \langle \bm{a} \bm{\eta'} \rangle \Big) \\ \times&\, p_a \cdot p_1\, p_2 \cdot p_3\, \frac{|\mathcal{A}_{\rm BaBar}| e^{i \phi_{\rm BaBar}} \big(\sqrt{m_{23}^2}\,\big)}{m_{K^\ast_0(1430)} \Gamma_{K^\ast_0(1430)}}\, \mathcal{F}(m_a) + \{1 \leftrightarrow 2\}. \nonumber
\end{align}
This expression is obtained through the following logic: the measured $S(K\pi)$ amplitude is dominated by the $I = 1/2$ scalar $K^\ast_0(1430)$~\cite{Lees:2015zzr}, whose mass and width are found to be $m_{K^\ast_0(1430)} = 1.438$~GeV and $\Gamma_{K^\ast_0(1430)} = 0.210$~GeV~\cite{Lees:2014iua}, while also displaying other important effects, in particular a high-mass structure attributed to $K^\ast_0(1950)$. We then approximate the couplings of the $K^\ast_0(1430)$ with those of the $\kappa(900)$ obtained in Ref.~\cite{Fariborz:1999gr},\footnote{Using the PDG values of the $K^\ast_0(1430)$ mass, total width and branching ratios~\cite{Zyla:2020zbs} we may extract $|\gamma_{K^\ast_0(1430) K\pi}| = 4.2\;\mathrm{GeV}^{-1}$ and $|\gamma_{K^\ast_0(1430) K\eta}| = 2.1\;\mathrm{GeV}^{-1}$, to be compared with $\gamma_{\kappa K\pi} = - 5.02\;\mathrm{GeV}^{-1}$ and $\gamma_{\kappa K\eta} = -0.94\;\mathrm{GeV}^{-1}$~\cite{Fariborz:1999gr}, but we cannot estimate $\gamma_{K^\ast_0(1430) K\eta'}$ which is large for $\kappa$, $\gamma_{\kappa K\eta'} = -9.68\;\mathrm{GeV}^{-1}$. Therefore we use the $\kappa$ couplings.} write the amplitude as the sum of two diagrams with $K^\ast_0(1430)$ exchange, and finally arrive at Eq.~\eqref{eq:S_Kpi} by replacing the $\mathrm{BW}$ of $K^\ast_0(1430)$ with the phenomenological BaBar amplitude (we use the amplitude that was extracted from $\eta_c \to K^+ K^- \pi^0$, given in the last two columns of Table V in Ref.~\cite{Lees:2015zzr}) times the normalization factor $(m_{K^\ast_0(1430)} \Gamma_{K^\ast_0(1430)})^{-1}$. A similar, although likely not identical, procedure was adopted in Ref.~\cite{Aloni:2018vki}. Notice that the $K^\ast$ amplitude in Eq.~\eqref{eq:Kstar_K0K0barpi0} includes the exchange of the longitudinal mode, which in general contributes also to $S(K\pi)$, leading in principle to a double counting. However, the empirical fact that $S(K\pi)$ is dominated by the scalar $K^\ast_0(1430)$ supports our simplified prescription to just sum the VMD and $S(K\pi)$ amplitudes. 

Finally, the tensor meson exchange amplitude reads
\begin{equation}
\mathcal{M}_{f_2}  = - \frac{1}{2} g_{f_2 \pi\pi}^2  \frac{f_\pi}{f_a} \langle \bm{a} \bm{\pi_0} \rangle \mathrm{BW}_{f_2}(m_{12}^2) \widehat{\mathcal{M}}_{f_2} \Theta \big[ m_{12}^2 - (m_{f_2} - \Gamma_{f_2})^2 \big] \mathcal{F}(m_a).
\end{equation}
The appropriate symmetry factor for Eq.~\eqref{eq:width_3body_scalars} is $S = 1$.

\subsection{$a\to K^+ K^- \pi^0$}
The four components of $\mathcal{M} = \mathcal{M}_{\rm VMD} + \mathcal{M}_{a_0} + \mathcal{M}_{S(K\pi)} + \mathcal{M}_{f_2}$ read
\begin{align}
\mathcal{M}_{\rm VMD} =&\, \frac{g^2}{2}\frac{f_\pi}{f_a} \big( \tfrac{2}{\sqrt{6}} \langle \bm{a} \bm{\eta} \rangle - \tfrac{1}{2\sqrt{3}} \langle \bm{a} \bm{\eta'} \rangle + \tfrac{1}{2} \langle \bm{a} \bm{\pi_0} \rangle \big) \\ 
& \times\,\Big[ m_{12}^2 - m_{13}^2  - \frac{(m_a^2 - m_K^2)(m_K^2 - m_\pi^2)}{m_{K^\ast}^2} \Big] \mathrm{BW}_{K^\ast} (m_{23}^2) \mathcal{F}_V(m_a) + \{1 \leftrightarrow 2\}, \nonumber \\
 \mathcal{M}_{a_0} =&\, - \frac{\gamma_{a_0 KK}}{\sqrt{2}}\frac{f_\pi}{f_a} (\gamma_{a_0 \pi\eta} \langle \bm{a} \bm{\eta} \rangle + \gamma_{a_0 \pi\eta'} \langle \bm{a} \bm{\eta'} \rangle ) \,p_a \cdot p_3\, p_1 \cdot p_2\, \mathrm{BW}_{a_0}(m_{12}^2) \mathcal{F}(m_a), \\
 \mathcal{M}_{S(K\pi)} =&\, - \frac{\gamma_{\kappa K \pi}}{\sqrt{2}} \frac{f_\pi}{f_a} \Big(\frac{\gamma_{\kappa K\pi}}{\sqrt{2}} \langle \bm{a} \bm{\pi_0} \rangle + \gamma_{\kappa K \eta} \langle \bm{a} \bm{\eta} \rangle + \gamma_{\kappa K \eta'} \langle \bm{a} \bm{\eta'} \rangle \Big) \\ &\qquad\quad\times \, p_a \cdot p_1\, p_2 \cdot p_3\, \frac{|\mathcal{A}_{\rm BaBar}| e^{i \phi_{\rm BaBar}} \big(\sqrt{m_{23}^2}\,\big)}{m_{K^\ast_0(1430)} \Gamma_{K^\ast_0(1430)}}\, \mathcal{F}(m_a) + \{1 \leftrightarrow 2\}, \nonumber\\ 
 \mathcal{M}_{f_2}  =&\, - \frac{1}{2} g_{f_2 \pi\pi}^2  \frac{f_\pi}{f_a} \langle \bm{a} \bm{\pi_0} \rangle  \mathrm{BW}_{f_2}(m_{12}^2) \widehat{\mathcal{M}}_{f_2} \Theta \big[ m_{12}^2 - (m_{f_2} - \Gamma_{f_2})^2 \big] \mathcal{F}(m_a).
\end{align}
The above amplitudes differ from those for $K^0 \overline{K}^0 \pi^0$ only in some (important) signs. The symmetry factor is $S = 1$. 

\subsection{$a\to K^+ \overline{K}^0 \pi^-, K^- K^0 \pi^+$}
In this channel we have $\mathcal{M} = \mathcal{M}_{\rm VMD} + \mathcal{M}_{a_0} + \mathcal{M}_{S(K\pi)}$, where the vector meson contribution includes $\rho_\pm$ exchange in addition to diagrams with $K^\ast$. Focusing on $K^+ \overline{K}^0 \pi^-$ we find
\begin{align}
&\mathcal{M}_{\rm VMD} = \Big\{ - \frac{g^2}{\sqrt{2}} \frac{f_\pi}{f_a} \langle \bm{a} \bm{\pi_0} \rangle (m_{13}^2 - m_{23}^2) \mathrm{BW}_{\rho} (m_{12}^2) \\ +&\, \frac{g^2}{\sqrt{2}}\frac{f_\pi}{f_a} \big( \tfrac{2}{\sqrt{6}} \langle \bm{a} \bm{\eta} \rangle - \tfrac{1}{2\sqrt{3}} \langle \bm{a} \bm{\eta'} \rangle + \tfrac{1}{2} \langle \bm{a} \bm{\pi_0} \rangle \big) \Big[ m_{12}^2 - m_{13}^2  - \frac{(m_a^2 - m_K^2)(m_K^2 - m_\pi^2)}{m_{K^\ast}^2} \Big] \mathrm{BW}_{K^\ast} (m_{23}^2) \nonumber \\
+&\, \frac{g^2}{\sqrt{2}}\frac{f_\pi}{f_a} \big( \tfrac{2}{\sqrt{6}} \langle \bm{a} \bm{\eta} \rangle - \tfrac{1}{2\sqrt{3}} \langle \bm{a} \bm{\eta'} \rangle - \tfrac{1}{2} \langle \bm{a} \bm{\pi_0} \rangle \big) \Big[ m_{12}^2 - m_{23}^2  - \frac{(m_a^2 - m_K^2)(m_K^2 - m_\pi^2)}{m_{K^\ast}^2} \Big] \mathrm{BW}_{K^\ast} (m_{13}^2) \Big\} \mathcal{F}_V(m_a), \nonumber
\end{align}
and
\begin{align}
&\quad\mathcal{M}_{ a_0} = -\, \gamma_{a_0 KK} \frac{f_\pi}{f_a} \big(\gamma_{a_0\pi\eta} \langle \bm{a} \bm{\eta} \rangle + \gamma_{a_0\pi\eta'} \langle \bm{a} \bm{\eta'} \rangle \big) p_a \cdot p_3\, p_1 \cdot p_2\, \mathrm{BW}_{a_0} (m_{12}^2) \mathcal{F}(m_a), \\
&\hspace{-2mm}\mathcal{M}_{S(K\pi)} = - \Big\{ \gamma_{\kappa K \pi} \frac{f_\pi}{f_a} \Big(\frac{\gamma_{\kappa K\pi}}{\sqrt{2}} \langle \bm{a} \bm{\pi_0} \rangle + \gamma_{\kappa K \eta} \langle \bm{a} \bm{\eta} \rangle + \gamma_{\kappa K \eta'} \langle \bm{a} \bm{\eta'} \rangle \Big) p_a \cdot p_1\, p_2 \cdot p_3\, \frac{|\mathcal{A}_{\rm BaBar}| e^{i \phi_{\rm BaBar}} \big(\sqrt{m_{23}^2}\,\big)}{m_{K^\ast_0(1430)} \Gamma_{K^\ast_0(1430)}} \nonumber \\
 +&\, \gamma_{\kappa K \pi} \frac{f_\pi}{f_a} \Big(\hspace{-1mm} - \frac{\gamma_{\kappa K\pi}}{\sqrt{2}} \langle \bm{a} \bm{\pi_0} \rangle + \gamma_{\kappa K \eta} \langle \bm{a} \bm{\eta} \rangle + \gamma_{\kappa K \eta'} \langle \bm{a} \bm{\eta'} \rangle \Big) p_a \cdot p_2\, p_1 \cdot p_3\, \frac{|\mathcal{A}_{\rm BaBar}| e^{i \phi_{\rm BaBar}} \big(\sqrt{m_{13}^2}\,\big)}{m_{K^\ast_0(1430)} \Gamma_{K^\ast_0(1430)}} \Big\} \mathcal{F}(m_a).
\end{align}
The symmetry factor is $S = 1$, and for the conjugate channel we have $\Gamma(a\to K^- K^0 \pi^+) = \Gamma(a\to K^+ \overline{K}^0 \pi^-)$.

\subsection{$a\to \omega\omega, \phi\phi, K^{\ast+} K^{\ast -}, K^{\ast0} \overline{K}^{\ast 0}$}
Since these vector resonances are narrow we take the two-body approximation, finding
\begin{align}
\Gamma(a\to VV) =&\, \frac{N_c^2}{1024\pi^5}  \frac{m_a^3}{f_a^2} \left| g^2 \langle \bm{a} \bm{V} \bm{V} \rangle \mathcal{F}(m_a) \right|^2 \Big( 1 - \frac{4m_V^2}{m_a^2} \Big)^{3/2},\quad (V = \omega, \phi) \\ 
\Gamma(a\to K^{\ast +} K^{\ast -}) =&\,  \frac{N_c^2}{2048\pi^5}  \frac{m_a^3}{f_a^2} \left| g^2 \langle \bm{a} \{ \bm{K^{\ast+}}, \bm{K^{\ast -}} \} \rangle \mathcal{F}(m_a) \right|^2 \Big( 1 - \frac{4m_{K^\ast}^2}{m_a^2} \Big)^{3/2}, \label{eq:aKstarpKstarm}
\end{align}
whereas for $a\to K^{\ast0} \overline{K}^{\ast 0}$ the trace in Eq.~\eqref{eq:aKstarpKstarm} is replaced with $\langle \bm{a} \{ \bm{K^{\ast0}}, \bm{\overline{K}^{\ast 0}} \} \rangle$.

\subsection{$a\to \pi^+ \pi^- \omega$}
This mode proceeds through $a\to (\rho_0 \to \pi^+ \pi^-)\,\omega$. The spin-summed squared matrix element is
\begin{align}
&\,\overline{|\mathcal{M}|^2} = \Big(2 g_{VVP} \frac{f_\pi}{f_a} g_{\rho \pi\pi}   \Big)^2 \left| \langle \bm{a} \bm{\rho_0} \bm{\omega} \rangle \mathrm{BW}_{\rho} (m_{12}^2) \mathcal{F}(m_a) \right|^2 \\ 
\times&\,\Big\{m_{12}^2 \big[ (m_{13}^2 - m_\pi^2 - m_\omega^2)(m_{23}^2 - m_\pi^2 - m_\omega^2)  - m_{12}^2 m_\omega^2 + 4m_\pi^2 m_\omega^2 \big] - m_\pi^2 (m_a^2 - m_{12}^2 - m_\omega^2)^2 \Big\}\,, \nonumber
\end{align}
yielding the decay width
\begin{equation}
\Gamma(a\to \pi^+ \pi^- \omega) = \frac{1}{2S m_a} \int \overline{|\mathcal{M}|^2} d\Phi_3
\end{equation}
with $S = 1$. The four-body decay $a\to \rho\rho \to 4\pi$ is neglected.

\subsection{Comparison with previous studies}
\begin{figure}[t]
\begin{center}
\includegraphics[width=0.495\textwidth]{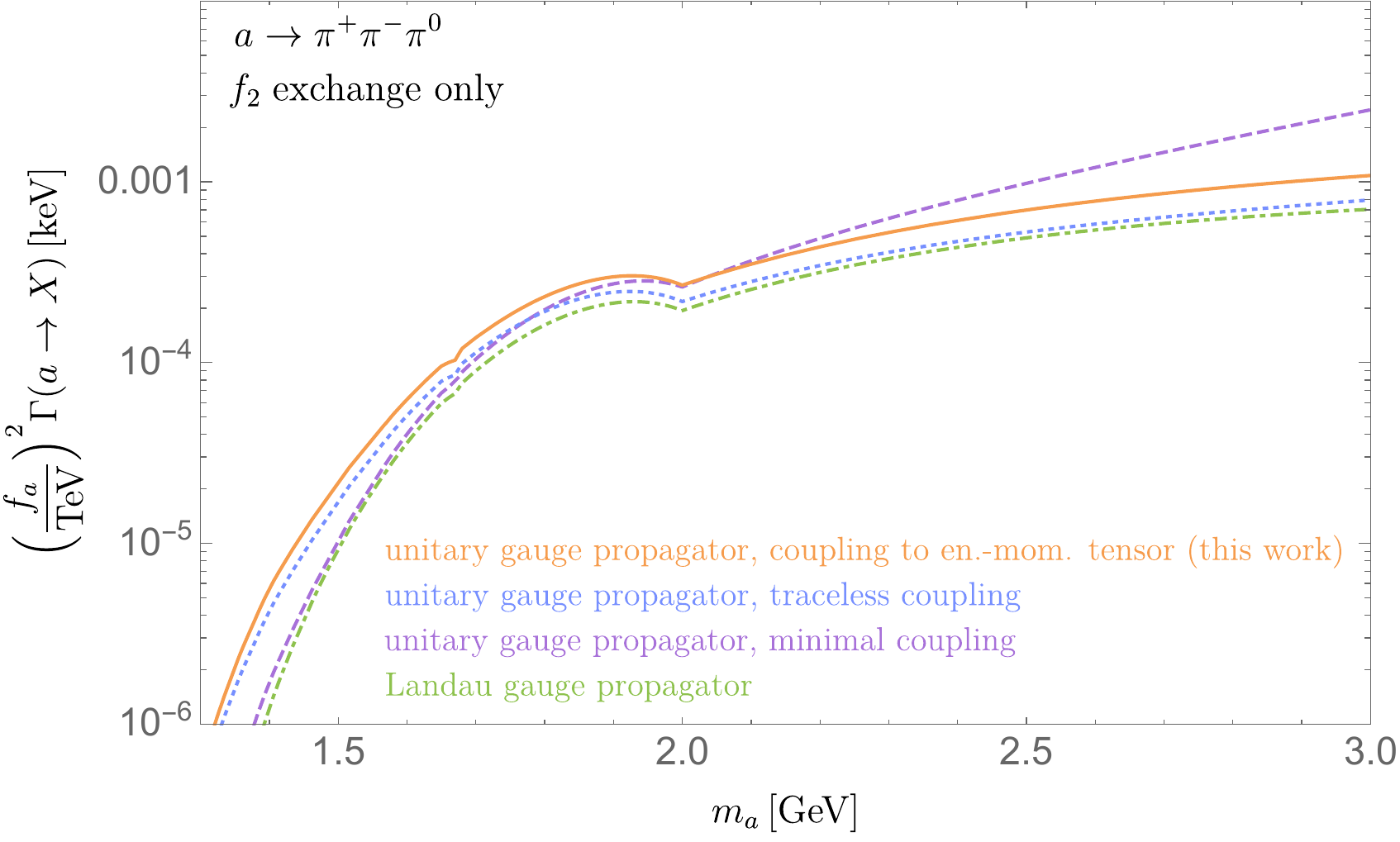}
\includegraphics[width=0.485\textwidth]{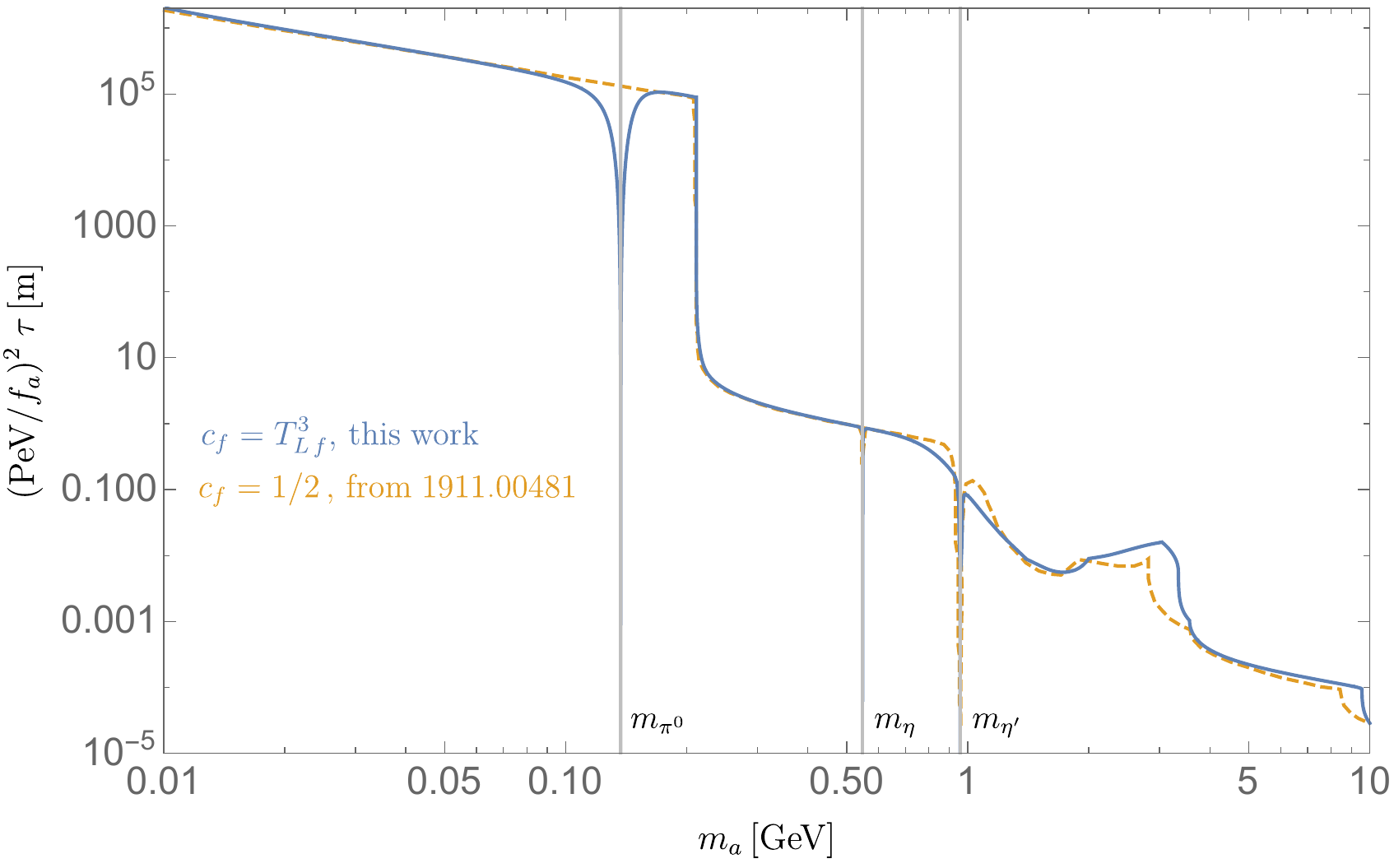}
\caption{\label{fig:f2_comparison} {\it (Left)} comparison of different treatments of the $f_2$ exchange. For all curves the amplitude is given only by Eq.~\eqref{eq:f2_pi+pi-pi0} with $c_f = T_{L f}^3$, but different prescriptions for the $f_2$ couplings and propagator lead to different expressions for $\widehat{\mathcal{M}}_{f_2}$. In solid orange, the choice made in this work: $f_2$ couples to the energy-momentum tensor and has a unitary gauge propagator, leading to Eq.~\eqref{eq:f2_pi+pi-pi0_reduced}. In dotted blue (dashed purple), alternative versions where the coefficient of the $g^{\mu\nu}$ piece in Eq.~\eqref{eq:f2_coupling} is set to $-1/4$~(0), still with unitary gauge propagator. In dot-dashed green, the version used in Ref.~\cite{Aloni:2018vki} where the propagator has the Landau gauge expression, i.e. in Eq.~\eqref{eq:B_tensor} one replaces $m_{f_2}^2 \to k^2$. For this choice, which does not seem justified, the result is independent of the coefficient of the $g^{\mu\nu}$ piece in the coupling. {\it (Right)} total lifetime obtained from our calculation with ALP-fermion couplings proportional to weak isospin, compared to the lifetime for universal couplings~\cite{Aielli:2019ivi}.} 
\end{center}
\end{figure}

The main predecessor in the study of light ALP hadronic decays is Ref.~\cite{Aloni:2018vki}, with which our analysis shares several important aspects. In particular, we adopt their choice of vertex form factors in Eq.~\eqref{eq:form_factors} to suppress the resonance exchange amplitudes at large $m_a$. There are, however, some major differences that we wish to summarize here:

\begin{itemize}
\item The key distinction is that, as we consider scenarios where the couplings to SM fermions dominate, in general the ALP has a non-trivial $U(3)$ representation for all masses up to $\approx 3$~GeV (where we match to perturbative QCD). By contrast, Ref.~\cite{Aloni:2018vki} focused on the case where the coupling to gluons dominates, therefore $\mathcal{C}_u = \mathcal{C}_d = \mathcal{C}_s$ was assumed for $m_a \gtrsim 1\;\mathrm{GeV}$. The nontrivial $U(3)$ representation of ALPs with mass above $1$~GeV implies that here the $a\to P (V\to P P)$ decays are in general unsuppressed and play a crucial role. This is clearly demonstrated by our benchmark model $c_f = T_{Lf}^3$, where $a\to \pi^\pm (\rho^\mp \to \pi^\mp \pi^0)$ dominates not only the $a\to \pi^+ \pi^- \pi^0$ amplitude, but also the {\it total} ALP width for $m_a \gtrsim 1\;\mathrm{GeV}$, as shown in Fig.~\ref{fig:ALP_widths}. This is a consequence of the sizable ALP mixing with $\pi_0$ and the strong coupling $g_{V\pi\pi} \approx 6$. Other effects of the nontrivial $U(3)$ charges include strong relative suppressions for certain channels, such as e.g., $\Gamma_{3\pi_0} \ll \Gamma_{\pi^+ \pi^- \pi^0}$ and $\Gamma_{K^{\ast +} K^{\ast -}} \ll \Gamma_{K^{\ast 0} \overline{K}^{\ast 0}}$ (see Fig.~\ref{fig:ALP_widths}). 
\item We do not assume $U(3)$ invariance to determine the scalar nonet contributions to $a\to 3P$ decays, as the results of Ref.~\cite{Fariborz:1999gr} show this to be a rather poor approximation. Instead, we make use of all the couplings fitted to data in Ref.~\cite{Fariborz:1999gr}, taking into account all relevant $a\,$-$\,P$ mixings. As a result, our amplitudes for scalar mediation agree in kinematic structure with Ref.~\cite{Aloni:2018vki}, but differ in the values of the couplings.
\enlargethispage{10pt}
\item For the tensor meson $f_2$ we assume $U(3)$ invariance with $\bm{f_2} = \mathrm{diag}\,(1,1,0)/2$ and determine the $g_{f_2 \pi\pi}$ coupling from data, as in Ref.~\cite{Aloni:2018vki}. However, we differ from that reference in that we use the unitary gauge propagator for the massive spin-$2$ field, leading to corrections to the $f_2$ contribution to $a\to 3P$ amplitudes. In addition, we fix the coefficient of the $g^{\mu\nu}$ piece in the $\partial \Sigma^\dagger \partial \Sigma f_2$ interaction (this piece does not enter the calculation of on-shell $f_2 \to \pi\pi$, so its coefficient has to be fixed from other considerations) to the value corresponding to $f_2$ coupled to the energy-momentum tensor~\cite{Katz:2005ir}, see Eq.~\eqref{eq:f2_coupling}. Finally, we turn off the $f_2$ exchange amplitudes for $m_{ij}^2 < (m_{f_2} - \Gamma_{f_2})^2$, to avoid unphysical contributions to the $O(p^4)$ terms in the chiral Lagrangian. The impact of different prescriptions for the $f_2$ couplings and propagator is shown in the left panel of Fig.~\ref{fig:f2_comparison}, considering for illustration the $a\to \pi^+ \pi^- \pi^0$ decay.
\item Other differences compared to Ref.~\cite{Aloni:2018vki} are described above for each process. These include a different treatment -- with several new contributions -- for $a\to \pi^+ \pi^- \gamma$ and the addition of further decay channels such as $a\to \eta\eta \pi^0$ and $a\to (\rho_0 \to \pi^+ \pi^-)\,\omega$.
\end{itemize}

\noindent The ALP lifetime for the scenario with universal couplings to fermions, derived from the methods of Ref.~\cite{Aloni:2018vki}, has also appeared before in the literature~\cite{Aielli:2019ivi}. In the right panel of Fig.~\ref{fig:f2_comparison} we compare it to our determination for $c_f = T_{Lf}^3$. While the results are qualitatively compatible, important quantitative differences appear for $m_a \sim m_{\pi}$ and in the region $m_\eta \lesssim m_a \lesssim 2m_c$.

\section{Chiral perturbation theory for dark pions}\label{app:dark_ChPT}
At energies below the scale of resonances, the dark pions are described using ChPT. To lowest order for $N = 2$,
\begin{equation}
\mathcal{L}_{\pid}^{(2)} \supset \frac{\fpid^2}{4} {\rm Tr}[ (D^\mu U)^\dagger D_\mu U] + \frac{\hat{B}_0 \fpid^2}{2} {\rm Tr}[ U  {\bm{ \widehat{m}}}_{\psi'}^\dagger  +  {\bm{\widehat{m}}}_{\psi'} U^\dagger]\,,
\label{eq:cpt}
\end{equation} 
where $U$ is the pion matrix transforming as $U\to L \,U R^\dagger$ under $SU(2)_L \times SU(2)_R$, $\bm{\widehat{m}}_{\psi'}$ is the generalized quark mass matrix containing also the interactions with the Higgs, and
 $\hat{B}_0$ is a non-perturbative constant that determines the dark pion masses, 
\begin{equation} \label{eq:pion_mass_chpt}
U = \exp\bigg( i\frac{\sigma_a \pid^a }{\fpid} \bigg)\,,  \qquad \bm{\widehat{m}}_{\psi'} = \bm{m}_{\psi'} - \bm{B} h \,, \qquad  m_{\pid}^2 = \hat{B}_0 {\rm Tr}( {\bm m}_{\psi'} )  ,
\end{equation}
where the form of $ \bm{\widehat{m}}_{\psi'}$ follows from Eq.~\eqref{eq:pi2_decay}. According to Eq.~\eqref{eq:zint}, the covariant derivative of $U$ takes the form
\begin{equation}
D_\mu U = \partial_\mu U - i\frac{g_Z}{2} (\bA U - U \bAt) Z_\mu\,.
\end{equation}
The above equations allow us to derive, in particular, the linear mixing between the dark pions and the $Z$, $\mathcal{L}_{\pid}^{(2)} \supset - g_Z \fpid {\rm Tr} [\sigma_a  (\bA - \bAt) ] \partial^\mu \pid^a Z_\mu/4$, and the linear mixing between the dark pions and the $h$, $\mathcal{L}_{\pid}^{(2)} \supset \hat{B}_0 f_{\hat{\pi}} \mathrm{Tr}[i\sigma_a (\bm{B} - \bm{B}^\dagger)] \hat{\pi}^a h/2$, both of which are of course consistent with the current algebra results given in Section~\ref{sec:theory_hadrons}. If $SU(2)_V$ is exact and therefore $\bA, \bAt, \bm{B} \propto \mathbf{1}_2$, all interactions of the dark pions with the $Z$ in Eq.~\eqref{eq:cpt} vanish. For the single$\,$-$\,Z$ terms this is a consequence of $\mathrm{Tr}(U^\dagger \partial^\mu U) = 0$, valid for any $N$ (see e.g., Ref.~\cite{Scherer:2002tk}).

\bibliographystyle{JHEP}
\bibliography{darkpionbib}

\end{document}